\documentclass[a4paper,11pt]{article}	
\pdfoutput=1 
\usepackage{jheppub} 

\usepackage[table]{xcolor}                  
\usepackage{colortbl}
\usepackage{booktabs}                       
\usepackage{multirow,bigdelim}              
\usepackage{longtable}                      
\usepackage{arydshln}                       

\usepackage{units}                          
\usepackage{amsmath,amssymb,mathtools}      
\usepackage{slashed}                        
\usepackage{amsthm}
\usepackage{bbm,dsfont}                     

\usepackage{bm}                             
\usepackage{upgreek}                        
\usepackage[vcentermath]{youngtab}          
\usepackage{ytableau}                       
\usepackage{empheq}                         
\usepackage{suffix}	                        
\usepackage{lscape}	                        

\usepackage{graphicx}                       
\usepackage{subfig}	                        
\usepackage{float}
\usepackage[export]{adjustbox}              

\usepackage{url}                            
\usepackage{makeidx}                       
\usepackage[numbers,sort&compress]{natbib}  
\usepackage[colorlinks=true,urlcolor=blue,anchorcolor=blue,citecolor=blue,filecolor=blue,linkcolor=blue,menucolor=blue,pagecolor=blue,linktocpage=true,pdfproducer=medialab,pdfa=true]{hyperref}	

\usepackage{ifthen}	                        

\usepackage{tikz}                           
\usetikzlibrary{calc,positioning,intersections,arrows,shapes,fit,tikzmark}
\tikzstyle{roundbox}=[rectangle, draw=black, rounded corners, text
centered, text width=3.4cm, rectangle split part align={center}]
\tikzstyle{arrow}=[->,>=stealth]
\tikzstyle{split3box} = [rectangle,rounded
corners,draw=black,rectangle split,rectangle split parts=3]

\DeclareMathOperator{\Tr}{Tr}

\DeclareMathOperator{\diag}{diag}

\def\mc{\mathcal}
\def\md{\mathbf}
\def\mf{\mathfrak}

\def\IC{\mathbb{C}}
\def\IE{\mathbb{E}}
\def\IF{\mathbb{F}}

\def\IP{\mathbb{P}}

\def\IR{\mathbb{R}}
\def\IZ{\mathbb{Z}}

\def\cL{\mathcal{L}}

\def\cO{\mathcal{O}}

\def\cS{\mathcal{S}}

\def\fg{\mathfrak{g}}

\def\fn{\mathfrak{n}}

\def\vev#1{\left\langle #1 \right\rangle}

\DeclarePairedDelimiterX\opm[3]{\langle}{\rangle}{#1 \delimsize\vert #2 \delimsize\vert #3}
\DeclarePairedDelimiterX\ip[2]{\langle}{\rangle}{#1 \delimsize\vert #2}

\def\({\left(}
\def\){\right)}
\def\[{\left[}
\def\]{\right]}

\newcommand{\be}{\begin{equation}}
\newcommand{\ee}{\end{equation}}
\newcommand{\ba}{\begin{aligned}}
\newcommand{\ea}{\end{aligned}}
\newcommand{\ben}{\begin{eqnarray}\displaystyle}
\newcommand{\een}{\end{eqnarray}}

\newcommand{\pd}{\partial}

\newcommand{\re}{{\rm e}}
\newcommand{\ri}{{\mathsf{i}}}
\newcommand{\rd}{{\rm d}}

\newcommand{\nn}{\nonumber \\}

\newcommand{\wt}[1]{\widetilde{#1}}
\newcommand{\wh}[1]{\widehat{#1}}
\newcommand{\wb}[1]{\overline{#1}}

\newcommand{\eq}{\epsilon_1}
\newcommand{\et}{\epsilon_2}

\newcommand{\und}[1]{\underline{#1}}
\newcommand{\w}{\omega}

\newtheorem{theorem}{Theorem}
\newtheorem{lemma}[theorem]{Lemma}
\newtheorem{prop}[theorem]{Proposition}

\makeindex

\title{\boldmath Elliptic Blowup Equations for 6d SCFTs. III:\\ E--strings, M--strings and Chains}
\author{Jie Gu${}^{a}$, Babak Haghighat${}^{b}$, Albrecht Klemm${}^{c,d}$, Kaiwen Sun${^e}$, Xin
  Wang${}^{c,f}$}

\affiliation{
  ${}^a$ D\'epartement de Physique Th\'eorique et Section
  de Math\'ematiques\\
  Universit\'e de Gen\`eve, Gen\`eve, CH-1211 Switzerland\\
  \\
  ${}^b$ Yau Mathematical Sciences Center, Tsinghua University, Beijing, 100084, China\\
  \\
  ${}^c$ Bethe Center for Theoretical Physics and ${}^d$Hausdorff
  Center for Mathematics
  \\Universit\"at Bonn, D-53115 Bonn\\
  \\
  ${}^e$ Scuola Internazionale Superiore di Studi Avanzati (SISSA)\\
  via Bonomea 265, 34136, Trieste, Italy\\
  \\
  ${}^f${Max Planck Institute for Mathematics\\
    Vivatsgasse 7, D-53111 Bonn, Germany} \\}

\emailAdd{jie.gu@unige.ch}
\emailAdd{babak@math.tsinghua.edu.cn}
\emailAdd{aklemm@th.physik.uni-bonn.de}
\emailAdd{ksun@sissa.it}
\emailAdd{wxin@mpim-bonn.mpg.de}

\preprint{Bonn-TH-2019-06}

\abstract{We establish the elliptic blowup equations for E--strings and
  M--strings and solve elliptic genera and refined BPS invariants from
  them. Such elliptic blowup equations can be derived from a path
  integral interpretation. We provide toric hypersurface construction
  for the Calabi-Yau geometries of M--strings and those of E--strings
  with up to three mass parameters turned on, as well as an approach
  to derive the perturbative prepotential directly from the local
  description of the Calabi-Yau threefolds. We also demonstrate how
  to systematically obtain blowup equations for all rank one 5d SCFTs
  from E--string by blow-down operations. Finally, we present blowup
  equations for E--M and M string chains.}

\keywords{keywords}

\begin{document}

\setcounter{tocdepth}{2}
\maketitle
\flushbottom

\section{Introduction}
\label{sc:intro}

Non-Lagrangian superconformal theories without gravity in various
dimensions are among the most remarkable discoveries made in the
framework of string theory.  Four dimensional examples have been
realised in limits of the Coulomb branch of $N=2$ supersymmetric
theories called Argyres-Douglas loci where electric--- and magnetic
charged states become simultaneously
massless~\cite{Argyres:1995jj,Argyres:1995xn,Argyres:2007cn}.  An
interesting sequence of superconformal rank one theories with $E_n$,
$n=6,7,8$, extended flavour symmetries has been described
in~\cite{Minahan:1996fg,Minahan:1996cj} that arises in conformal
limits of Seiberg-Witten geometries from theories which do have a
Lagrangian description~\cite{Argyres:2007cn}.  In~\cite{Kachru:1995fv,
  Klemm:1996bj,Katz:1996fh} it was understood how Seiberg-Witten
geometries emerge in Type II String compactifications in non-compact
limits of Calabi-Yau 3-folds M that decouple gravity. The local
Calabi-Yau geometry from which the $E_n$ theories
of~\cite{Minahan:1996fg,Minahan:1996cj} derive was discovered
in~\cite{Morrison:1996pp,Klemm:1996hh} in the context of F-theory. It
is a local elliptic fibred surface with 12 Kodaira $I_1$ singularities
over a base $\mathbb{P}^1$ with self intersection $(-1)$.  Shrinking
this surface in an F-theory compactification gives rise to the six
dimensional E--string theory with $(1,0)$ super conformal symmetry and
an affine $E_8$ symmetry. The sequence of 5d $E_n$
theories\footnote{This sequence can be enlarged to include all the
  other rank one SCFTs with lower rank flavor symmetries which
  correspond to lower degree del Pezzo surfaces.} is associated to
M-theory on non-compact Calabi-Yau constructed as the canonical line
bundle over del Pezzo surfaces. The latter as well as their 4d limit
can all be obtained by blowing down the E--string geometry.  A
non-critical string with $(2,0)$ super conformal symmetry arises if
the elliptic fibre is smooth over $\mathbb{P}^1$ with self
intersection $(-2)$.

In the absence of Lagrangian descriptions the BPS spectrum of these
theories is of particular interest and certain BPS indices for the
E--string compactified on $S^1$ have been obtained
in~\cite{Klemm:1996hh}, using mirror symmetry and the M-theory
interpretation of 5d BPS invariants~\cite{Gopakumar:1998jq}. Due to an
additional $U(1)_R$ symmetry in the local limit, the BPS indices can
be refined to the actual count of BPS states whose quantum numbers are
described by the central charge, in the geometric context identified
with the class $\beta\in H_2(M,\mathbb{Z})$, their mass as well as
their representation w.r.t.\ the 5d little group
$su(2)_l\times su(2)_r$.

Four methods have been developed to calculate the refined BPS spectrum
for the $S^1$ compactifications of E-- and M--string as well as a
class of interacting E-- and M--string theories which emerge in
elliptic fibrations over chains of intersecting base curves with self
intersection $(-1)$ and $(-2)$, which was described
in~\cite{Heckman:2013pva,Heckman:2015bfa} and referred to as
E--M string chains.

The B-model topological string approach calculates the BPS indices by
solving the all genus topological string amplitude using the
holomorphic anomaly equations~\cite{Bershadsky:1993cx}.  For elliptic
fibered Calabi-Yau manifolds a more powerful version of the
holomorphic anomaly equations can be solved using a modular bootstrap
ansatz in terms of meromorphic Jacobi forms~\cite{Huang:2015ada}. In
the local models this can be refined and generalised to Jacobi forms
with many elliptic
elements~\cite{Gu:2017ccq,DelZotto:2016pvm,DelZotto:2017mee} and the
ambiguities in the ansatz can be fixed by geometric vanishing
conditions on the BPS
invariants~\cite{Gu:2017ccq,DelZotto:2016pvm,DelZotto:2017mee}. Another
method solves the E--string by a modification of the elliptic genera
of auxiliary two dimensional $SO(16)$ quiver models~\cite{Kim:2014dza}
for each winding of the base $\mathbb{P}^1$.  Different quiver
descriptions whose elliptic genera solve the E--string have been found
in~\cite{Kim:2015fxa}. The third method is to use the topological
vertex with identified legs. It was applied to the M--string
in~\cite{Haghighat:2013gba} and to the E--string
in~\cite{Kim:2015jba}.

The purpose of this work is to extend the fourth method namely the
elliptic blowup equations~~\cite{Gu:2017ccq,Gu:2018gmy,Gu:2019dan}
(based on \cite{Sun:2016obh,Grassi:2016nnt,Huang:2017mis}) to the E--
and M--string and the E--M string chains.  The elliptic blowup
equations extend a method developed by Nakajima and
Yoshioka~\cite{Nakajima:2003pg,Nakajima:2005fg,Nakajima:2011} (see
later development \cite{Keller:2012da,Kim:2019uqw}) to solve the $N=2$
supersymmetric gauge theory partitions function by blowup
equations\footnote{It was already noticed in
  \cite{Edelstein:1998sp,Edelstein:1999xk} that the blowup formulas of
  Donaldson invariants \cite{MR1394968,Moore:1997pc,Marino:1998bm}
  could be used to solve the prepotential of Seiberg-Witten theory.},
that emerge from a relation between localisation results before and
after blowing up a $\mathbb{P}^1$ in space time.

The main results of the paper are the explicit form of the unity
elliptic blowup equations for the M--string (\ref{eq:M-ublowup}) the
E--string (\ref{eq:E-ublowup}) as well as the E--M string chains
(\ref{eq:EM-ublowup}). Only for the E--string one has vanishing blowup
equations given in (\ref{eq:E-vblowup}). For the E-- and M--string we
discuss their solutions and show that the elliptic blowup equations
can be solved from the set of all possible $\und{r}$-vectors using for
example genus zero BPS invariants at base degree one as input.  The
information about the input data that are sufficient to solve the
blowup equations is encoded in (\ref{eq:rd-cond}).  For the M-- and
E--string the minimal inputs for given $\und{r}$-vectors are
summarised in (\ref{Mstringdegreeinput}) and
(\ref{Estringdegreeinput}) respectively.  We also outline a new method
to solve the blowup equations using the Weyl orbit expansion which
respects the Weyl symmetries of the E--string flavour group in Section
\ref{ssec:Weylorbit}.

The elliptic generalisation of the blowup equations of Nakajima and
Yoshioka have not been derived so far. We give a physical derivation
of the elliptic blowup equations using the path integral approach to
the partition function of the 6d SCFT on
$M_6=T^2\times_{\epsilon_1,\epsilon_2} {\widehat{ \mathbb{C}^2}}$.
Here ${\widehat{ \mathbb{C}^2}}$ is the blown up target space used
in~\cite{Nakajima:2003pg} to obtain the original $N=2$ 4d gauge
theoretical blowup equations.  A virtue of our derivation presented in
Section \ref{sc:der} is that it gives a natural explanation of how the
$\Theta$-functions in the elliptic blowup equations arise in the path
integral.  As mentioned above the solutions for the BPS states of the
E--string encodes the solutions for the BPS states of the 5d $E_n$
theories that are geometrically engineered on local del Pezzo surfaces
by successive blow down limits.  We follow this approach to derive
blowup equations for refined invariants on the del Pezzo surfaces.

This paper is organized as follows: in Section \ref{sc:geom}, we give
the geometric construction of elliptic non-compact Calabi-Yau
threefolds associated to E--string and M--string theories. In Section
\ref{sc:blowups}, we present the elliptic blowup equations for
E--strings, M--strings and their higher rank theories known as E--M
string chain and M string chain. We also demonstrate how the blowup
equations for the chains reduce to those of E--strings and M--strings
themselves. In Section \ref{sc:BPS}, we show how to solve elliptic
genera of E--, M--strings from blowup equations. We use two methods
which are refined BPS expansion in Section \ref{sc:41} and the
Weyl orbit expansion in Section \ref{ssec:Weylorbit}. In Section
\ref{sc:delPezzo}, we derive the blowup equations for local del Pezzo
surfaces from those of E--strings. In Section \ref{sc:der}, we give a
derivation of the elliptic blowup equations of E--strings and M--strings
from path integral interpretation. The readers only interested in the
form of the blowup equations for E--, M--strings can directly go to
equations (\ref{eq:M-ublowup}), (\ref{eq:E-ublowup}) and
(\ref{eq:E-vblowup}).

\section{Geometry of E--,M--string theories}
\label{sc:geom}

Here we construct the embeddings of the non-compact Calabi-Yau
threefolds associated to E--, M--string theories on $T^2$ into compact
Calabi-Yau threefolds, which are hypersurfaces in toric varieties.
There are two purposes of this construction.  The first purpose is the
semiclassical partition function encoding classical geometric data
like triple intersection numbers of divisors, which are needed for the
formulation of the blowup equations, can be readily computed from such
a compact construction.  Secondly and more importantly, the construction
of compact Calabi-Yau threefolds as hypersurfaces in toric varieties
allows us to compute the instanton corrected prepotential, which as we
will see in Section~\ref{sc:BPS} is needed as extra input in order to
extract refined BPS invariants of E--, M--string theories on $T^2$
from blowup equations.  We would like to point out that in the case of
E--string, both purposes can be realised by using the mirror curve
constructed by Sakai \cite{Eguchi:2002fc,Eguchi:2002nx,Sakai:2011xg}.
But the method of toric hypersurface construction presented here can
be applied to generic rank one 6d SCFTs \cite{IV} as well.

The first purpose needs a little more explanation.  The semiclassical
partition function has the form
\begin{equation}\label{eq:Z-cls}
  Z^{\text{cls}} = \exp\(\frac{F_{(0,0)}}{\eq\et} +
  F_{(1,0)}+\frac{(\eq+\et)^2}{\eq\et} F_{(0,1)} \)
\end{equation}
with\footnote{In general the perturbative $F_{(0,0)}$ can also have
  quadratic and linear terms.  However, the quadratic terms are
  ambiguous for local geometries, and the linear terms do not
  contribute to blowup equations.}
\begin{equation}\label{eq:F-cls}
  F_{(0,0)} = \frac{1}{6}\sum_{i,j,k} \kappa_{ijk} t_i t_j t_k,\quad F_{(1,0)}
  = \sum_{i} b_i^{\text{GV}}t_i ,\quad F_{(0,1)} = \sum_{i} b_i^{\text{NS}} t_i .
\end{equation}
Here $t_i$ are K\"ahler moduli of curve classes $\Sigma_i$.  The
coefficients $\kappa_{ijk}$ of the semiclassical prepotential
$F_{(0,0)}$ are the triple intersection numbers of divisors $\Gamma_i$
Poincar\'e dual to the curves $\Sigma_i$.  They are usually defined
rigorously and computed on a compact threefold $\wt{X}$ as integration
of $(1,1)$-forms $\omega_i$ dual to $\Gamma_i$
\begin{equation}
  \kappa_{ijk} = \int_{\wt{X}} \omega_i \wedge\omega_j \wedge\omega_k,
\end{equation}
which is the reason that for the non-compact CY3 associated to E--,
M--string theories on $T^2$ we would like to construct a compact
embedding, compute $F_{(0,0)}$ there and then do a proper
decompactification limit.  We hasten to clarify that the $F_{(0,0)}$
of a 6d SCFT, at least its relevant components, does not depend on the
way the associated non-compact CY3 is embedded into a compact one,
which may not be unique.  The non-compact CY3 can locally be seen as the
neighborhood of a union of connected compact surfaces
$S = \cup_i S_i$.  It is important to distinguish between curves with
non-trivial and vanishing intersection numbers with $S$.  The curves in
the first category are either fixed components of some surfaces or
intersect with them, while the curves in the second category, which we
will call \emph{free curves}, can be moved freely away from $S$.  We
call the corresponding K\"ahler moduli the ``true'' moduli and the mass
parameters respectively, as the latter usually correspond to flavor
masses in the field theory engineered by the geometry.  The
\emph{relevant} components of $F_{(0,0)}$ are those involving at least
one true modulus, and the terms with only mass parameters decouple in
all types of calculations.  We will be only concerned with the
relevant $F_{(0,0)}$, and we will give an argument in
Section~\ref{sc:direct} that it is inherently the property of the
non-compact CY3 and does not depend on the way of its compact
embedding.  Finally we comment that the linear coefficients
$b_i^{\text{GV}},b_i^{\text{NS}}$ in \eqref{eq:F-cls} also encode some
topological information.  Although they can be computed once a compact
embedding is constructed, they are most easily fixed by consistency
condition of blowup equations as we will see in
Section~\ref{sc:blowups}.

The idea of constructing the compact embedding is very simple.  The
E--, M--string theories are 6d SCFTs with no gauge symmetry but
non-trivial flavor symmetry $E_8$, $SU(2)$, respectively.  They also
have a tensor multiplet with the labelling $\fn=1,2$, which is the
coefficient of the Dirac pairing of the non-critical strings coupled
to the tensor multiplet.  Each of the theories can be constructed by
F-theory compactified on a non-compact Calabi-Yau threefold, which are
roughly speaking elliptic fibrations over $\cO(-1)$ and $\cO(-2)$
bundles of $\IP^1$ respectively.  It is natural to regard the two
non-compact Calabi-Yau threefolds as the decompactification limit of the
familiar compact Calab-Yau threefolds, namely the ellipic fibration over $\IF_1,\IF_2$,
along the direction of the (0)-curve in the base
\cite{Haghighat:2014vxa}.  The flavor symmetry can then be realised as
weakly coupled gauge symmetry corresponding to singularity over the
(0)-curve, which becomes ungauged in the decompactification limit.  We
will illustrate this idea in the following subsections, and write down
the compact Calabi-Yau as hypersurfaces in toric varieties.

\subsection{M--string geometry}
\label{sc:M-geom}

The geometry $X$ associated to the M--string theory was constructed
concretely in
\cite{Bhardwaj:2018yhy,Bhardwaj:2018vuu,Bhardwaj:2019jtr}.  It is
locally the neighborhood of a complex surface $S$, which can be
described as follows.  We start with the Hirzebruch surface
$\IF_{0} \cong \IP^1\times \IP^1$ blown up at two generic points and
denote the resulting surface by $\IF_0^{1+1}$.  We take the
independent curves in $\IF_0^{1+1}$ to be the $\IP^1$ base $e$ and
the $\IP^1$ fiber $f$ of the Hirzebruch surface as well as the two
exceptional curves $x,y$.  Their mutual intersection numbers inside
the surface are
\begin{equation}
  e^2 = f^2 = 0, \; x^2 = y^2 = -1,\; e.f = 1,\; e.x=e.y=f.x=f.y=x.y=0.
\end{equation}
The surface $S$ is obtained by gluing two ($-1$)-curves $e-x$ and
$e-y$ together.  It can be graphically represented as
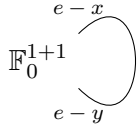
\begin{center}
  \begin{tikzpicture}
    \node (0) at (0,0) {$\IF_0^{1+1}$}; %
    \draw (0.north east) node[above=0.1]{\scriptsize$e-x$}.. controls +(1,1) and
    +(1,-1) .. (0.south east) node[below=0.1]{\scriptsize$e-y$}; %
  \end{tikzpicture}
\end{center}

The canonical class of the self-glued surface $S$ is
\cite{Bhardwaj:2018yhy}
\begin{equation}
  K_S = K_{\IF_0^{1+1}} + (e-x) + (e-y) = K_{\IF_0} + x + y + (e-x) +
  (e-y) = -2f.
\end{equation}
One can then use the adjunction formula
\begin{equation}
  2g(C)- 2 = C.K_S +  C.C
\end{equation}
to compute the genera of curves.  After the gluing, the curve class
$f$ becomes a genus one curve, and it is identified with the elliptic
fiber of $X$, while the $(0)$-curve $e$ remains rational and it is the
base of the elliptic fibration.  On the other hand, neither of the two
exceptional curves $x,y$ alone is irreducible as they have fractional
genus and they merge into a single irreducible rational $(-2)$-curve
$x+y$.  We tabulate the independent curve classes of $S$ in
Table~\ref{tb:M-curves}.  We denote the K\"ahler moduli of these curve
classes by
\begin{equation}\label{eq:M-kahler}
  t_b = \ri\text{Vol}(e),\quad \tau = \ri\text{Vol}(f),\quad 2m = \ri\text{Vol}(x+y).
\end{equation}
Inside the Calabi-Yau $X$, the intersection numbers of these curves
with the surface $S$ can be computed with the formula
\begin{equation}
  C.S = (C. K_S)_S,
\end{equation}
and the results are also given in Table~\ref{tb:M-curves}.  We find
that only $t_b$ is a true K\"ahler modulus, while both $\tau$ and $m$
are mass parameters.  We will sometimes call $m$ the flavor mass, as it
is the holonomy of the $SU(2)$ flavor symmetry on a circle.

\begin{table}
  \centering
  \begin{tabular}{*{4}{>{$}c<{$}}}
    \toprule
    & C. S & (C.C)_S & g(C)\\
    \midrule
    e & -2 & 0 & 0 \\
    f & 0 & 0 & 1 \\
    x+y & 0 & -2 & 0 \\
    \bottomrule
  \end{tabular}
  \caption{Curve classes in the M--string geometry.}
  \label{tb:M-curves}
\end{table}

\subsubsection{Toric hypersurface construction}
\label{sc:M-toric}

The M--string geometry $X$ can be embedded into a compact CY3 $\wt{X}$,
which is the anti-canonical divisor of a toric variety $\wt{Y}$.  The
toric data of $\wt{Y}$ including toric divisors, Mori cone generators,
and their intersection numbers are given in Table~\ref{tb:M-toric-1}.
This is based on the well-known toric variety whose anti-canonical
divisor is the elliptic fibration over $\IF_2$
\cite{Haghighat:2014vxa}, where in addition, we insert the exceptional
divisor $D'_v$ that results from blowing up the intersection point of
the three divisors $D_x,D_y,D_v$ (as well as the anti-canonical
divisor).  This operation effectively creates a resolved $A_1$
singularity over the $(0)$-curve in the base \cite{Esole:2019ynq}, and
it is equivalent to constructing an $A_1$ type toric top over the
$(0)$-curve \`a la \cite{Candelas:1996su,Bouchard:2003bu}.  The
non-compact CY3 $X$ is obtained by decompactifying along the
$l_{(de)} = l^{(2)}$ curve, which is the $(0)$-curve in the base.  We
identify the base curve $l_b$, the elliptic fiber $l_f$, and the curve
associated to the mass paramter $m$ to be
\begin{equation}
  l_b=l^{(3)},l_f=l^{(1)}+3l^{(3)}+3l^{(4)},\ l_m=l^{(3)}+l^{(4)}.
\end{equation}

\begin{table}
  \begin{center}
    \begin{tabular}{>{$}c<{$} *{4}{>{$}r<{$}}|*{4}{>{$}r<{$}}|}
      &\multicolumn{4}{c}{$\nu_i^*$}
      &l^{(1)}&l^{(2)}&l^{(3)}&l^{(4)}\\
      D_0  &  0& 0& 0& 0&  0& 0& 0&-2\\
      D_x  & -1& 0& 0& 0& -1& 0& 0& 1\\
      D_y  &  0&-1& 0& 0&  0& 0& 0& 1\\
      D_z  &  2& 3& 0& 0&  1&-2& 0& 0\\
      D_u  &  2& 3& 1& 0&  0& 0& 1&-1\\
      S    &  2& 3 &0&-1&  0& 1&-2& 2\\
      D_v  &  2& 3&-1&-2& -3& 0& 1& 0\\
      D_v' &  1& 2&-1&-2&  3& 0& 0&-1\\
      D_t  &  2& 3& 0& 1&  0& 1& 0& 0\\
    \end{tabular}
    \caption{Toric divisors and Mori cone generators of the toric
      variety whose anti-canonical divisor is the compact embedding of
      massive M--string geometry.}\label{tb:M-toric-1}
  \end{center}
\end{table}

The intersection ring of the compact CY3 $\wt{X}$ can be computed to be
\begin{align}
  \mathcal{R}=
  &8 J_1^3+4 J_2 J_1^2+26 J_3 J_1^2+24 J_4 J_1^2+2 J_2^2
    J_1+80 J_3^2 J_1+68 J_4^2 J_1+13
    J_2 J_3 J_1\nn
  &+12 J_2 J_4 J_1+74 J_3 J_4 J_1+242 J_3^3+189 J_4^3+40 J_2 J_3^2+34 J_2
    J_4^2+206 J_3 J_4^2+6 J_2^2 J_3\nn
  &+6 J_2^2 J_4+224 J_3^2 J_4+37 J_2 J_3 J_4
\end{align}
from which we can write down the perturbative prepotential.  The genus
0 Gopakumar-Vafa invariants
$n_0^{\und{d} = (d_b,d_f,d_m,d_{\text{de}})}$ of $\wt{X}$ can also be
computed with techniques of mirror symmetry, and the GV invariants of
$X$ are those of $\wt{X}$ with $d_{\text{de}} = 0$.  They agree with
the results of refined topological vertex, with one exception.  The
invariant\footnote{The apperance of negative curve degree is because
  $l_b,l_f,l_m$ do not constitute the Mori cone basis.}
$n_0^{(1,-1,0)}=1$ is missing, while we have a new invariant
$n_0^{(-1,1,0)}=1$.  It means that the non-compact geometry we
constructed through the toric method is slightly off in the Mori cone and
we have to bring it to the correct Mori cone chamber by flopping the curve
$-l_b+l_m$.

In addition, when we perform the decompactification limit
$t_2 \rightarrow \ri\infty$, the only B-period which remains finite is
\begin{equation}
  \Pi_B=\frac{\partial F_{(0,0)}}{\partial t_2}-2\frac{\partial
    F_{(0,0)}}{\partial t_3}+2\frac{\partial F_{(0,0)}}{\partial t_4}=t_1
  t_3+2t_3^2+t_3 t_4,
\end{equation}
In terms of the K\"ahler moduli of the non-compact geometry it reads
\begin{equation}
  \Pi_B= -2\frac{\pd F_{(0,0)}}{\pd t_b} + \frac{\pd F_{(0,0)}}{\pd t_{de}}
  =t_b^2+t_b(\tau-2 m).
\end{equation}
The relevant perturbative prepotential of the massive M--string
geometry can be computed by integrating this B-period.  After taking
into account the flop of $-l_b+l_m$ by adding $(t_b -m)^3$, we find
\begin{equation}\label{eq:M-F0}
  F_{(0,0)} = -\frac{1}{4} t_b^2 \tau+ \frac{1}{2} t_b m^2.
\end{equation}

\subsection{E--string geometry}


The non-compact CY3 associated to the E--string theory is locally the
neighborhood of a compact surface $S$, which is $\IP^2$ blown up at
nine points, also known as the half K3 surface.  The independent curve
classes include the hyperplane class $h$ of $\IP^2$, and the nine
exceptional curves $x_i$ ($i=1,\ldots,9$).  Their intersection number
within $S$ are
\begin{equation}
  h^2 = 1,\quad x_i^2 = -1, \quad h.x_i = x_i.x_j = 0,\quad i\neq j,
\end{equation}
while they intersect with $S$ by
\begin{equation}
  h.S = -3,\quad x_i.S = -1.
\end{equation}
The half K3 surface is an elliptic rational surface.  The base of the
elliptic fibration can be chosen as $b = x_9$, while the elliptic
fiber is the anti-canonical class
\begin{equation}
  f = - K_S = 3h - x_1-\ldots - x_9.
\end{equation}
This is a free curve with trivial intersection of $\cS$.
There are eight other linearly independent free curves, which we
choose to be
\begin{equation}\label{eq:ax}
  \alpha_i = x_i - x_{i+1},\; i=1,\ldots,7, \;\;\text{and}
  \;\; \alpha_8 = h-x_1-x_2-x_3.
\end{equation}
They are all $(-2)$ rational curves.  They actually correspond to the
simple roots of the Lie algebra $E_8$, as their mutual intersections
within $S$ form the negative Cartan matrix of $E_8$
\begin{equation}
  (\alpha_i.\alpha_j)_S = -A^{E_8}_{ij},\quad i,j=1,\ldots,8.
\end{equation}
Therefore they generate the $E_8$ lattice $\Lambda_{E_8}$ in
$H_2(S,\IZ)$.  The $E_8$ lattice can be embedded into $\IR^8$, which
has a standard basis $e_i$ ($i=1,\ldots 8$) with orthonormal inner
product.  In terms of this basis, we can write the $E_8$ simple roots
as
\begin{gather}
  \alpha_1 = \frac{e_1+e_8}{2} - \frac{e_2+e_3+e_4+e_5+e_6+e_7}{2}, \nn
  \alpha_i = -e_{i-1}+ e_i,\quad
  i=2,\ldots,7,\quad\text{and}\;\;\alpha_8 = e_1 + e_2.
  \label{eq:ae}
\end{gather}
We denote the K\"ahler moduli of these curves by
\begin{equation}\label{eq:E-kahler}
  t_b = \ri\text{Vol}(b),\quad \tau = \ri\text{Vol}(f),\quad  m_i =
  \ri\text{Vol}(e_i),\;i=1,\ldots,8,
\end{equation}
Here $m_i$ parameterise a vector $\und{m}$ in
$\Lambda_{E_8}\otimes\IR$, where one can define a Weyl invariant
bilinear form
\begin{equation}
  (\und{m},\und{m})_{E_8} = \sum_{i=1}^8 m_i^2 .
\end{equation}

\subsubsection{Toric hypersurface construction}
\label{sc:E-toric}


\begin{table}
 \begin{center}
   \begin{tabular}{>{$}c<{$} *{4}{>{$}r<{$}}|*{4}{>{$}r<{$}}|}
     &\multicolumn{4}{c}{$\nu_i^*$}
     &l^{(1)}&l^{(2)}&l^{(3)} &l^{(4)}\\
     D_0  &  0& 0& 0& 0&  0& 0&-2& 0\\
     D_x  & -1& 0& 0& 0& -1& 0& 1& 0\\
     D_y  &  0&-1& 0& 0&  0& 0& 1& 0\\
     D_z  &  2& 3& 0& 0&  1&-1& 0&-2\\
     D_u  &  2& 3& 1& 0&  0& 1& 0& 0\\
     S    &  2& 3& 0&-1&  0&-1& 0& 1\\
     D_v  &  2& 3&-1&-1& -3& 1& 1& 0\\
     D'_v &  1& 2&-1&-1&  3& 0&-1& 0\\
     D_t  &  2& 3& 0& 1&  0& 0& 0& 1\\
\end{tabular}
\caption{Toric divisors and Mori cone generators of the toric variety
  whose anti-canonical divisor is the compact embedding of the
  E--string geometry with one flavor mass turned on breaking
  $SU(2)\subset E_8$.}\label{tb:Estring1}
 \end{center}
\end{table}

\begin{table}
  \begin{center}
    \begin{tabular}{>{$}c<{$} *{4}{>{$}r<{$}}|*{5}{>{$}r<{$}}|}
      &\multicolumn{4}{c}{$\nu_i^*$}
      &l^{(1)}&l^{(2)}&l^{(3)}&l^{(4)}&l^{(5)}\\
      D_0  &  0& 0& 0& 0&  0& 0& 0&-3& 0\\
      D_x  & -1& 0& 0& 0&  0& 0& 0& 1& 0\\
      D_y  &  0&-1& 0& 0& -1& 0& 1& 0& 0\\
      D_z  &  2& 3& 0& 0&  1&-1& 0& 0&-2\\
      D_u  &  2& 3& 1& 0&  0& 1& 0& 0& 0\\
      S    &  2& 3& 0&-1&  0&-1& 0& 0& 1\\
      D_v  &  2& 3&-1&-1& -2& 1& 0& 1& 0\\
      \wt{D}'_v &  2& 3&-2&-2&  0& 0& 1&-2& 0\\
      D''_v&  1& 1&-1&-1&  2& 0&-2& 3& 0\\
      D_t  &  2& 3& 0& 1&  0& 0& 0& 0& 1\\
    \end{tabular}
    \caption{Toric divisors and Mori cone generators of the toric
      variety whose anti-canonical divisor is the compact embedding of
      the E--string geometry with two flavor masses turned on breaking
      $SU(3)\subset E_8$.}\label{tb:Estring2}
  \end{center}
\end{table}

\begin{table}
  \begin{center}
    \begin{tabular}{>{$}c<{$} *{4}{>{$}r<{$}}|*{6}{>{$}r<{$}}|}
      &\multicolumn{4}{c}{$\nu_i^*$}
      &l^{(1)}&l^{(2)}&l^{(3)}&l^{(4)}&l^{(5)}&l^{(6)}\\
      D_0   &  0& 0& 0& 0&  0& 0& 0& 0& 0&-1\\
      D_x   & -1& 0& 0& 0& -2& 0& 0& 0& 0& 1\\
      D_y   &  0&-1& 0& 0&  0&-1& 0& 0& 1& 0\\
      D_z   &  2& 3& 0& 0&  0& 1&-2&-1& 0& 0\\
      D_u   &  2& 3& 1& 0&  0& 0& 0& 1& 0& 0\\
      S     &  2& 3& 0&-1&  0& 0& 1&-1& 0& 0\\
      D_v   &  2& 3&-1&-1&  1&-2& 0& 1& 0& 0\\
      \wt{D}'_v  &  2& 3&-2&-2& -2& 0& 0& 0& 1& 0\\
      D''_v &  1& 1&-1&-1&  0& 2& 0& 0&-2& 1\\
      D'''_v&  0& 1&-1&-1&  3& 0& 0& 0& 0&-1\\
      D_t   &  2& 3& 0& 1&  0& 0& 1& 0& 0& 0\\
    \end{tabular}
    \caption{Toric divisors and Mori cone generators of the toric
      variety whose anti-canonical divisor is the compact embedding of
      the E--string geometry with three flavor masses turned on breaking
      $SO(7)\subset E_8$.}\label{tb:Estring3}
  \end{center}
\end{table}

Instead of the fully massive E--string geometry, we consider here the
embedding of the non-compact CY3 $X^{(n)}$ associated to the E--string
theory with $n=1,2,3$ flavor masses turned on.  It corresponds to
breaking a subgroup $SU(2), SU(3), SO(7)$ of the $E_8$ flavor group.
The non-compact geometry can also be embedded into a compact CY3
$\wt{X}^{(n)}$, which is the anti-canonical divisor of a toric variety
$\wt{Y}^{(n)}$, a blow-up of the well-known toric variety whose
anti-canonical divisor is the elliptic fiberation over $\IF_1$.  The
$\wt{Y}^{(1)}$ is obtained by blowing up along the intersection of
$D_x,D_y,D_v$ with exceptional divisor $D_v'$, while $\wt{Y}^{(2)}$ is
obtained by further blowing up along the intersection of $D_y,D_v'$
with exceptional divisor $D_v''$, $\wt{Y}^{(3)}$ by blowing up in
addition along the intersection of $D_x,D_v''$ with exceptional
divisor $D_v'''$.  Similar to the M--string geometry construction, the
result here is to create a resolve $A_1, A_2, B_3$ singularity over
the $(0)$-curve in the base \cite{Esole:2019ynq} or a $A_1, A_2, B_3$
type toric top \cite{Candelas:1996su,Bouchard:2003bu}.  See
Tables~\ref{tb:Estring1},\ref{tb:Estring2},\ref{tb:Estring3} for
details of the toric construction\footnote{In
  Tables~\ref{tb:Estring2},\ref{tb:Estring3} we have defined
  $\wt{D}'_v = D'_v + D''_v$.}.

In the case of one flavor mass, the non-compact CY3 $X^{(1)}$ is
obtained from $\wt{X}^{(1)}$ by decompactifying along the
$l_{(de)} = l^{(4)}$ curve.  The base curve $l_b$, the elliptic fiber
$l_f$, and the curve $l_{m_1}$ associated to the one flavor mass are
identified to be
\begin{equation}
  l_b = l^{(2)},\quad l_f = l^{(1)} + 3l^{(3)},\quad l_{m_1} = 2l^{(3)}.
\end{equation}
and their volumes are $t_b,t_f,t_{m_1}$.  From the intersection of
$l_{m_1}$ with $D_v,D_v'$, we find that it correponds to the simple root
$\alpha_1$ of $A_1$.  Any curve corresponding to the weight
$\w = d_{m_1} \alpha_1$ has volume $d_{m_1} t_{m_1}$.
Similarily with two flavor masses, the non-compact CY3 $X^{(2)}$ is
obtained from $\wt{X}^{(2)}$ by decompactifying along the $l_{(de)} =
l^{(5)}$ curve.
We have the identification
\begin{equation}
  l_b = l^{(2)},\quad l_f = l^{(1)}+4l^{(3)}+2l^{(4)},\quad l_{m_1} =
  l^{(4)},\quad l_{m_2} = 3l^{(3)}+l^{(4)},
\end{equation}
whose volumes are $t_b,t_f,t_{m_1},t_{m_2}$.  From the intersection of
$l_{m_1},l_{m_2}$ with $D_v,D_v',D_v''$, we find they correpond to the
simple roots $\alpha_1,\alpha_2$ of $A_2$, and thus the volume of any
curve corresponding to the weight
$\w = d_{m_1} \alpha_1 +d_{m_2} \alpha_2$ is
$d_{m_1} t_{m_1} + d_{m_2}t_{m_2}$.
Finally with three flavor masses turned on, the non-compact CY3
$X^{(3)}$ is obtained from $\wt{X}^{(3)}$ by decompactifying along
$l_{(de)} = l^{(3)}$, and we have
\begin{equation}
  l_b = l^{(4)}, \;\; l_f = 2l^{(1)}+l^{(2)}+4l^{(5)}+6l^{(6)},
  \;\; l_{m_1} = l^{(5)}+2l^{(6)},\;\; l_{m_2} = l^{(1)} + 2l^{(6)},
  \;\; l_{m_3} = l^{(5)}.
\end{equation}

We use mirror symmetry techniques to compute the genus 0 GV invariants
of $\wt{X}^{(n)}$ and extract the GV invariants of $X^{(n)}$ by
choosing those with degree zero along $l_{(de)}$.  They agree well
with the genus 0 GV invariants of the E--string theory
\cite{Huang:2013yta}.
For instance, the massless E--string theory at
$(d_b,d_f) = (1,1)$ has the invariant
\begin{equation}
  n_0^{(1,1)} = 252 = 240+12\cdot 1,
\end{equation}
which becomes the characters of the Weyl orbit $\cO_{2,240}$ and
$\cO_{0,1}$ when all flavor masses are turned on.  Here we denote by
$\cO_{n,p}$ the Weyl orbit whose size is $p$ and the norm square of
whose elements is $n$ and its character by
$\chi^{\fg}_{n,p}(\und{m})$.
We expect that when only one flavor
mass is turned on, given the branching rule of
$E_8\supset E_7\oplus \mf{su}(2)$
\begin{equation}
  \cO_{2,240} = (\cO_{2,126},1) \oplus
  (\cO_{\frac{3}{2},56},\cO_{\frac{1}{2},2}) \oplus (1,\cO_{2,2}),
\end{equation}
the $(d_b,d_f) = (1,1)$ invariant should be broken to
\begin{equation}
  n_0^{(1,1)}: 252 \to 138+ 56\chi_{\frac{1}{2},2}^{\mf{su}_2}(\und{m}) +
  \chi_{2,2}^{\mf{su}(2)} (\und{m}) =
  138+56(Q_{m_1}^{1/2}+Q_{m_1}^{-1/2})+(Q_{m_1}+Q_{m_1}^{-1}),
\end{equation}
where $Q_{m_i} = \md{e}[t_{m_i}]$ with the notation
$\md{e}[x] = \exp(2\pi\ri x)$.
When two flavor masses are turned on, with the branching rule of
$E_8\supset E_6\oplus \mf{su}(3)$\footnote{The weights in
  $\cO_{\frac{2}{3},\wb{3}}$ are opposite of those in
  $\cO_{\frac{2}{3},3}$; it is the same as the weight space of the
  irrep $\wb{\md{3}}$.}
\begin{equation}
  \cO_{2,240} = (\cO_{2,72},1)\oplus
  (\cO_{\frac{4}{3},27},\cO_{\frac{2}{3},\wb{3}})
  \oplus
  (\cO_{\frac{4}{3},\wb{27}},\cO_{\frac{2}{3},3})
  \oplus (1,\cO_{2,6}),
\end{equation}
the $(d_b,d_f) = (1,1)$ invariant should be broken to
\begin{align}
  n_0^{(1,1)}:
  252 \to
  &\,84+ 27(\chi_{\frac{2}{3},\wb{3}}^{\mf{su}_3}(\und{m}) +
    \chi_{\frac{2}{3},3}^{\mf{su}(3)} (\und{m})) +
    \chi_{2,8}^{\mf{su}(3)}(\und{m})\nn =
  &\,84+27(Q_{m_1}^{2/3}Q_{m_2}^{1/3}+Q_{m_1}^{1/3}Q_{m_2}^{2/3}+Q_{m_1}^{1/3}Q_{m_2}^{-1/3}
    +\{ Q_{m_i}\to 1/Q_{m_i}\})\nn
  &+(Q_{m_1}+Q_{m_2}+Q_{m_1}Q_{m_2}
    + \{ Q_{m_i}\to 1/Q_{m_i}\})
    .
\end{align}

When three flavor masses are turned on, with the branching rule of
$E_8\supset SO(16)\supset SO(9)\oplus SO(7)$
\begin{equation}
  \cO_{2,240} = (\cO_{2,24},1)\oplus 2(\cO_{\frac{3}{4},8},1)
  \oplus (\cO_{1,8}+2*1, \cO_{1,6})
  \oplus (\cO_{1,16},\cO_{\frac{3}{4},8})
  \oplus(1,\cO_{2,12}),
\end{equation}
the $(d_b,d_f)=(1,1)$ invariant should be broken to
\begin{align}
  n_0^{(1,1)}:
  252 \to
  &\,52+ 10\chi_{1,6}^{\mf{so}(7)}(\und{m}) +
    16\chi_{\frac{4}{3},8}^{\mf{so}(7)} (\und{m}) +
    \chi_{2,12}^{\mf{so}(7)}(\und{m}).
\end{align}
This is precisely what we find.

In addition, we computed the perturbative prepotential of
$\wt{X}^{(n)}$, and obtain that of $X^{(n)}$ by integrating over the
B-periods that remain finite in the limit $\text{Vol}(l_{(de)}) \to
\infty$, as in Section~\ref{sc:M-toric}.
We find that for $X^{(1)}$
\begin{equation}\label{eq:F0-Em1}
  F_{(0,0)} = -\frac{1}{2}t_b^2\tau - \frac{1}{2}t_b\tau^2 +
  \frac{1}{3}t_bt_{m_1}^2,
\end{equation}
for $X^{(2)}$
\begin{equation}\label{eq:F0-Em2}
  F_{(0,0)} = -\frac{1}{2}t_b^2\tau - \frac{1}{2}t_b\tau^2 +
  \frac{1}{4}t_b(t_{m_1}^2+t_{m_2}^2+t_{m_1}t_{m_2}).
\end{equation}
and for $X^{(3)}$
\begin{equation}\label{eq:F0-Em3}
  F_{(0,0)} = -\frac{1}{2}t_b^2\tau - \frac{1}{2}t_b\tau^2 +
  \frac{1}{4}t_b(t_{m_1}^2+2t_{m_2}^2+3t_{m_2}^2
  +2t_{m_1}t_{m_2}+2t_{m_1}t_{m_3}+4t_{m_2}t_{m_3}).
\end{equation}
All of them can be put in the form
\begin{equation}
  F_{(0,0)} = - \frac{1}{2}t_b^2\tau - \frac{1}{2}t_b\tau^2 +
  \frac{1}{2}t_b(\und{m},\und{m})_{\mf{g}},
\end{equation}
where
\begin{equation}
  (\und{m},\und{m})_{\mf{g}} =
  \sum_{i,j=1}^{r_{\mf{g}}} t_{m_i}t_{m_j}(\w_i,\w_j)_{\mf{g}},
\end{equation}
with
$\w_i$ being the fundamental weights and
$(\bullet,\bullet)_{\mf{g}}$ the Weyl invariant bilinear form on the
complexified weight lattice.  Generalising this, we conclude that the
perturbative prepotential of the massive E--string theory with all
eight flavor masses turned on should be
\begin{equation}\label{eq:E-F0}
  F_{(0,0)} = -\frac{1}{2}t_b^2\tau - \frac{1}{2} t_b\tau^2
  +\frac{1}{2}t_b(\und{m},\und{m})_{E_8},
\end{equation}
up to irrelevant terms.

\subsection{Direct computation of intersection numbers}
\label{sc:direct}

We argue here that the relevant perturbative prepotential is
inherently well-defined for the non-compact CY3 associated to our 6d SCFT
on the torus and it does not depend on the compact embedding of the
non-compact CY3.

A non-compact CY3 $X$ is locally the neighborhood of a union of $r$
connected compact surfaces $S = \cup_i S_i$
\cite{Bhardwaj:2018yhy,Bhardwaj:2018vuu}.  Let the number of
independent compact curve classes in $S$ be $n$, which is necessarily
greater than or equal to $r$.  Among these curves we can find $r$
independent curves $u_i$ which have non-trivial intersection numbers
with $S_i$, while the remaining $n-r$ curves $v_k$ are free curves and
have vanishing intersection numbers with any $S_i$.  We denote their
K\"ahler moduli by\footnote{Here the mass parameters $m_k$ include not
  only flavor masses but also the volume of elliptic fiber $\tau$.}
\begin{equation}
  t_i = \ri\text{Vol}(u_i),\quad m_k = \ri\text{Vol}(v_k).
\end{equation}

The $r\times r$ matrix of intersection numbers between $u_i$ and $S_j$
is of full rank since they generate dual lattices.  We would like to
enlarge the intersection matrix $(u_i. S_j)$ to an $n\times n$
full rank matrix which includes the $n-r$ free curves as well.  This
requires carefully choosing $n-r$ non-compact surfaces, and it can be
done as follows.  Among the $n-r$ free curves there is a unique genus
one curve $v_0 = f$, the elliptic fiber, and the others $v_k$
($k=1,\ldots,n-r-1$) are all rational curves.  Their self-intersection
inside $S$ can be determined by the adjunction formula to be
\begin{equation}
  (f^2)_S = 0,\quad (v_k^2)_S = -2,\quad k=1,\ldots,n-r-1.
\end{equation}
We choose the first non-compact surface $N_0$ to be the base of
elliptic fibration.  It only has non-trivial intersection numbers with
the base curve $b$ (not a free curve) and the elliptic fiber $f$
\begin{equation}\label{eq:fN0}
  b. N_0 = \fn-2,\quad f. N_0 = 1.
\end{equation}
We define additional $n-r-1$ non-compact surfaces $N_k$ by their
gluing curves with $S$
\begin{equation}
  N_k. S = v_k,\quad k =1,\ldots,n-r-1.
\end{equation}
The intersection matrix $(v_k.N_l)_{k,l=1,\ldots,n-r-1}$ is of full
rank because
\begin{equation}\label{eq:vN}
  v_k. N_l = S. N_k. N_l = (v_k.v_l)_S.
\end{equation}
The total curve-surface intersection matrix
\begin{equation}
  \begin{pmatrix}
    u_i.S_j & u_i.N_l \\
    0 & v_k.N_l
  \end{pmatrix}_{\substack{i,j=1,\ldots,r\\k,l=0,1,\ldots,n-r-1}}
\end{equation}
is block upper triangular, and its determinant factorises
\begin{equation}
  \det
  \begin{pmatrix}
    u_i.S_j & u_i.N_l \\
    0 & v_k.N_l
  \end{pmatrix}_{\substack{i,j=1,\ldots,r\\k,l=0,1,\ldots,n-r-1}}
  =
  \det
  \begin{pmatrix}
    u_i.S_j
  \end{pmatrix}_{i,j=1,\ldots,r}
  \det
  \begin{pmatrix}
    v_k.v_l
  \end{pmatrix}_{k,l=1,\ldots,n-r-1} \neq 0,
\end{equation}
where we have used \eqref{eq:fN0}.  Therefore it is also of full rank.

The K\"ahler class of $X$ can be decomposed as
\begin{equation}
  J = \sum_{i=1}^r \phi_i S_i + \sum_{k=0}^{n-r-1} \varphi_k N_k,
\end{equation}
where the coefficients $\phi_i,\varphi_k$ are related to $t_i,m_k$ by
\begin{equation}\label{eq:tm}
\begin{aligned}
  &t_i = \sum_{j=1}^r (u_i. S_j)\phi_j + \sum_{l=0}^{n-r-1}
  (u_i. N_l) \varphi_l ,\\
  &m_k = \sum_{l=0}^{n-r-1} (v_k. N_l) \varphi_l.
\end{aligned}
\end{equation}

The perturbative prepotential $F_{(0,0)}$ can be computed by integrating
B-periods.  B-periods measure in the semiclassical limit volumes of
surfaces.  Therefore in a local CY3, the only well-defined B-periods
are those associated to compact surfaces $S_j$
\begin{equation}\label{eq:sp-geom}
  \Pi_j:= \frac{\pd F_{(0,0)}}{\pd \phi_j} =
  \sum_{i=1}^r\frac{\pd F_{(0,0)}}{\pd t_i} (u_i. S_j) .
\end{equation}
Given that the matrix $(u_i. S_\alpha)$ is of full rank, if
$\frac{\pd F_{(0,0)}}{\pd \phi_j}$ are known, $\frac{\pd F_{(0,0)}}{\pd t_i}$ can
be solved from \eqref{eq:sp-geom} and further be integrated to produce
all relevant terms in $F_{(0,0)}$.  To compute $\frac{\pd F_{(0,0)}}{\pd \phi_j}$,
recall that $F_{(0,0)}$ can also be written as
\begin{equation}
  F_{(0,0)} = \frac{1}{6} J^3 = \frac{1}{6}\(\sum_{i=1}^r \phi_i
  S_i + \sum_{k=0}^{n-r-1} \varphi_k N_k\)^3,
\end{equation}
which leads to
\begin{equation}\label{eq:dF0dphi}
  \frac{\pd F_{(0,0)}}{\pd \phi_j} = \frac{1}{2}D_j.
  \(\sum_{i=1}^r \phi_i
  S_i+\sum_{k=0}^{n-r-1} \varphi_k N_k\)^2.
\end{equation}
The triple intersection numbers on the RHS of \eqref{eq:dF0dphi}
involve at least one compact surface, and they can all be converted to
intersection of curves in $S$ and thus are computable (see Section 2.6
of \cite{Bhardwaj:2018vuu}).  We then substitute $t_i,m_k$ for
$\phi_i,\varphi_k$ by inverting \eqref{eq:tm}.

To summarise, we provide here a prescription to compute
$\frac{\pd F_{(0,0)}}{\pd t_i}$ and therefore all relevant terms in $F_{(0,0)}$
from the local description of a non-compact CY3, thus proving that the
former does not depend on the compact embedding of the non-compact
CY3.  We illustrate this idea with examples.

\subsubsection{M--string}

Following the prescription in the previous section, we choose the
non-compact surface $N_0$ to be the base of the elliptic fibration, and
define an additional non-compact surface $N_1$ by its gluing
\begin{equation}
  S. N_1 = x+y.
\end{equation}
The full curve-surface intersection matrix is given in
Table~\ref{tb:M-geom-DN}, which is indeed of full rank.
We define the K\"ahler class of the M--string geomtry to be
\begin{equation}
  J = \phi S + \varphi_0 N_0 + \varphi_1 N_1
\end{equation}
whose coefficients are related to the volumes of $e,f,x+y$ by
\begin{equation}
  \(\begin{array}{c}
    t_b\\ \tau \\ 2m
  \end{array}\)
  =
  \begin{pmatrix}
    -2 & 0 & 0 \\
    0 & 1 & 0 \\
    0 & 0 & -2
  \end{pmatrix}\cdot
  \(\begin{array}{c}
    \phi \\ \varphi_1 \\\varphi_2
  \end{array}\).
\end{equation}
The only well-defined B-period measures semiclassically the volume of
$S$, and it reads
\begin{equation}\label{eq:M-F0phi}
  \frac{\pd F_{(0,0)}}{\pd \phi} = -2 \frac{\pd F_{(0,0)}}{\pd t_b}.
\end{equation}
Using the K\"ahler form representation of $F_{(0,0)}$ we can compute this B-period
\begin{align}
  \frac{\pd F_{(0,0)}}{\pd \phi} =
  &-\frac{1}{2} D. \( \phi D + \varphi_1
    N_1 +\varphi_2 N_2\)^2 \\ =
  &-2 \phi \varphi_1 - \varphi_2^2,
\end{align}
where we have use the following triple intersection numbers
\begin{equation}
  \begin{aligned}
    &S^3 = (K_S.K_S)_S = 0, \\
    &S^2.N_0 = (K_S.e)_S = -2, \\
    &S^2.N_1 = (K_S.(x+y))_S = 0, \\
    &S.N_0^2 = (e.e)_S = 0,\\
    &S.N_1^2 = ((x+y).(x+y))_S = -2,\\
    &S.N_0.N_1 = (e.(x+y))_S = 0.
  \end{aligned}
\end{equation}
Substituting $t_b,\tau,m$ for $\phi,\varphi_0,\varphi_1$, we find
\begin{equation}\label{eq:M-Pi}
  \frac{\pd F_{(0,0)}}{\pd \phi} = t_b \tau -m^2,
\end{equation}
which, together with \eqref{eq:M-F0phi}, integrates to \eqref{eq:M-F0}
up to irrelevant terms.

\begin{table}
  \centering
  \begin{tabular}{*{4}{>{$}c<{$}}}
    \toprule
    & S & N_0 & N_1\\
    \midrule
    e & -2 & 0 & 0 \\
    f &  0 & 1 & 0 \\
    x+y& 0 & 0 &-2 \\
    \bottomrule
  \end{tabular}
  \caption{The full curve-divisor intersection matrix in the M--string
    geometry.}
  \label{tb:M-geom-DN}
\end{table}

\subsubsection{E--string}

There are nine linearly independent free curves in the E--string
geometry, the elliptic fiber $f$ and the eight $(-2)$ curves
$\alpha_i$ ($i=1,\ldots,8$).
Following the prescription in the beginning of the section, we need to
first choose nine non-compact surfaces.

We choose
the base of the elliptic fibration to be the first non-compact surface
$N_0$
and define the remaining eight non-compact surfaces $N_i$ by
\begin{equation}
  S.N_i = \alpha_i ,\quad i=1,\ldots,8.
\end{equation}
$N_0$ only intersects non-trivially with the base curve and the
elliptic fiber of the elliptic fibration by
\begin{equation}
  b.N_0 = -1,\quad f.N_0 = 1.
\end{equation}
The other eight surfaces $N_i$ only intersect non-trivally with
$\alpha_i$ and the intersection numbers form the negative Cartan $E_8$
lattice according to \eqref{eq:vN}.
We display the full curve-surface intersection matrix in
Table~\ref{tb:E-DN}.

We introduce the K\"ahler class of the E--string geometry
\begin{equation}
  J = \phi S + \sum_{k=0}^{8} \varphi_i N_i,
\end{equation}
whose coefficients are related to the K\"ahler moduli $t_b,\tau,m_i$ by
\eqref{eq:tm}.
The only well-defined B-period is the one associated to $S$ and it
reads
\begin{equation}
  \frac{\pd F_{(0,0)}}{\pd \phi} = -\frac{\pd F_{(0,0)}}{\pd t_b}.
\end{equation}
Plugging in the K\"ahler class form of $F_{(0,0)}$ and using the following
triple intersection numbers
\begin{equation}
\begin{aligned}
  &S^3 = (K_S.K_S)_S = 0,\\
  &S^2.N_0 = (K_S.x_9)_S = 1,\\
  &S^2.N_i = (K_S.\alpha_i)_S = 0,\quad i=1,\ldots,8,\\
  &S.N_0^2 = (x_9.x_9)_S = -1, \\
  &S.N_i.N_j = (\alpha_i.\alpha_j)_S = -A^{E_8}_{ij},\quad i,j=1,\ldots,8,\\
  &S.N_0.N_i = (x_9.\alpha_i)_S = 0,\quad i=1,\ldots,8.
\end{aligned}
\end{equation}
and subtituting $t_b,\tau,m_i$ for $\phi,\varphi_i$, we find
\begin{equation}\label{eq:E-Pi}
  \frac{\pd F_{(0,0)}}{\pd \phi} = t_b\tau + \frac{1}{2}\tau^2 -
  \frac{1}{2}(\und{m},\und{m})_{E_8},
\end{equation}
which integrates to \eqref{eq:E-F0} up to irrelevant terms.

\begin{table}
  \centering
  \begin{tabular}{*{11}{>{$}c<{$}}}\toprule
    & S & N_0 & N_1 & N_2 & N_3 & N_4 & N_5 & N_6 & N_7 & N_8 \\
    \midrule
    b
    &-1&-1&0&0&0&0&0&0&0&0 \\
    f
    &0&1&0&0&0&0&0&0&0&0 \\
    \alpha_1
    &0&0&-2&1&0&0&0&0&0&0 \\
    \alpha_2
    &0&0&1&-2&1&0&0&0&0&0 \\
    \alpha_3
    &0&0&0&1&-2&1&0&0&0&1 \\
    \alpha_4
    &0&0&0&0&1&-2&1&0&0&0 \\
    \alpha_5
    &0&0&0&0&0&1&-2&1&0&0 \\
    \alpha_6
    &0&0&0&0&0&0&1&-2&1&0 \\
    \alpha_7
    &0&0&0&0&0&0&0&1&-2&0 \\
    \alpha_8
    &0&0&0&0&1&0&0&0&0&-2 \\
    \bottomrule
  \end{tabular}
  \caption{The full curve-surface intersection matrix of the E--string geometry.}
  \label{tb:E-DN}
\end{table}

\subsubsection{Higher rank E--,M--string theories}
\label{sc:rEM-F0}

We further illustrate the power of the direct computation to derive
the perturbative prepotentials for two higher rank 6d SCFTs, the
higher rank E--,M--string theories.

A higher rank 6d SCFT corresponds to in the F-theory compactification
multiple $\IP^1$'s in the base of elliptic fibration, and it is
characterised by the negative-definite intersection matrix $-\Omega$
of the base curves as well as the singular elliptic fibers over the
base curves.  A higher rank 6d SCFT can be obtained by properly gluing
rank one 6d SCFTs corresponding to a single base curve following
certain consistency rules.  We consider in this paper two simple
higher rank theories, higher rank M--string theory and higher rank
E--string theory.  The former corresponds to a chain of $(-2)$ curves
in the base, and the latter corresponds to in addition a $(-1)$ curve
glued to one end of the chain \cite{Gadde:2015tra}.  Both of these two
theories have no gauge symmetry. Together with rank one M--string and E--string theories, they constitute
the full list of relatively simple 6d SCFTs with no gauge symmetry. Since the geometry of higher rank M--string can be obtained from that
of higher rank E--string by decompactifying the $(-1)$ curve in the
base, we will first compute the perturbative prepotential of the
higher rank E--string, and then deduce the prepotential of higher rank
M--string as a special limit.

Consider a rank $r$ E--string theory with one $(-1)$ curve $u_0$
attached to a chain of $r-1$ $(-2)$ curves $u_i$ for $i=1,\ldots,r-1$
in the base with the intersection matrix
\begin{equation}
  -\Omega=-\(
    \begin{array}{ccccccc}
      1 & -1 & 0  & 0 & 0 & 0 & 0  \\
      -1 & 2  & -1  & 0 & 0 & 0 & 0   \\
      0 & -1 & 2 & -1 & 0 & 0 & 0  \\
      0 & 0 & -1 & \cdots & \cdots & 0 & 0 \\
      0 & 0 & 0  &\cdots & \cdots & -1 & 0 \\
      0 & 0  & 0 & 0 & -1 & 2 & -1\\
      0 & 0  & 0 & 0 & 0 & -1 & 2 \\
    \end{array}
  \).
\end{equation}
Let the K\"ahler parameters of the these base curves be
\begin{equation}
  t_0 = \ri\text{Vol}(u_0),\quad t_i = \ri\text{Vol}(u_i),\quad i=1,\ldots,r-1.
\end{equation}
There are in addition ten mass parameters: the K\"ahler parameter of
the elliptic fiber $\tau$, the flavor mass $\und{m}$ of rank eight for
the flavor symmetry $E_8$ associated to the $(-1)$ curve, and the
flavor mass $m$ for the $SU(2)$ flavor symmetry associated to the
$(-2)$ chain.  There are $r$ connected compact surfaces $S_0,S_i$ for
$i=1,\ldots,r-1$, which are pull-backs of the elliptic fibration from
the base curves $u_0,u_i$ ($i=1,\ldots,r-1$).  $S_0$ is the half K3
surface associated to the rank one E--string, and each of $S_i$
($i=1,\ldots,r-1$) is a self-glued $\IF_0^{1+1}$ associated to the
rank one M--string.  Therefore, using \eqref{eq:E-Pi} and
\eqref{eq:M-Pi}, we can write down the B-periods that measure the
volume of each compact surface
\begin{align}
  &\Pi_0 = \text{Vol}(S_0) = t_0\tau + \frac{1}{2}\tau^2
    -\frac{1}{2}(\und{m},\und{m})_{E_8}, \\
  &\Pi_i = \text{Vol}(S_i) = t_i\tau - m^2 ,\quad i=1,\ldots,r-1.
\end{align}
On the other hand, following \eqref{eq:sp-geom} the B-periods should
be related to the perturbative prepotenital by
\begin{equation}
  \Pi_j = -\sum_{i=0}^{r-1} \frac{\pd F_{(0,0)}}{\pd t_i} \Omega_{ij}.
\end{equation}
It follows that up to irrelevant terms the perturbative prepotential
of the rank $r$ E--string should be
\begin{equation}\label{eq:rE-F0}
  F_{(0,0)} = -\frac{1}{2}\sum_{i,j=0}^{r-1} \tau t_i t_j (\Omega^{-1})_{ij}
  -\frac{1}{2}\sum_{j=0}^{r-1}\tau^2 t_j(\Omega^{-1})_{0j} +
  \frac{1}{2}\sum_{i=0}^{r-1}t_i(\und{m},\und{m})_{E_8}(\Omega^{-1})_{0i}
  + m^2\sum_{j=0}^{r-1}\sum_{i=1}^{r-1}t_j(\Omega^{-1})_{ij}.
\end{equation}

To obtain the perturbative prepotential of the rank $r$ M--string, we
remove everything related to the index 0 in \eqref{eq:rE-F0} and
increase the summation upper bound to $r$:
\begin{equation}\label{eq:rM-F0}
  F_{(0,0)} = -\frac{1}{2}\sum_{i,j=1}^{r} \tau t_i t_j (\Omega^{-1})_{ij}
  + m^2\sum_{i,j=1}^{r}t_j(\Omega^{-1})_{ij}.
\end{equation}
Here $\Omega$ is the negative intersection matrix of the $(-2)$ curve
chain, and it coincides with the Cartan matrix of the Lie algebra $A_r$.

\section{Blowup equations}\label{sc:blowups}
Let us quickly review the formalism of the generalised blowup
equations \cite{Gu:2017ccq,Huang:2017mis,Gu:2018gmy,Gu:2019dan} (see
also \cite{Sun:2016obh,Grassi:2016nnt}).  Given a non-compact
Calabi-Yau threefold $X$, we denote by $C = (C_{ij})$ the matrix of
intersections between compact curve classes $[\Sigma_i]$,
$i=1,\ldots, b^c_2$ and compact divisor classes $[\mf D_j]$,
$j=1,\ldots,b_4^c$ of $X$.  We also define the vector
\begin{equation}\label{eq:R}
  \und{R}(\und{n})=C \cdot \und{n}+\und{r}/2,
\end{equation}
which parameterises the shift of K\"ahler parameters with
$\und{n}\in \mathbb{Z}^{b_4^c}$ and $\und{r}\in \mathbb{Z}^{b_4^c}$.
The integral vector $\und{r}$, which we call the $\und{r}$-field, is
consistent with the checkerboard pattern of non-vanishing refined BPS
invariants $N_{j_l,j_r}^{\beta}$, in other words, they satisfy
\begin{equation}
  2j_l + 2j_r+ 1 \equiv \und{r}\cdot\beta \quad \text{mod}\;\;2.
\end{equation}
The claim of the generalised blowup equations is that there exists a
non-empty set $\cS$ of $\und{r}$-fields such that the twisted
partition function of topological string defined by
\begin{equation}\label{eq:twist}
  \wh{Z}(\und{t},\eq,\et) = Z^{\text{cls}}(\und{t},\eq,\et)
  Z^{\text{inst}}(\und{t}+\und{r},\eq,\et) \ .
\end{equation}
satisfies the following identity\footnote{Practically one can also
  shift the $\und{r}$ field in the polynomial part but keep the
  instanton part unshifted.}
\begin{equation}
  \! \sum_{\und{n}\in \mathbb{Z}^{b_4^c}} \! (-1)^{|\und{n}|} \!
  \wh{Z}({\und{t}}+\epsilon_1 \und{R}(\und{n}),\epsilon_1,
  \epsilon_2-\epsilon_1)  \wh{Z}(\und{t}+\epsilon_2 \und{R}(\und{n})
  ,\epsilon_1-\epsilon_2, \epsilon_2) =
  \Lambda(\epsilon_1,\epsilon_2,\und{m}, \und{r})
  \wh{Z}(\und{t},\epsilon_1, \epsilon_2).\ \label{eq:blowupI}
\end{equation}
with $|\und{n}|=\sum_{i=1}^{b_4^c} n_i$.  Here we have separated the
K\"ahler parameters $\und{m}$ from the K\"ahler parameters $\und{t}$
to denote those curve classes that do not intersect with compact
divisors $[\mf D_k]$, $k=1,\ldots,{b_4^c}$, and these $\und{m}$
correspond to mass parameters in the gauge theory context, while the
other K\"ahler parameters correspond to Coulomb branch parameters,
thus are also called ``true'' parameters.  It is important here that
the coefficient $\Lambda(\epsilon_1,\epsilon_2,\und{m},\und{r})$ on
the RHS of \eqref{eq:blowupI} depends in addition to $\epsilon_{1,2}$
only on mass parameters but not true K\"ahler parameters.  We will
also make the distinction between the equations where $\Lambda$
vanishes identically and those where $\Lambda$ is non-trivial.  We
call the former the \emph{vanishing} blowup equations, and the latter
the \emph{unity} blowup equations.

For any non-compact Calabi-Yau threefold, the first interesting
question is whether such $\und{r}$-fields exist so that
\eqref{eq:blowupI} holds, and if they do, how to find all of them and
write down the corresponding blowup equations.  This is the goal of
the current section for the geometries of E--, M--string theories and
their higher rank brethrens.  In the next section, we discuss the next
question: the computation of refined BPS invariants from these
equations.

The first thing one immediately notices is that one only has to
consider the set of $\und{r}$-fields in (\ref{eq:blowupI}) modulo
\begin{equation}\label{eq:r-equiv}
  \und{r} \sim \und{r}' = \und{r} + 2 C\cdot \und{n}',\quad
  \und{n}'\in \IZ^{b_4^c},
\end{equation}
as any such a shift can be absorbed into the summation index vector
$\und{n}$.  Next, we can constrain the $\und{r}$-fields by looking at
the contribution of $Z^{\text{cls}}$ to the blowup equations.
If we divide both hand sides of \eqref{eq:blowupI} by
$\wh{Z}(\und{t},\eq,\et)$, the leading contribution of each summand is
\begin{align}
  &(-1)^{|\und{n}|}
    \frac{Z^{\rm cls}(\epsilon_1,\epsilon_2-\epsilon_1)
    Z^{\rm cls}(\epsilon_1-\epsilon_2,\epsilon_2)}{Z^{\rm cls}(\epsilon_1,\epsilon_2)}
    =
    (-1)^{|\und{n}|}\exp\Big(f_0(\und{n})(\epsilon_1+\epsilon_2) + \sum_{k=1}^{b_2^c}
    f_k(\und{n}) t_k\Big) \ ,
    \label{eq:fk}
\end{align}
where
\begin{align}
  &f_0(\und{n}) = -\frac{1}{6}
    \sum_{i,j,k=1}^{b_2^c}\kappa_{ijk}R_iR_jR_k
  +\sum_{i=1}^{b_2^c}(b_i^{\rm GV} + b_i^{\rm NS})R_i,\\
  &f_k(\und{n}) = b_k^{\rm GV} - b_k^{\rm NS}
    -\frac{1}{2}\sum_{i,j=1}^{b_2^c}\kappa_{ijk}R_iR_j ,\quad k=1,\ldots,b_2^c.
\end{align}
Let us assume that we have chosen a basis of $t_k$ such that the first
$b_4^c$ of them are true K\"ahler parameters, while the remaining
$b_2^c - b_4^c$ are mass parameters.
The function $\Lambda(\eq,\et,\und{m})$ on the RHS should be
\begin{equation}\label{eq:L-defn}
  \Lambda(\eq,\et,\und{m},\und{r}) = \sum_{\und{n}\in \mc{I}}
  (-1)^{|\und{n}|}\exp\Big(f_0(\und{n})(\epsilon_1+\epsilon_2)
  + \sum_{k=1}^{b_2^c}f_k(\und{n}) t_k\Big).
\end{equation}
Here $\mc{I}$ is the set of $\und{n}$ that ``locally'' minimize
$f_k(\und{n})$ for $k=1,\ldots,b_4^c$, which means that one cannot
find any $\und{n}'$ such that
\begin{equation}
  f_k(\und{n}') \leq f_k(\und{n}) ,\;\; k=1,\ldots,b_4^c;
  \;\; \text{at least one inequality is not saturated}.
\end{equation}
One necessary condition is that every $f_k(\und{n})$ for
$k=1,\ldots,b_4^c$ is positive semi-definite.  In addition, for
vanishing blowup equations, the RHS of \eqref{eq:L-defn} should
cancel.  For unity blowup equations, the set $\mc{I}$ should contain a
single element that minimizes $f_k(\und{n})$ for $k=1,\ldots,b_4^c$
simultaneously, and the minimal values should all be zero.  Using the
equivalence relation \eqref{eq:r-equiv}, we can assume
$\mc{I} = \{\und{0}\}$.  We find that these conditions constrain the
possible $\und{r}$-fields to a finite set $\cS'$.  For the theories
considered here, we verify that the corresponding blowup equations are
valid up to very high orders of exponentiated K\"ahler parameters; in
other words, the set $\cS'$ determined in this way is correct.

One problem of this approach of fixing the $\und{r}$-fields is that it
depends on the values of the topological data
$b_k^{\text{GV}}, b_k^{\text{NS}}$ for $k=1,\ldots,b_4^c$ associated
to true K\"ahler parameters.  Let us first write down the components
of the $\und{r}$-field: $\und{r} = (r_b,r_\tau,\und{r}_m)$, which
correspond to the base curve $b$, the fiber curve $f$, and the flavor
curves.  Recall that the component of the $\und{r}$-field associated
to a rational curve $C$ with normal bundle $\cO(-n)\oplus\cO(-2+n)$ is
$n$ modulo 2 \cite{Gu:2018gmy}.  Then $r_b,r_\tau$ must be $-\fn,0$
modulo 2, and $\und{r}_m$ twice a weight vector of $SU(2)$ or $E_8$
for M--string and E--string theories respectively\footnote{See
  sections~\ref{sc:M-eblowup},\ref{sc:E-eblowup} for detailed
  explanation for the last statement.}.  The component $r_b$ can be
further reduced by the equivalence condition \eqref{eq:r-equiv} to
within the range
\begin{equation}\label{eq:rb}
  r_b = -\fn + 2j, \quad j=0,\ldots,\fn-1.
\end{equation}
In the case of E--,M--string theories, only $t_b$ is a true K\"ahler
parameter.  The positive semi-definite condition then implies that
\begin{equation}\label{eq:rtau0}
  r_\tau = 0
\end{equation}
In addition, for unity equations, the condition that the minimum of
$f_b$ is 0 is equivalent to the solution of $\und{r}_m$ in terms of
the data $b_k^{\text{GV}},b_k^{\text{NS}}$
\begin{equation}\label{eq:rm2}
  (\frac{1}{2}\und{r}_m,\frac{1}{2}\und{r}_m)_{\mf{g}} = b^-_b.
\end{equation}
Here $(\bullet,\bullet)_{\mf{g}}$ is the Weyl invariant bilinear form
on the weight lattice of the Lie algebra $\mf{g}$ of the flavor
symmetry $G$, and we have used the notation
\begin{equation}
  b_i^{\pm} = b_i^{\text{GV}} \pm b_i^{\text{NS}}.
\end{equation}
The coefficients $b_k^{\text{GV}}$ are the second Chern class
evaluated at divisors and they can be computed once we have a toric
compact embedding of the non-compact CY3.  $b_k^{\text{NS}}$, on the
other hand, are more elusive as they are only defined in the refined
holomorphic anomaly equations \cite{Huang:2010kf}.  For the E--string
theory, the refined holomorphic anomaly equations can be formulated
using the mirror curve of Sakai \cite{Sakai:2011xg} and
$b_i^{\text{NS}}$ can be computed.  But for a generic 6d SCFT this is
difficult to do, and we solve this problem from a different angle.

Let us expand the instanton partition function in terms of the
exponentiated tensor modulus $Q_b=\md{e}[t_b]$, which is identified
with the volume of the base curve\footnote{Usually the instanton
  partition function also includes a nontrivial ``1-loop''
  contribution coming from BPS states wrapping only fibral curves.
  For theories with no gauge symmetry, this contribution is either
  absent or can be factored out of the blowup equations.}
\begin{equation}
  Z^{\text{inst}}(\und{t},\eq,\et) = 1 + \sum_{k=1}^\infty Q_b^k
  Z_k(\tau,\und{m},\eq,\et) = 1+\sum_{k=1}^\infty Q_{\text{ell}}^k
  \IE_k(\tau,\und{m},\eq,\et) .
\end{equation}
In the last equality we use the fact that the instanton partition
function $Z_k$ can be identified with the $k$-string elliptic genus
$\IE_k$ with
\begin{equation}
  Q_{\text{ell}} = Q_b Q_{\tau}^{-\frac{\fn-2}{2}}.
\end{equation}
We can expand the blowup equations in terms of $Q_{\text{ell}}$ and
obtain the following equations of elliptic genera of the E--,M--string
theories, which we call the elliptic blowup equations
\begin{align}
  \sum_{k_1+k_2=k}
  &\theta_{3,4}^{[a]}(\fn\tau,\frac{1}{2}\und{r}_m\cdot
    \und{m}+y(\eq+\et)-\fn(k_1\eq+k_2\et)) \nn[-3mm] \times
  &\IE_{k_1}(\tau,\und{m}+ \frac{1}{2}\und{r}_m\eq,\eq,\et-\eq)
    \IE_{k_2}(\tau,\und{m}+ \frac{1}{2}\und{r}_m\et,\eq-\et,\et) \nn
    =&\,\theta_{3,4}^{[a]}(\fn\tau,\frac{1}{2}\und{r}_m\cdot
    \und{m}+ y(\eq+\et))
    \IE_{k}(\tau,\und{m},\eq,\et) \label{eq:e-blowup}
\end{align}
Here the subscript of theta function is 3 if $\fn$ is even and 4 if
$\fn$ is odd.  The characteristic $a$ is given by the $r_b$ component
\begin{equation}
  a = -\frac{r_b}{2\fn} =\frac{1}{2} - \frac{j}{\fn},\quad j =0,\ldots,\fn-1.
\end{equation}
We also have
\begin{equation}
  y = \frac{1}{2}(\und{r}_m/2,\und{r}_m/2)_{\mf{g}} - \fn b^+_b.
\end{equation}

One important property of these equations is that every component is a
Jacobi form.  In particular the elliptic genus $\IE_k$ is a
meromorphic Jacobi form with modular weight 0 and modular index
polynomial \cite{Haghighat:2013gba,Gu:2017ccq}
\begin{equation}
  \text{ind}_k = -\frac{3-\fn}{4}k(\eq+\et)^2 + \frac{k(\fn k+2-n)}{2}
  \eq\et + \frac{k}{2}(\und{m},\und{m})_{\mf{g}}
\end{equation}
One natural consistency condition for \eqref{eq:e-blowup} is that
every term has the same modular index polynomial depending only on $k$
but not on $k_1,k_2$ individually, which we will call the
\emph{modularity} condition.  This imposes the constraint that
\begin{equation}\label{eq:y}
  y = \frac{\fn-1}{4} + \frac{1}{2}(\und{r}_m/2,\und{r}_m/2)_{\mf{g}},
\end{equation}
from which $b^+_b$ can be read off.  Combined with $b_b^\text{GV}$
computed using the toric embedding we constructed in
Section~\ref{sc:geom}, one can write down $b_b^{-}$ and proceed to
constrain $\und{r}_m$ of unity blowup equations with \eqref{eq:rm2}.
Alternatively, the $\und{r}$-fields of unity equations can be directly
constrained by the modularity condition to be
\begin{equation}\label{eq:rm2-mod}
  (\und{r}_m/2,\und{r}_m/2)_{\mf{g}} = \frac{7-3\fn}{2}.
\end{equation}
We point out that vanishing blowup equations can arise if the theta
function on the RHS of \eqref{eq:e-blowup} vanishes identically for
certain value of $\und{r}_m$, as we will see in
Section~\ref{sc:E-eblowup} for the E--string theory.  We write down the
elliptic blowup equations for the E--,M--string theories explicitly in
the following subsections.

\subsection{M--strings}
\label{sc:M-eblowup}


We first argue that $\frac{1}{2}\und{r}_m$ is a $\mf{su}(2)$ weight
vector.  The fact that $2m$ is associated to a rational $(-2$) curve
($x+y$), cf.~Section~\ref{sc:M-geom}, has two implications.  First the
associated $\und{r}$-field component should satisfies
\begin{equation}
  2r_{m} \equiv 0\;\;\text{mod}\;\;2,
\end{equation}
in other words, $r_m$ is an integer.  Second, it is natural to treat
$2m$ as a component of a one dimensional vector
$\und{m}\in P(\mf{su}(2))\otimes_{\IZ} \IR$ through
\begin{equation}
  2m = (\alpha_1,\und{m})_{\mf{su}(2)} \Longleftrightarrow
  \und{m} = m\,\alpha_1,
\end{equation}
where $\alpha_1$ is the simple root of $\mf{su}(2)$.  Likewise, we can
promote $r_m$ to a one-dimensional vector $\und{r}_m = r_m\alpha_1$.
Consequently, $\frac{1}{2}\und{r}_m$ must be a weight vector of
$\mf{su}(2)$.

Using the modularity constraint \eqref{eq:rm2-mod}, we conclude that
the $\und{r}_m$ for unity blowup equations can only be
\begin{equation}
  \frac{1}{2}\und{r}_m = \pm\w_1,
\end{equation}
where $\w_1$ is the fundamental weight of $\mf{su}(2)$.
The corresponding unity blowup equations read
\begin{align}
  \sum_{k_1+k_2=k}
  &\theta_{3}^{[a]}(2\tau,2(\pm\frac{m}{2}
    +\frac{\eq+\et}{4}-k_1\eq-k_2\et))
    \IE_{k_1}(\tau,m \pm \frac{\eq}{2},\eq,\et-\eq)
    \IE_{k_2}(\tau,m \pm \frac{\et}{2},\eq-\et,\et)\nn
    =
  &\,\theta_{3}^{[a]}(2\tau,2(\pm\frac{m}{2}+\frac{\eq+\et}{4}))
    \IE_{k}(\tau,m,\eq,\et) \label{eq:M-ublowup},
\end{align}
where $a = 0,-1/2$.

These equations can be checked very explicitly.  From the domain wall
picture, we know that the elliptic genera of the M--string theory are
\cite{Haghighat:2013gba}\footnote{We suppress the modular parameter of
  theta function if it is $\tau$.}
\begin{equation}
  \label{eq:Mstrpfres}
  \IE_k(\tau,m,\eq,\et)  = \sum_{|\nu|=k}
  \prod_{(i,j)\in \nu} \frac{\theta_1(z_{ij})
    \theta_1(v_{ij})}{\theta_1(w_{ij})\theta_1(u_{ij})},
\end{equation}
where
\begin{eqnarray}
  z_{ij} = -m +(\nu_i - j +1/2)\eq+(i-1/2)\et,
  & \quad
  & v_{ij} = -m-(\nu_i-j+1/2)\eq-(i-1/2)\et,  \nonumber\\
  w_{ij} = (\nu_i -j +1)\eq -(\nu_j^t -i) \et,
  & \quad
  & u_{ij} = (\nu_i-j)\eq -(\nu^t_j - i +1) \et.
\end{eqnarray}
In particular, the one-string elliptic genus is
\cite{Haghighat:2013gba}
\begin{equation}\label{eq:MZ1}
  \IE_1(\tau,m,\eq,\et)=
  \frac{\theta_1(\tfrac{1}{2}(\eq+\et)+m)\theta_1(\tfrac{1}{2}(\eq+\et)-m)}
  {\theta_1(\eq)\theta_1(\et)}.
\end{equation}
Substituting (\ref{eq:MZ1}) into \eqref{eq:M-ublowup}, we find the
unity blowup equations at base degree one is equivalent to
\begin{align}
  &\frac{\theta_3^{[a]}(2\tau,\mp{m}+(3\eq-\et)/2)\theta_1(\et/2+(m\pm{\eq}/{2}))\theta_1(\et/2-(m\pm{\eq}/{2}))}{\theta_1(\eq)\theta_1(\et-\eq)}\nn+
  &\frac{\theta_3^{[a]}(2\tau,\mp{m}+(-\eq+3\et)/2)\theta_1(\eq/2+(m\pm{\et}/{2}))\theta_1(\eq/2-(m\pm{\et}/{2}))}{\theta_1(\eq-\et)\theta_1(\et)}\nn=
  &\frac{\theta_3^{[a]}(2\tau,\mp{m}-(\eq+\et)/2)\theta_1((\eq+\et)/2+m)\theta_1((\eq+\et)/2-m)}{\theta_1(\eq)\theta_1(\et)},
\end{align}
which we have checked up to $Q_\tau^{10}$.  Using
(\ref{eq:Mstrpfres}), we have also checked the unity blowup equations
of base degrees two and three up to $Q_\tau^{10}$.

We note that M--string theory has no vanishing blowup equations, as
there is no value of $\und{r}_m$ with which the theta function on the
RHS of \eqref{eq:M-ublowup} vanishes identically.

\subsection{E--strings}
\label{sc:E-eblowup}


We start by arguing that the sub-vector $\und{r}_m$ of the
$\und{r}$-field associated to the flavor mass $\und{m}$ is twice a
weight vector of $E_8$.  Let $r_{m_i}$ ($i=1,\ldots,r$) be the
components of $\und{r}_m$ associated to $e_i$ just like $m_i$,
cf.~\eqref{eq:E-kahler}.  Since the curves $\alpha_i$ ($i=1,\ldots$)
given by \eqref{eq:ae} are rational $(-2)$-curves, the components
$r_{m_i}$ must satisfy
\begin{equation}
  \begin{aligned}
    r_{m_1} + r_{m_2}\equiv 0 \;\;
    &\text{mod}\;\; 2, \\
    r_{m_i} - r_{m_{i+1}}\equiv 0\;\;
    &\text{mod}\;\; 2,\quad i=1,\ldots,7, \\
    \frac{r_{m_1}+r_{m_8}}{2} - \frac{r_{m_2}+\ldots+r_{m_7}}{2}
    \equiv 0\;\; &\text{mod}\;\;2.
\end{aligned}
\end{equation}
These conditions are equivalent to that $\frac{1}{2}r_{m_i}$ are
either all integers or all half integers and that
$\sum_{i=1}^8 \frac{1}{2}r_{m_i}$ is an even number; in other words
\begin{equation}
  \frac{1}{2}\und{r}_m = \(\frac{1}{2}r_{m_1},\ldots,\frac{1}{2}r_{m_8}\)
\end{equation}
is a vector in the $E_8$ weight lattice.

Using the modularity condition \eqref{eq:rm2-mod}, we find that
$\frac{1}{2}\und{r}_m$ for unity blowup equations can be any of the
240 $E_8$ weight vectors whose norm square is 2;

in other words, we have
\begin{equation}
  \frac{1}{2}\und{r}_m \in \cO_{2,240}
\end{equation}
for unity blowup equations of E--string theory.  The equation itself
reads
\begin{align}
  \sum_{k_1+k_2=k}
  &\theta_{1}(\tau,\frac{1}{2}\und{r}_m\cdot
    \und{m}+\eq+\et-k_1\eq-k_2\et)
    \IE_{k_1}(\tau,\und{m}+ \frac{\eq}{2}\und{r}_m,\eq,\et-\eq)
    \IE_{k_2}(\tau,\und{m}+ \frac{\et}{2}\und{r}_m,\eq-\et,\et)\nn
    =
  &\,\theta_{1}(\tau,\frac{1}{2}\und{r}_m\cdot \und{m}+\eq+\et)
    \IE_{k}(\tau,\und{m},\eq,\et). \label{eq:E-ublowup}
\end{align}

These equations can be checked explicitly.  Using the expression of
the one-string elliptic genus \cite{Gu:2017ccq}
\begin{equation}\label{eq:Z1E}
  \IE_1(\tau,\und{m},\eq,\et) = -\left(\frac{A_1(\und{m})}{\eta^8}\right)
  \frac{\eta^{2}}{\theta_1(\epsilon_1)\theta_1(\epsilon_2)},
\end{equation}
where $A_1(\und{m}) = \Theta_{E_8}(\tau,\und{m})$ is the $E_8$ theta
function, the unity blowup equation at base degree one reads
\begin{equation}
  \frac{\theta_1(\frac{1}{2}\und{r}_m\cdot\und{m}+\et)
    A_1(\und{m}+\frac{1}{2}\und{r}_m\eq)}
  {\theta_1(\eq)\theta_1(\et-\eq)}
  +\frac{\theta_1(\frac{1}{2}\und{r}_m\cdot\und{m}+\eq)
    A_1(\und{m}+\frac{1}{2}\und{r}_m\et)}
  {\theta_1(\eq-\et)\theta_1(\et)}
  =\frac{\theta_1(\frac{1}{2}\und{r}_m\cdot\und{m}+\eq+\et)
    A_1(\und{m})}{\theta_1(\eq)\theta_1(\et)},
\end{equation}
which we have verified to very high orders of $Q_\tau$.
We have also verified the unity blowup equation at base degree two.

On the other hand, if we choose $\und{r}_m = 0$, the parameter $y$ and
then the entire RHS of \eqref{eq:e-blowup} vanishes.  The resulting
vanishing blowup equation has been presented and verified in
\cite{Gu:2017ccq,Huang:2017mis}.  We include it here as well for
completeness
\begin{align}
  \sum_{k_1+k_2=k}
  \theta_{1}(\tau,k_1\eq+k_2\et) \IE_{k_1}(\tau,\und{m},\eq,\et-\eq)
  \IE_{k_2}(\tau,\und{m},\eq-\et,\et)
  = 0.
  \label{eq:E-vblowup}
\end{align}
It has been verified up to base degree $k=3$ for high orders of $Q_\tau$. Since there is no shift for the $E_8$ parameters, it is easy to see the above equation is also the vanishing blowup equations for massless E--string theory. In fact, it is the unique blowup equation for massless E--string.

\subsection{Higher rank E-- and M--strings}

Here we construct the first instances of blowup equations for higher
rank 6d SCFTs.  The blowup equations of a higher rank 6d SCFT can in
principle be obtained by gluing those of rank one theories.

We construct blowup equations for the simplest higher rank 6d SCFTs,
the higher rank M--,E--strings, whose perpurbative prepotentials have
been computed in Section~\ref{sc:rEM-F0}, using the blowup equations
of M--,E--string theories as constituents.  We will report the progress
on constructing blowup equations for generic higher rank 6d SCFTs
through gluing in the near future \cite{V}.

We first introduce a central ingredient of our construction.  Given
the lattice $\Gamma$ of $r$ base curves and a symmetric positive
definite bilinear form on $\Gamma$ defined by matrix $\Omega$, we can
naturally define a family of generalised theta functions with
$\und{z}\in\IC^r$
\begin{equation}\label{eq:Theta}
  \Theta_{\Omega}^{[\und{a}]}(\tau,\und{z}) =
  \sum_{\und{n}\in\und{a}+\IZ^r} (-1)^{\und{n}\cdot\diag(\Omega)}
  \md{e}\[\frac{1}{2}\und{n}\cdot\Omega\cdot\und{n}\;\tau
    + \und{n}\cdot\Omega\cdot\und{z}\]
\end{equation}
where $\diag(\Omega)$ is the diagonal vector, and
\begin{equation}\label{eq:a}
  \und{a} \in \Omega^{-1}\cdot\(\frac{1}{2}\diag(\Omega)+\IZ^r\)
\end{equation}
It is clear that any two $\und{a}$ differ by an integer vector are
equivalent.  The number of inequivalent $\und{a}$-vectors is
$\det(\Omega)$.  The generalised theta function has the modular index
polynomial
\begin{equation}\label{eq:ind-Omega}
  \text{ind}_\Omega = \frac{1}{2}\und{z}\cdot\Omega\cdot\und{z}.
\end{equation}

Consider the rank $r>1$ M--string theory whose matrix $\Omega$ is the
Cartan matrix of $SU(r+1)$.  Let $m$ be the $SU(2)$ flavor symmetry,
and $\IE_{\und{k}}(\tau,m,\eq,\et)$ be the elliptic genus with
wrapping numbers $\und{k}=(k_1,\ldots,k_r)$ of the base curves.  The
idea to construct blowup equations of this theory is to ``glue'' the
blowup equations for each individual $(-2)$ base curves by merging the
theta functions $\theta_3^{[a]}$ in those equations into
$\Theta_\Omega$.  The type of the resulting new equations can be
determined by the following simple rule.  We obtain a unity blowup
equation if all the constituent blowup equations are of the unity
type, and a vanishing blowup equation if one of the constituent blowup
equations is of the vanishing type.  Schematically we have
\begin{equation}\label{eq:UV}
  \mathrm{U}\star\mathrm{ U }= \mathrm{U},\quad \mathrm{U} \star \mathrm{V} = \mathrm{V},\quad \mathrm{V}\star \mathrm{V} = \mathrm{V}.
\end{equation}
Since the rank one M--string theory has only unity blowup equations,
the higher rank M--string has also only unity blowup equations, whose
exact form can be derived from the semiclassical data \eqref{eq:rM-F0}
to be
\begin{align}
  \sum_{\und{k}'+\und{k}''=\und{k}}
  &\Theta_{\Omega}^{[\und{a}]}(\tau,\und{M}_u-\und{k}'\eq-\und{k}''\et)
    \IE_{\und{k}'}(\tau,m+\frac{s}{2}\eq,\eq,\et-\eq)
    \IE_{\und{k}''}(\tau,m+\frac{s}{2}\et,\eq-\et,\et) \nn
    =
  &\,\Theta_{\Omega}^{[\und{a}]}(\tau,\und{M}_u-\und{k}'\eq-\und{k}''\et)
    \IE_{\und{k}}(\tau,m,\eq,\et),
    \label{eq:M2-ublowup}
\end{align}
where
\begin{equation}
  \und{M}_u = \Omega^{-1}\cdot (sm+\frac{\eq+\et}{2},\ldots,sm+\frac{\eq+\et}{2}),
\end{equation}
with $s=\pm 1$.  The characteristic $\und{a}$ takes the value in
\eqref{eq:a}, and their total number is $\det(\Omega) = r+1$.

These blowup equations can be checked in various ways.  Using the
modular index polynomial of $\IE_{\und{k}}$ of the higher rank
M--string \cite{Haghighat:2013gba,Gu:2017ccq,Haghighat:2017vch}
\begin{equation}
  \text{ind}_{\und{k}}^{\text{M}^r} = -\frac{(\eq+\et)^2}{4}\sum_{i=1}^rk_i +
  \frac{\eq\et}{2} \und{k}\cdot\Omega\cdot\und{k} + m^2\sum_{i=1}^r k_i
\end{equation}
and \eqref{eq:ind-Omega}, one can find easily that the modularity
condition is satisfied.  Furthermore, we have verified these equations
at $\und{k}=(1,1)$ to high degrees of $Q_\tau$ with the explicit
expressions of $\IE_{\und{k}}$ in \cite{Haghighat:2013gba}.  Finally,
it is possible to demonstrate that these equations reduce properly to
the blowup equations of rank one M--string theory.  We will use the
shorthand notation that for a theory $T$,
\begin{equation}
  \mathrm{V}_T^{[\und{a}]}=0,\quad \mathrm{U}_T^{[\und{a}]} = 0
\end{equation}
denote the vanishing and the unity blowup equations with
characteristic $\und{a}$ respectively, where in the latter case we
have moved the two sides of the equation together.  Let us consider
the M-M chain and decompactify the $(-2)$ curve on the right.  We can
choose the inequivalent characteristics $\und{a}$ of the unity blowup
equations to be $\und{a}=(0,0)$, $(1/3,2/3)$, $(2/3,1/3)$, with the
corresponding equations denoted by
\begin{equation}
  \mathrm{U}_{\rm MM}^{[0,0]}=0,\quad
  \mathrm{U}_{\rm MM}^{[\frac{1}{3},\frac{2}{3}]}=0,\quad
  \mathrm{U}_{\rm MM}^{[\frac{2}{3},\frac{1}{3}]}=0.
\end{equation}
We can decompactify the $(-2)$ curve on the right by setting
$k_2,k_2',k_2''$ to zero.  Then the two dimensions in the summation in
$\Theta_{\Omega}^{[a]}$ decouple.  It is easy to deduce that in this
limit
\begin{equation}
\begin{aligned}
  0=\mathrm{U}_{\rm MM}^{[0,0]}
  &=\theta_3^{[0]}(6\tau,3z)\mathrm{U}_{\rm M}^{[0]}+\theta_3^{[-\frac{1}{2}]}(6\tau,3z)\mathrm{U}_{\rm M}^{[\frac{1}{2}]},\\
  0=\mathrm{U}_{\rm MM}^{[\frac{1}{3},\frac{2}{3}]}
  &=\theta_3^{[\frac{1}{3}]}(6\tau,3z)\mathrm{U}_{\rm M}^{[0]}+\theta_3^{[-\frac{1}{6}]}(6\tau,3z)\mathrm{U}_{\rm M}^{[\frac{1}{2}]},\\
  0=\mathrm{U}_{\rm MM}^{[\frac{2}{3},\frac{1}{3}]}
  &=\theta_3^{[\frac{2}{3}]}(6\tau,3z)\mathrm{U}_{\rm M}^{[0]}+\theta_3^{[\frac{1}{6}]}(6\tau,3z)\mathrm{U}_{\rm M}^{[\frac{1}{2}]},\\
\end{aligned}
\end{equation}
where $z=sm+(\eq+\et)/2$, $s=\pm 1$.  Since this is clearly a
full-rank system for $\mathrm{U}_{\rm M}^{[0]}$ and
$\mathrm{U}_{\rm M}^{[1/2]}$, we conclude
\begin{equation}
  \mathrm{U}_{\rm M}^{[0]}=0,\quad\quad \mathrm{U}_{\rm M}^{[\frac{1}{2}]}=0.
\end{equation}
These are exactly the unity blowup equations for M--string, as we
already know.  Similar situation happens when $\mathrm{M}^r$ chain
reduces to $\mathrm{M}^{r-1}$ chain.

Let us move onto the rank $r>1$ E--string theory.
The matrix $\Omega$ is
\begin{equation}
  \Omega=\left(
    \begin{array}{ccccccc}
      1 & -1 & 0  & 0 & 0 & 0 & 0  \\
      -1 & 2  & -1  & 0 & 0 & 0 & 0   \\
      0 & -1 & 2 & -1 & 0 & 0 & 0  \\
      0 & 0 & -1 & \cdots & \cdots & 0 & 0 \\
      0 & 0 & 0  &\cdots & \cdots & -1 & 0 \\
      0 & 0  & 0 & 0 & -1 & 2 & -1\\
      0 & 0  & 0 & 0 & 0 & -1 & 2 \\
    \end{array}
  \right).
\end{equation}
where the lower right $(r-1)\times (r-1)$ submatrix is the Cartan
matrix of $SU(r)$, which will be denoted by $\hat{\Omega}$.  Let
$\und{m}$ and $m$ be the $E_8$ and $SU(2)$ flavor masses respectively,
and $\IE_{\und{k}}(\tau,\und{m},m,\eq,\et)$ with
$\und{k}=(k_0,k_1,\ldots,k_{r-1})$ be the elliptic genus with wrapping
number $k_0$ on the $(-1)$ base curve and wrapping numbers
$\hat{\und{k}}=(k_1,\ldots,k_{r-1})$ on the $(-2)$ curves.  The blowup
equations of this theory is again constructed by merging the theta
functions in the constituent blowup equations of rank one E--,M--string
theories to $\Theta_\Omega$.  Following the rule \eqref{eq:UV}, we
expect vanishing blowup equations constructed from vanishing equations
of the E--string theory and unity equations of the M--string theory, and
unity blowup equations constructed from unity equations of both the
E--,M--string theories.  The exact forms of these blowup equations can
be derived from the semiclassical data \eqref{eq:rE-F0}.  The
vanishing blowup equations of the rank $r$ E--string theory read
\begin{equation}
  \sum_{\und{k}'+\und{k}''=\und{k}}
  \Theta_{\Omega}^{[\und{a}]}(\tau,\und{M}_v-\und{k}'\eq-\und{k}''\et)
  \IE_{\und{k}'}(\tau,\und{m},m+\frac{s}{2}\eq,\eq,\et-\eq)
  \IE_{\und{k}''}(\tau,\und{m},m+\frac{s}{2}\et,\eq-\et,\et)
  = 0
  \label{eq:EM-vblowup}
\end{equation}
where $s=\pm 1$ and
\begin{equation}
  \und{M}_v = \Omega^{-1}\cdot (0,sm+\frac{\eq+\et}{2},\ldots,sm+\frac{\eq+\et}{2}).
\end{equation}
The unity blowup equations read
\begin{align}
  \sum_{\und{k}'+\und{k}''=\und{k}}
  &\Theta_{\Omega}^{[\und{a}]}(\tau,\und{M}_u-\und{k}'\eq-\und{k}''\et)
    \IE_{\und{k}'}(\tau,\und{m}+\alpha\eq,m+\frac{s}{2}\eq,\eq,\et-\eq)
    \IE_{\und{k}''}(\tau,\und{m}+\alpha\et,m+\frac{s}{2}\et,\eq-\et,\et)\nn
    =
  &\,\Theta_{\Omega}^{[\und{a}]}(\tau,\und{M}_u)
    \IE_{\und{k}'}(\tau,\und{m},m\eq,\et)
    \label{eq:EM-ublowup}
\end{align}
where $s=\pm 1$ and $\alpha$ is one of the 240 roots of $E_8$, and
\begin{equation}
  \und{M}_v = \Omega^{-1}\cdot
  (\alpha\cdot\und{m}+\eq+\et,sm+\frac{\eq+\et}{2},\ldots,sm+\frac{\eq+\et}{2}).
\end{equation}
In both equations, $\und{a}$ is unique and it can be writte as
\begin{equation}
  \und{a} = \Omega^{-1}\cdot (\frac{1}{2},0,\ldots,0).
\end{equation}

We verify these blowup equations in the following ways.
First of all, using the modular index polynomial of $\IE_{\und{k}}$
\begin{equation}
  \text{ind}_{\und{k}}^{\text{EM}^{r-1}} =
  -\frac{(\eq+\et)^2}{4}(2k_0 + \sum_{i=1}^{r-1}k_i)
  +\frac{\eq\et}{2}(\hat{\und{k}}\cdot\hat{\Omega}\cdot\hat{\und{k}}+k_0)
  +\frac{k_0}{2}(\und{m},\und{m})_{E_8} + m^2\sum_{i=1}^{r-1}k_i.
\end{equation}
and \eqref{eq:ind-Omega}, we find
\eqref{eq:EM-vblowup},\eqref{eq:EM-ublowup} satisfy the modularity
condition.  Furthermore, we verified these equations at
$\und{k}=(1,1)$ up to high orders of $Q_\tau$ with the explicit
expressions of $\IE_{\und{k}}$ given in \cite{Gadde:2015tra}.
Finally, we demonstrate that the blowup equations of the rank two
E--string, or the E--M chain, can be reduced to the blowup equations of
E--,M--string theories by decompactifying base curves.  The blowup
equations of the E--M chain all have a unique characteristic which we
choose to be $\und{a}=(0,1/2)$.  Let us first decompactify the $(-1)$
curve by setting $k_0 = 0$.  The vanishing blowup equations of the E--M
chain become in this limit
\begin{equation}
  0=\mathrm{V}_{\rm EM}^{[0,\frac{1}{2}]}=
  \theta_3(2\tau,sm+(\eq+\et)/2)\cdot
  \mathrm{U}_{\rm M}^{[-\frac{1}{2}]}-
  \theta_3(2\tau,sm+(\eq+\et)/2)\cdot\mathrm{U}_{\rm M}^{[0]},
\end{equation}
while the unity blowup equations become
\begin{equation}
  0=\mathrm{U}_{\rm EM}^{[0,\frac{1}{2}]}=
  \theta_2(2\tau,2\und{m}\cdot
  \alpha+sm+5(\eq+\et)/2)\cdot
  \mathrm{U}_{\rm M}^{[-\frac{1}{2}]}
  -\theta_3(2\tau,2\und{m}\cdot\alpha
  +sm+5(\eq+\et)/2)\cdot\mathrm{U}_{\rm M}^{[0]}.
\end{equation}
Since $s=\pm 1$ and $\alpha\in\Delta(E_8)$, we have a full rank system
for $\mathrm{U}_{\rm M}^{[-\frac{1}{2}]}$ and
$\mathrm{U}_{\rm M}^{[0]}$, and therefore
$\mathrm{U}_{\rm M}^{[-\frac{1}{2}]}=0$ and
$\mathrm{U}_{\rm M}^{[0]}=0$, which are the unity blowup equations of
the M--string as we know.
Next we decompactify the $(-2)$ curve by setting $k_1 = 1$.  The
vanishing blowup equations of the E--M chain become in this limit
\begin{equation}
  0=\mathrm{V}_{\rm EM}^{[0,\frac{1}{2}]}=
  \theta_2(\tau,sm+(\eq+\et)/2)\cdot\mathrm{V}_{\rm E}^{[-\frac{1}{2}]}.
\end{equation}
Thus $\mathrm{V}_{\rm E}^{[-\frac{1}{2}]}=0$, which is the vanishing
elliptic blowup equation for E--string as we know.  The unity blowup
equations of E--M chain become
\begin{equation}
  0=\mathrm{U}_{\rm EM}^{[0,\frac{1}{2}]}=
  \theta_2(\tau,\und{m}\cdot\alpha+sm+3(\eq+\et)/2)
  \cdot\mathrm{U}_{\rm E}^{[-\frac{1}{2}]}.
\end{equation}
Thus $\mathrm{U}_{\rm E}^{[-\frac{1}{2}]}=0$, which are the 240 unity
elliptic blowup equations for E--string.

We comment that in general, the blowup equations of a higher rank
theory do not necessarily reduce to all the blowup equations of the
blowup equations of lower rank blowup equations due to the gluing
rules.  We will give a more detailed discussion in our future work
\cite{V}.

\section{Solution of blowup equations}
\label{sc:BPS}

Here we would like to argue that the elliptic genera of the
E--,M--string theories, or equivalently the refined BPS invariants of
the associated geometries, can be solved completely from their blowup
equations.

To see this quickly, we quote the following statement from
\cite{Huang:2017mis}, obtained by counting the number of equations
satisfied by the refined topological string free energies $F_{(n,g)}$
extracted from blowup equations through expansion in terms of
$\eq,\et$.  Given a generic local Calaib-Yau threefold $X$ with
$b_2^c$ K\"ahler parameters, and let $w_u, w_v$ be the numbers of
inequivalent unity and vanishing blowup equations
respectively.\footnote{In \cite{Huang:2017mis} $w_u,w_v$ are defined
  to be the numbers of inequivalent $\und{r}$-fields which are in
  addition \emph{not} reflective to each other.  This additional
  condition is actually not necessary.} Then all the refined BPS
invariants on $X$ can be completely solved from the blowup equations
with the input of $F_{(0,0)}$ if
\begin{equation}\label{eq:wuv}
  w_u \geq 1 \quad\text{and}\; w_u+w_v \geq b_2^c.
\end{equation}
For the E--string theory with $b_2^c = 10$ and
\begin{equation}
  w_u = 240, \quad w_v = 1,
\end{equation}
and the M--string theory with $b_2^c = 3$ and
\begin{equation}
  w_u = 4,\quad w_v = 0,
\end{equation}
the condition \eqref{eq:wuv} is clearly satisfied.

Next we corroborate the solvability of the E--,M--string theories by
presenting two explicit algorithms to solve the two theories from the
blowup equations.

\subsection{Extraction of refined BPS invarianys}
\label{sc:41}

The first method is to extract relations of refined BPS invariants
from unity blowup equations.  The instanton partition function
$Z^{\text{inst}}$ encodes the BPS invariants by
\cite{Huang:2010kf,Iqbal:2007ii}
\begin{equation}
  Z^{\text{inst}} = \exp{\Bigg[
  \sum_{j_L,j_R=0}^\infty\sum_{\beta}\sum_{w=1}^\infty(-1)^{2(j_L+j_R)}
  \frac{N^{\beta}_{j_L,j_R}}{w} f_{(j_L,j_R)}(q_1^w,q_2^w)Q^{w\beta}
  \Bigg]}.
\end{equation}
Here we define
\begin{equation}
  f_{(j_L,j_R)}(q_1,q_2) = \frac{\chi_{j_L}(q_L)\chi_{j_R}(q_R)}
  {(q_1^{1/2}-q_1^{-1/2})(q_2^{1/2}-q_2^{-1/2})},
\end{equation}
where $\chi_j(q)$ is the $SU(2)$ character given by
\begin{equation}
  \chi_j(q) = \frac{q^{2j+1}-q^{-2j-1}}{q-q^{-1}}.
\end{equation}
In addition, we define
\begin{equation}
  Q^\beta = \md{e}[\und{d}(\beta)\cdot\und{t}]
\end{equation}
with $\und{d}(\beta)$ the degree of the curve class $\beta$ in certain
basis of $H_2(X,\IR)$\footnote{Sometimes the basis we choose is not
  integral and the curve degrees can be fractional numbers.}.  We
choose a basis which we call \emph{positive} basis so the
$\und{d}(\beta)$ is a non-negative vector\footnote{In the case that
  the Mori cone of $X$ is simplicial, one can choose the basis to
  consist of Mori cone generators.  When the Mori cone is not
  simplicial, one can choose the generators of a larger but simplicial
  cone that covers the Mori cone. } for any effective curve $\beta$ in
$X$.  The contribution of a spin $(j_L,j_R)$ BPS state wrapping curve
class $\beta$ to the blowup equation is then
\begin{equation}
  Bl_{(j_L,j_R,R)}(q_1,q_2) = f_{(j_L,j_R)}(q_1,q_2/q_1)q_1^R +
  f_{(j_L,j_R)}(q_1/q_2,q_2)q_2^R - f_{(j_L,j_R)}(q_1,q_2),
\end{equation}
where $R = R(\beta,\und{n})$ is the component of the $\und{R}$-vector
\eqref{eq:R} associated to $\beta$, and the blowup equations
\eqref{eq:blowupI} can be written in the following form
\begin{equation}\label{eq:detail}
  \sum_{\und{n}\in \IZ^{b_4^c}}(-1)^{|\und{n}|}
  \re^{f_0(\und{n})(\eq+\et)+\sum_{i=1}^{b_2^c} f_i(\und{n})t_i}
  \exp\Bigg[-\sum_{j_L,j_R,\beta}\sum_{w=1}^\infty
  N^\beta_{j_L,j_R}
  \frac{Q^{w\beta}}{w}Bl_{(j_L,j_R,R)}(q_1^w,q_2^w)\Bigg]
  =\Lambda(\eq,\et,\und{m}).
\end{equation}
Then if there exists a unity $\und{r}$-field\footnote{For a vanishing
  $\und{r}$-field, the solutions $\und{n}$ that minimize
  $f_i(\und{n})$ are all in pairs.} so that for a positive basis of
$H_2(X,\IR)$
\begin{equation}\label{eq:f-cond}
  *: \text{all $f_i(\und{n})$ can be minimised at the same time with a unique
    solution $\und{n} = \und{n}_0$}
\end{equation}
the coefficients of $Q^\beta$ can be written as
\begin{equation}\label{eq:Qbeta-coefs}
  \sum_{j_L,j_R} N^\beta_{j_L,j_R}
  Bl_{(j_L,j_R,R(\beta,\und{n}_0))}(q_1,q_2)
  = I^\beta(q_1,q_2),
\end{equation}
where $I^{\beta}(q_1,q_2)$ collects contributions of multi-wrapping
$w>1$, summands with $\und{n}\neq \und{n}_0$, and higher order
expansions of the exponential, and it only contains BPS invariants of
curve degrees $\und{d} < \und{d}(\beta)$, which means
\begin{equation}\label{eq:dless}
  \und{d} < \und{d}(\beta): \;\;
  d_i \leq d_i(\beta), \;\; i=1,\ldots,b_2^c;\;\;
  \text{at least one inequality is not saturated}.
\end{equation}
Since the unique minimal solution $\und{n}_0$, if exists, can be
shifted to $\und{0}$ using the equivalence relation
\eqref{eq:r-equiv}, we always assume that $\und{n}_0=\und{0}$ and thus
$R(\beta,\und{n}_0) = \und{r}\cdot\und{d}(\beta)/2$.  Let us now
assume that the RHS of \eqref{eq:Qbeta-coefs} has already be computed.
Then given the lemma proved in appendix~\ref{sc:lemma} that, with a
few exceptions, all $Bl_{(j_L,j_R,R)}$ with fixed $R$ and different
$j_L,j_R$ (which are finite in number) are linearly independent to
each other and non-vanishing, the BPS invariants $N^\beta_{j_L,j_R}$
can be extracted from the LHS of \eqref{eq:Qbeta-coefs} and thus
recursively computed.  The only exceptions are the spin (0,0) or
$(0,\frac{1}{2})$ BPS invariants with curve classes $\beta$ satisfying
\begin{equation}\label{eq:rd-cond}
  | \und{r}\cdot \und{d}(\beta) | \leq 1.
\end{equation}
These invariants cannot be solved from the LHS of
\eqref{eq:Qbeta-coefs} due to the fact that
\begin{equation}
  Bl_{(0,0,\pm\frac{1}{2})} = Bl_{(0,\frac{1}{2},0)} = 0.
\end{equation}
In practice, we can use all the unity blowup equations whose
$\und{r}$-fields satisfy the condition \eqref{eq:f-cond} to solve all
the refined BPS invariants, provided that we have some extra data
which can help determine the spin (0,0), $(0,\frac{1}{2})$ BPS
invariants whose curve class $\beta$ satisfy \eqref{eq:rd-cond} for
all the unity $\und{r}$-fields that we have used.  In an elliptic
non-compact CY3, the number of such curve classes can be infinite.
Nevertheless in the examples where the toric hypersurface construction
or the mirror curve is known, the genus 0 GV invariants of $\beta$ can
be computed and used as extra input data, from which, together with
the BPS invariants of $\beta$ with spins other than $(0,0)$ or
$(0,\frac{1}{2})$ recursively extracted from \eqref{eq:Qbeta-coefs},
the unknown invariants $N^\beta_{0,0}$ or $N^\beta_{0,\frac{1}{2}}$
can be deduced\footnote{We remind the readers that due to the
  checkerboard pattern BPS states wrapping a fixed curve classs can
  have either spin (0,0) or $(0,\frac{1}{2})$ but not both.}.  We have
successfully used this approach to compute the refined BPS invariants
of the massive M--string and the E--string with up to three flavor
masses turned on, where the genus 0 GV invariants as extra input data
are computed using the toric hypersurface construction in
Sections~\ref{sc:M-toric}, \ref{sc:E-toric}.

In the following, we illustrate for the examples of E--,M--string
theories the finding of unity $\und{r}$-fields satisfying
\eqref{eq:f-cond} and all the ambiguous BPS invariants with $\beta$ of
the type \eqref{eq:rd-cond}.

\subsubsection{M--string}

We choose the positive basis of $H_2(X,\IR)$ to be the generators of
the Mori cone, whose K\"ahler parameters are \cite{Haghighat:2013gba}
\begin{equation}
  t_b - m, \;\; \tau-m,\;\; m.
\end{equation}
The corresponding $f_i(n)$ functions are
\begin{equation}\label{eq:fi-M}
\begin{aligned}
  &f_m(n) = (n-\frac{r_b}{4}+\frac{r_m}{2})^2 + \text{const.},\\
  &f_{\tau-m}(n) = (n-\frac{r_b}{4})^2 + \text{const.},\\
  &f_{b-m}(n) = \text{const.}.
\end{aligned}
\end{equation}
The unity $\und{r}$-fields of the M--string theory associated to the
elliptic blowup equations \eqref{eq:M-ublowup} are\footnote{We do not
  mod out the shift $2 C\cdot n$ equivalence here.}
\begin{equation}
  \und{r} = (r_b,r_m,r_\tau) = (2k,\pm 1,0),\quad k\in\IZ
\end{equation}
among which, the ones which minimise \eqref{eq:fi-M} with the unique
solution $\und{n} = \und{0}$ are
\begin{equation}
  \und{r} = (0,\pm 1,0),(2,1,0),(-2,-1,0).
\end{equation}
The only BPS invariants which cannot be determined from the unity
blowup equations with these $\und{r}$-fields are the spin (0,0)
invariants with
\begin{equation}
  \und{d} =  (d_b,d_m,d_\tau) = (1,-1,d_\tau),\quad d_\tau=0,1,2,\ldots.
  \label{Mstringdegreeinput}
\end{equation}

\subsubsection{E--string}

In the case of E--string theory, we cannot find a positive basis of
$H_2(X,\IR)$\footnote{This is due to the fact that the half K3 surface
  associated to the E--string theory has a highly non-simplicial Mori
  cone whose generators are infinitely many.}.  However we find that
the following basis
\begin{equation}
  t_b,\quad \tau,\quad \hat{m}_{i} = (\alpha_i,\und{m}),\;\; i=1,\ldots,8
\end{equation}
where $\alpha_i$ are the simple roots of $E_8$, is good enough with
which a similar statement concerning the solvability of BPS invariants
from unity blowup equations can be made.  Using the perturbative
prepotential \eqref{eq:E-F0}, we find the corresponding $f_i(n)$
functions are
\begin{equation}
\begin{aligned}
  &f_b(n) = \text{const.},\;\; \\
  &f_\tau(n) = \frac{1}{2}(n-\frac{r_b}{2})^2+\text{const.},\;\; \\
  &f_{\hat{m},i}(n) =
  \frac{1}{2}(n-\frac{r_b}{2})r_{\hat{m},i}+\text{consts.},\;\;i=1,\ldots,8,
\end{aligned}
\end{equation}
as well as
\begin{equation}
  f_0(n) = \frac{1}{2}(n-\frac{r_b}{2})(\und{r}_m/2,\und{r}_m/2)_{E_8}+\text{const.}
\end{equation}

Let us recall that in a unity $\und{r}$-field the $\und{r}_m$ vector
when divided by two is an element in $\cO_{2,240}$ so that when
expanded in terms of $\alpha_i$ the coefficients $r_{\hat{m},i}$ are
either all non-negative or all non-positive.  We denote these two
cases by $\und{r}_m>0$ and $\und{r}_m<0$ respectively.  We then choose
240 unity $\und{r} = (r_b, \und{r}_m,r_\tau)$ fields to be
\begin{equation}\label{eq:E-r-choice}
  \und{r} = (+1,\und{r}_m>0, 0), (-1,\und{r}_m<0, 0),\quad
  \frac{1}{2}\und{r}_m\in \cO_{2,240}
\end{equation}
so that
\begin{equation}
  r_b r_{\hat{m},i}\geq 0,\;\;i=1,\ldots,8,\;\;\text{and}
  \;r_b \sum_i r_{\hat{m},i}>0.
\end{equation}

The BPS invariants with $d_b = 0$ decouple from the true K\"ahler
modulus $t_b$ and are thus factored out of the blowup equations.  For
the remaining BPS invariants, using the fact that $d_b>0,d_\tau\geq 0$
and that $f_\tau(n)$ is minimised by $n\in\{0,r_b\}$, we can mimic
\eqref{eq:Qbeta-coefs} and by expanding \eqref{eq:detail} find
\begin{align}
  &\sum_{j_L,j_R}N^{d_b,d_\tau,d_{\hat{m},i}}_{j_L,j_R}
    Bl_{(j_L,j_R,R(\beta,0))}(q_1,q_2)\nn
  &-(q_1q_2)^{\frac{r_b}{2}(\frac{\und{r}_m}{2},\frac{\und{r}_m}{2})_{E_8}}
    \sum_{j_L,j_R}N^{d_b,d_\tau,d_{\hat{m},i}-\frac{1}{2}r_b
    r_{\hat{m},i}}_{j_L,j_R}
    Bl_{(j_L,j_R,R(\beta,\frac{r_b}{2}))} =
    I^{d_b,d_\tau,d_{\hat{m},i}}(q_1,q_2),
\end{align}
where again $I^{d_b,d_\tau,d_{\hat{m},i}}(q_1,q_2)$ collects
contributions of multi-wrapping $w>1$, summands with $n\neq 0,r_b$,
and etc so that it only constrains BPS invariants with
$(d_b',d_\tau',d_{\hat{m},i}')<(d_b,d_\tau,d_{\hat{m},i})$ in the
sense of \eqref{eq:dless}.  On the other hand, since
$(d_b,d_\tau,d_{\hat{m},i}-\frac{1}{2}r_b r_{\hat{m},i}) <
(d_b,d_\tau,d_{\hat{m},i})$, we can move the second term on the LHS to
the RHS and absorb it into $I^{d_b,d_\tau,d_{\hat{m},i}}(q_1,q_2)$.
We find the same recursive expression as \eqref{eq:Qbeta-coefs} and
the same procedures following \eqref{eq:Qbeta-coefs} apply.

We also find that the only BPS invariants which cannot be determined
from the unity blowup equations associated to \eqref{eq:E-r-choice}
are the spin (0,0) BPS invariants with
\begin{equation}
  \und{d} = (d_b,d_{\hat{m},i},d_\tau) = (1,0,d_\tau),\quad d_\tau=0,1,2,\ldots.
  \label{Estringdegreeinput}
\end{equation}

\subsection{Weyl orbit expansion}
\label{ssec:Weylorbit}
In this section, we show how to directly solve elliptic genera from
blowup equations rather than computing the refined BPS invariants. The
basic idea is to express an elliptic genus as an expansion with
respect to $q=Q_{\tau}$ and $v=\sqrt{q_1q_2}$ with coefficients as
Weyl orbits of the flavor symmetry $F$ and $SU(2)_x$ where
$x=\sqrt{q_1/q_2}$. It is well-known the reduced one-string elliptic
genus is independent from $SU(2)_x$. This makes solving the one-string
elliptic genus from blowup equations particularly simple. In the
following we mainly demonstrate how this method works for
E--strings. As a byproduct, we obtain some interesting functional
equations for the $E_8$ theta function. The efficiency of this method
will be further illustrated in \cite{IV}.

The reduced one E--string elliptic genus is well-known to be
\begin{equation}\label{EstringE1red}
\begin{aligned}
\IE_1^{\rm red}&=\frac{\eta^2}{\theta_1(\eq)\theta_1(\et)}\IE_1={\eta^{-8}}\Theta_{E_8}(\und{m})={\eta^{-8}}\sum_{\mathcal{O}_{p,k}}q^{p/2}\cdot \mathcal{O}_{p,k}\\
=q^{-\frac{1}{3}}(1+&\mathbf{248}q+(\mathbf{3875}+\mathbf{248}+1)q^2+(\mathbf{30380}+\mathbf{3875}+2\times\mathbf{248}+1)q^3\\
&+(\mathbf{147250}+2\times\mathbf{30380}+\mathbf{3875}+5\times\mathbf{248}+1)q^4+\dots)
\end{aligned}
\end{equation}
where
\begin{equation}
  \Theta_{E_8}(\tau,\und{m}) = \sum_{\und{k}\in \Gamma_{E_8}}\exp(\pi \ri \tau \und{k}\cdot \und{k}+2\pi \ri \und{m}\cdot\und{k}) = \frac{1}{2}\sum_{k=1}^4\prod_{\ell=1}^8 \theta_{k}(\tau,m_{\ell}).
\end{equation}
The first few $E_8$ Weyl orbits are as follows:
\begin{eqnarray}\label{Weylorbit}
&& \mathcal{O}_{0,1}, ~~\mathcal{O}_{2,240},~~ \mathcal{O}_{4,2160},~~ \mathcal{O}_{6,6720}, ~~
 \mathcal{O}_{8,240}, \mathcal{O}_{8,17280}, ~~ \mathcal{O}_{10,30240} \nonumber \\
&& \mathcal{O}_{12,60480}, ~~   \mathcal{O}_{14,13440},
   \mathcal{O}_{14,69120},  ~~   \mathcal{O}_{16,2160},
   \mathcal{O}_{16,138240},  ~~  \mathcal{O}_{18,240},
   \mathcal{O}_{18,181440},    ~~  \nonumber \\
&&   \mathcal{O}_{20,30240},  \mathcal{O}_{20,241920},   ~~   \mathcal{O}_{22,138240},    \mathcal{O}_{22,181440},
~~  \mathcal{O}_{24,6720},   \mathcal{O}_{24,483840},  \nonumber \\
&& \mathcal{O}_{26,13440},   \mathcal{O}_{26,30240}, \mathcal{O}_{26,483840},
~~ \cdots  .
\end{eqnarray}

In unity blowup equations, each Weyl orbit breaks down due to the
shifts proportional to a root. For example, for $\mathcal{O}_{2,240}$,
\begin{equation}
  \sum_{w\in\mathcal{O}_{2,240}}\re^{w\cdot (m+\eq\alpha)}=
  q_1^{-2}\re^{- \alpha\cdot m}+\sum_{\alpha\cdot w=-1}q_1^{-1}\re^{w\cdot m}+
  \sum_{\alpha\cdot w=0}\re^{w\cdot m}+\sum_{\alpha\cdot w=1}q_1\re^{w\cdot m}+
  q_1^{2}\re^{ \alpha\cdot m}.
\end{equation}
This forces us to look into how every Weyl orbit splits under the
shift of a root. Due to the Weyl symmetry, all the elements in one
Weyl orbit intersect with any of the roots in the same way, i.e.\ for
any root the distribution of intersection numbers $R=\alpha\cdot w$
between the root and all Weyl orbit elements is the same. For example,
for any positive root $\alpha$, we list the distribution for some Weyl
orbits in Table~\ref{tb:rootWeyl}. Note the elements are all Weyl orbits of
$E_7$. Knowing (\ref{EstringE1red}), it is easy to check the unity
blowup equations \eqref{eq:E-ublowup} are correct.

\begin{table}[h]
  \centering
  \resizebox{\linewidth}{!}{
  \begin{tabular}{|c|ccccccccccc|}\hline
    $R=\alpha\cdot w$& $-5$ & $-4$ & $-3$ & $-2$ & $-1$ &0 & 1 & 2& 3 & 4 & 5 \\ \hline
    $\mathcal{O}_{0,1}$ &  &  &   &  &  & \textcolor{red}1 &  &  &  & &   \\ \hline
    $\mathcal{O}_{2,240}$ &  &  &   & \textcolor{red}1 & \textcolor{blue}{56} & \color{orange}{126} & \textcolor{blue}{56} & \textcolor{red}1 &  & &   \\ \hline
    $\mathcal{O}_{4,2160}$ & &   &   & \color{orange}{126} & \color{magenta}{576} & \color{cyan}{756} & \color{magenta}{576} & \color{orange}{126} & & &  \\ \hline
    $\mathcal{O}_{6,6720}$ & &   &  \textcolor{blue}{56} & \color{cyan}{756} & 1512 & 2072 & 1512 & \color{cyan}{756} & \textcolor{blue}{56}&  &  \\ \hline
    $\mathcal{O}_{8,240}$ & &   \textcolor{red}1 & &  56 &  & 126 & & 56 &  &\textcolor{red}1 &  \\ \hline
    $\mathcal{O}_{8,17280}$ & &   & \color{magenta}{576}&  2016 & 4032 & 4032 & 4032 & 2016 & \color{magenta}{576} &  &  \\ \hline
    $\mathcal{O}_{10,30240}$ & &  \color{orange}{126} & 1512&  4158  & 5544 & 7560 & 5544 & 4158 & 1512 &  \color{orange}{126} &   \\ \hline
    $\mathcal{O}_{12,60480}$ & &  \color{cyan}{756} & 4032&  7560 & 12096 & 11592 & 12096 & 7560 & 4032 & \color{cyan}{756}  &  \\ \hline
    $\mathcal{O}_{14,13440}$ &\textcolor{blue}{56} &  56 & 1512  & 1512  &  1568  &  4032  &  1568 & 1512  & 1512 & 56 & \textcolor{blue}{56} \\ \hline
    $\mathcal{O}_{14,69120}$ &  & 2016  & 4032 & 10080 & 12096 &  12672 & 12096 & 10080  & 4032 & 2016 &  \\ \hline
    $\mathcal{O}_{16,2160}$  & & 126& & 576& & 756 & & 576 & & 126 &  \\ \hline
    $\mathcal{O}_{16,138240}$ & \color{magenta}{576} & 4032 & 12096 & 16128 & 24192 & 24192 & 24192 & 16128 & 12096 & 4032 &\color{magenta}{576}                 \\ \hline
  \end{tabular}}
\caption{Intersection numbers between roots and elements of $E_8$ Weyl
  orbits.} \label{tb:rootWeyl}
\end{table}

Conversely it is possible to solve \eqref{EstringE1red} from the
blowup equations. Let us first write
$\IE_1^{\rm red}=f(q,v,\und{m})/\eta^8$.\footnote{The denominator
  $\eta^8$ can be later determined by requiring that $\IE_1^{\rm red}$
  be decomposed as representations of $E_8$, rather than just Weyl
  orbits. Besides, there is an overall constant in front of the whole
  elliptic genus $\IE_1$ that can not determined by blowup equations
  due to the lack of gauge symmetry. This is of course not
  surprising. Here we assume the overall constant is 1.}  The
vanishing blowup equation \eqref{eq:E-vblowup} gives
\begin{equation}
  f(q,{\eq\over2},\und{m})=f(q,{\et\over2},\und{m}).
\end{equation}
Thus $f(q,v,\und{m})$ is independent from $v$. We can simply write it
as
\begin{equation}
  f(q,\und{m})=\sum_{n=0}^\infty
  q^n\sum_{\mathcal{O}_{p,k}}x_{n,p,k}\mathcal{O}_{p,k}.
\end{equation}
The task is to determine all $x_{n,p,k}$.  It is convenient to write
the unity blowup equations as
\begin{equation}\label{fE1}
  \theta_1(\et)\theta_1(\alpha\cdot m+\et)f(q,m+\eq\alpha)
  -\theta_1(\eq)\theta_1(\alpha\cdot m+\eq)f(q,m+\et\alpha)
  =\theta_1(\et-\eq)\theta_1(\alpha\cdot m+\eq+\et)f(q,m),
\end{equation}
where $\alpha\in\Delta(E_8)$.
We conjecture the solution is uniquely
$f(\tau,\und{m})={\eta^{-8}}\Theta_{E_8}(\tau,\und{m})$ under the
conditions:
\begin{itemize}
\item The $q$ expansion coefficients of $\IE_1$ can be decomposed as
  sums of irreducible representations of $E_8$;
\item The leading $q$ expansion coefficient is 1, i.e.\ the trivial
  $E_8$ orbit $\cO_{0,1}$.
\end{itemize}
Note the blowup equations themselves only determine $f(q,\und{m})$ up
to a free function of $\tau$. The two assumptions assure the prefactor
is $\eta^{-8}$.\footnote{This agrees with the degrees of refined
  BPS invariants that need to be input for the E--string blowup
  equations in (\ref{Estringdegreeinput}).} In fact, it is proved
by Don Zagier that \eqref{fE1} has a unique solution which is the
$E_8$ theta function up to a free function of $\tau$, and similar
statements can be made for arbitrary positive definite even unimodular
lattices generated by roots, such as the $E_8\times E_8$ lattice and
the Barnes-Wall lattice $\Lambda_{16}$ in dimension 16 and the 23
Niemeier lattices in dimension 24. We give the proof in Appendix \ref{appB}.

Now we briefly show how the Weyl orbit recursion works. Given that we
have assumed the leading $q$ order of $f(q,\und{m})$ is the trivial
orbit $\mathcal{O}_{0,1}$, we find that in order for the subleading
order of (\ref{fE1}) to be satisfied, the subleading order of
$f(q,\und{m})$ should have two $\mathcal{O}_{0,1}$ with $R=\pm
2$. Looking up in Table~\ref{tb:rootWeyl}, one finds that in order to
store the $E_8$ symmetry, one has to add two $E_7$ orbits of length 56
at $R=\pm 1$ and one $E_7$ orbit of length 126 at $R=0$. Thus in the
subleading order, $f(q,\und{m})$ has the $E_8$ orbit
$\mathcal{O}_{2,240}$.  Next, for the subsubleading order of
(\ref{fE1}) to be satisfied, one needs to add two $E_7$ orbits of
length 126 at $R=\pm 2$ in the subsubleading order of $f(q,\und{m})$
to cancel the effect of the previous $E_7$ orbit of length 56. Then to
restore the $E_8$ symmetry, one needs to add two $E_7$ orbits of
length 576 at $R=\pm 1$ and one $E_7$ orbit of length 756 at
$R=0$. Repeating this process, we find each sub $E_7$ Weyl orbit in
Table~\ref{tb:rootWeyl} is in an infinite series of the ones in the
larger Weyl orbits. Moreover, the contributions from each infinite
series can be organized into one of the following two identities:
\begin{equation}\label{twoeq}
  \begin{aligned}
    \theta_1(\et)\theta_1(\lambda+\et)\theta_3(2\tau,\lambda+2\eq)-\theta_1(\eq)\theta_1(\lambda+\eq)\theta_3(2\tau,\lambda+2\et)
    &=\theta_1(\et-\eq)\theta_1(\lambda+\eq+\et)\theta_3(2\tau,\lambda)\\
    \theta_1(\et)\theta_1(\lambda+\et)\theta_2(2\tau,\lambda+2\eq)-\theta_1(\eq)\theta_1(\lambda+\eq)\theta_2(2\tau,\lambda+2\et)
    &=\theta_1(\et-\eq)\theta_1(\lambda+\eq+\et)\theta_2(2\tau,\lambda).
\end{aligned}
\end{equation}
With this in mind, one can directly write down the Weyl orbit
expansion satisfying the unity blowup equations by the following rule:
Each sub $E_7$ orbit in an $E_8$ Weyl orbit $\mathcal{O}_{p,k}$ with
intersection number $R$ generates an infinite series of sub $E_7$
orbits in $E_8$ Weyl orbits $\cO_{p',k'}$ with intersection numbers
$R'$ where $p'$ grows quadratically and $|R'|$ grows linearly.  To
be more explicit,
\begin{itemize}
\item If $R$ is even, the growth is based on $\theta_3(2\tau,2z)$,
  i.e.\ $p'$ increases by $2n^2$ and $|R'|$ increases by $2n$;
\item If $R$ is odd, the growth is based on $\theta_2(2\tau,2z)$,
  i.e.\ $p'$ increases by $2n(n+1)$ and $|R'|$ increases by $2n$.
\end{itemize}
We have marked some sub $E_7$ orbits in the same series with the same
color in Table~\ref{tb:rootWeyl}. It turns out all $E_8$ Weyl orbits
appear and just appear once in $f(q,\und{m})$, which means it is
indeed the $E_8$ theta function.

The same procedure can also be used to determine the two E--string
elliptic genus, where the flavor symmetry is effectively
$E_8\times SU(2)_x$. In this case, it is convenient to work with the
following reduced version of two E--strings elliptic genus
\begin{equation}
  f(q,v,x,\und{m})=Z_2^{\textrm{E-str}}
  (\tau,\eq,\et,\und{m})\theta_1(\eq)\theta_1(2\eq)\theta_1(\et)\theta_1(2\et)\eta^{-2}.
\end{equation}
From the unity blowup equations, one can solve it as
\begin{equation}
\begin{aligned}
  \phantom{-}& f(q,v,x,\und{m})=(v^{-1}+v)+(\mathbf{248}(x^{-1}+x)+\mathbf{248}(v^{-1}+v)-(v^{-3}+v^3)(x^{-2}+x^2))q\\
  &+(-(v^{-5}+v^5)(x^{-2}+x^2)-\mathbf{248}(v^{-4}+v^4)(x^{-1}+x)+(v^{-3}+v^3)(-\mathbf{248}(x^{-2}+x^2)+\mathbf{3875})\\
  &+(v^{-1}+v)((x^{-4}+x^4)+\mathbf{3875}(x^{-2}+x^2)+\mathbf{30380}+3\times \mathbf{248})+ \mathbf{248}(x^{-3}+x^3)\\
  &+(\mathbf{30380}+\mathbf{3875}+
  \mathbf{248}+1)(x^{-3}+x^3))q^2+\dots.
\end{aligned}
\end{equation}

For M--strings, it is completely parallel and actually easier to
solve elliptic genera from the unity blowup equations, as the flavor
$SU(2)$ is much simpler than $E_8$.

\section{From E--strings to del Pezzo surfaces}
\label{sc:delPezzo}

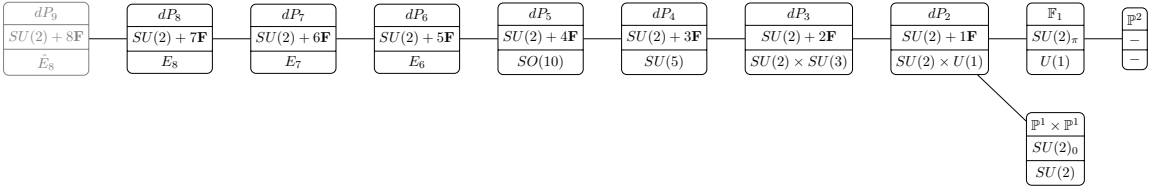
\begin{figure}
  \resizebox{\linewidth}{!}{
    \begin{tikzpicture}
      \node (0) at (0,0) [split3box]%
      {$\IP^2$ \nodepart{two} $-$ \nodepart{three} $-$};%
      \node (1) [split3box,left=of 0]%
      {$\IF_1$ \nodepart{two} $SU(2)_{\pi}$ \nodepart{three} $U(1)$};%
      \node (2) [split3box,left=of 1]%
      {$dP_2$ \nodepart{two} $SU(2) + 1\md{F}$ \nodepart{three}
        $SU(2)\times U(1)$};%
      \node (3) [split3box,left=of 2]%
      {$dP_3$ \nodepart{two} $SU(2)+2\md{F}$ \nodepart{three}
        $SU(2)\times SU(3)$};%
      \node (4) [split3box,left=of 3]%
      {$dP_4$ \nodepart{two} $SU(2)+3\md{F}$ \nodepart{three}
        $SU(5)$};%
      \node (5) [split3box,left=of 4]%
      {$dP_5$ \nodepart{two} $SU(2)+4\md{F}$ \nodepart{three}
        $SO(10)$};%
      \node (6) [split3box,left=of 5]%
      {$dP_6$ \nodepart{two} $SU(2)+5\md{F}$ \nodepart{three} $E_6$};%
      \node (7) [split3box,left=of 6]%
      {$dP_7$ \nodepart{two} $SU(2)+6\md{F}$ \nodepart{three} $E_7$};%
      \node (8) [split3box,left=of 7]%
      {$dP_8$ \nodepart{two} $SU(2)+7\md{F}$ \nodepart{three} $E_8$};%
      \node (9) [split3box,,draw=gray,left=of 8]%
      {\textcolor{gray}{$dP_9$} \nodepart{two}
        \textcolor{gray}{$SU(2)+8\md{F}$} \nodepart{three}
        \textcolor{gray}{$\hat{E}_8$}};%
      \node (10) [split3box,below=of 1]%
      {$\IP^1\times\IP^1$ \nodepart{two} $SU(2)_0$ \nodepart{three}
        $SU(2)$};%
      \draw (0) -- (1) -- (2) -- (3) -- (4) -- (5) -- (6) -- (7) --
      (8) -- (9)(2) -- (10);%
    \end{tikzpicture}}
  \caption{Rank one 5d SCFTs and E--string theory on $S^1$: the
    associated geometry, the IR effective gauge theory, and the global
    symmetry at the conformal point.}\label{fg:d5SCFTs}
\end{figure}

Rank one\footnote{The rank of a 5d SCFT is the dimension of its
  Coulomb branch.} 5d SCFTs have been classified \cite{Seiberg:1996bd,
  Morrison:1996xf, Douglas:1996xp, Jefferson:2018irk, Apruzzi:2019vpe,
  Apruzzi:2019opn}.  They can be constructed through M-theory
compactified on the canonical bundle over del Pezzo surfaces, which
include $\IP^2, \IF_0$ or $d$ ($d=1,\ldots,8$) points blow-ups of
$\IP^2$ denoted by $dP_d$, and their BPS spectra have been
computed in \cite{Huang:2013yta, Apruzzi:2019opn}.  The superconformal
limit is reached when the entire complex surface is shrunk to a point.
In particular the 5d SCFTs associated to $dP_d$ ($d=6,7,8$) are the
famous Minaham-Nemeschansky theories with $E_d$ flavor symmetry
\cite{Minahan:1996fg, Minahan:1996cj}.  All these 5d SCFTs can be
obtained from the E--string theory on $S^1$.  Indeed the del Pezzo
surfaces can be obtained from the half K3 surface by successively
blowing down the exceptional divisors.  From the physics point of
view, rank one 5d SCFTs are effectively 5d $SU(2)$ gauge theory in the
IR\footnote{The 5d SCFT associated to $\IP^2$ does not have a low
  energy effective gauge theory description.  See the diagram in the
  main text.} with up to 7 hypermultiplets transforming in the
fundamental represention of $SU(2)$, while the E--string theory on
$S^1$ has an effective description of a 5d $SU(2)$ gauge theory with 8
hypermultiplets.  The blow-down operation in geometry corresponds to
successively decoupling hypermultiplets by giving them infinitely
large mass.\footnote{To be exact, the mass of the hypermultiplet to be
  decoupled should be sent to negative infinity.}  The relation
between E--string theory and the family of rank one 5d SCFTs can be
summarised in Figure~\ref{fg:d5SCFTs}.

We demonstrate here that blowup equations for rank one 5d SCFTs in the
Coulomb branch can be obtained from the blowup equations of the
E--string theory by carefully carrying out the transformations that
correspond to the blow-down operations.

Our starting point is the unity and vanishing blowup equations of the
E--string theory \eqref{eq:E-ublowup},\eqref{eq:E-vblowup}, which we
can uniformly write as
\begin{align}
  \sum_{n\in\IZ}
  &(-1)^n\md{e}
  \[\frac{1}{2}(n+\frac{1}{2})^2\tau+
    (n+\frac{1}{2})(\frac{1}{2}\und{r}_m\cdot\und{m}
    +\frac{1}{8}\und{r}_m\cdot\und{r}_m(\eq+\et))\]
    \wh{Z}^{\text{inst}}_{dP_9}(\tau,\und{m},t_b,\eq,\et)^{-1}\nn\times
  &\wh{Z}^{\text{inst}}_{dP_9}(\tau,\und{m}+\frac{\eq}{2}\und{r}_m,
    t_b-(n+\frac{1}{2})\eq,\eq,\et-\eq)
    \wh{Z}^{\text{inst}}_{dP_9}(\tau,\und{m}+\frac{\et}{2}\und{r}_m,
    t_b-(n+\frac{1}{2})\et,\eq-\et,\et)\nn =
    \sum_{n\in\IZ}
  &(-1)^n\md{e}
  \[\frac{1}{2}(n+\frac{1}{2})^2\tau
    +(n+\frac{1}{2})(\frac{1}{2}\und{r}_m\cdot\und{m}
    +\frac{1}{8}\und{r}_m\cdot\und{r}_m(\eq+\et))\].
    \label{eq:dP9-ublowup}
\end{align}
Here we have massaged the equations a little bit by summing up
elliptic genera to the total instanton partition function, or
alternatively by refraining from expanding the instanton partition
function in terms of $t_b$ in \eqref{eq:blowupI}.  We use the hatted
notation of the instanton partition function to emphasize that the
K\"ahler moduli have been twisted in the sense of \eqref{eq:twist}, and
have chosen $r_b = -1$ following \eqref{eq:rb}.  Let us first derive
the blowup equation for the $E_8$ Minaham-Nemeschansky theory.  To
this end, we blow down the exceptional curve $x_9$ which also serves
as the base of the elliptic fibration and take the limit
\begin{equation}\label{eq:x9-lim}
  \lim_{dP_8}:\;t_b \to -\ri\infty, \quad \tau\to+\ri\infty,
  \quad t_{B_8}=t_b+\tau\;\;\text{finite}.
\end{equation}
Here we have defined $t_{B_8}$ which measures the volume of the
anti-canonical class of $dP_8$
\begin{equation}
  -K_{dP_8} = 3h - \sum_{i=1}^8 x_i
\end{equation}
and its corresponding $\und{r}$-component is
\begin{equation}
  r_{B_8} = r_b+r_\tau = -1.
\end{equation}
In this process, the terms in the instanton partition function of the
form $Q_b^{d_b}Q_\tau^{d_\tau}=\md{e}[d_b t_b+d_\tau\tau]$ with
$d_b>d_\tau$ become divergent, which need to resummed and analytically
continued.  It turns out that the only terms which diverge in
$\wh{Z}^{\text{inst}}_{dP_9}$ are those that come with the BPS
invariant $N_{(0,0)}^{(d_b,d_\tau,\und{d}_m)=(1,0,\und{0})} = 1$,
which corresponds to the base curve $b$, and the instanton partition
function factorises to the product of the contribution of this BPS
state $Z_{(0,0)}^{(1,0,\und{0})}$ and the partition function of the
daughter $E_8$ MN theory
\begin{equation}
  \lim_{dP_8} Z^{\text{inst}}_{dP_9}(\tau,\und{m},t_b,\eq,\et) =
  \(\lim_{dP_8} Z_{(0,0)}^{(1,0,\und{0})}(t_b,\eq,\et)\)
  Z^{\text{inst}}_{dP_8}(t_{B_8},\und{m},\eq,\et).
\end{equation}
In general, a spin (0,0) BPS state accompanying the K\"ahler parameter $t$
contributes to the instanton partition function by
\begin{equation}
  Z_{(0,0)}(t,\eq,\et) =
  \text{PE}\Big[f_{(0,0)}(q_1,q_2)Q_t\Big]
  :=\text{PE}\bigg[\frac{Q_t}{(q_1^{1/2}-q_1^{-1/2})(q_2^{1/2}-q_2^{-1/2})}\bigg]
\end{equation}
with
$\text{PE}[g(\bullet)] =
\exp\(\sum_{n=1}^\infty\frac{g(\bullet^n)}{n}\)$.  In each summand on
the LHS of \eqref{eq:dP9-ublowup}, the contribution of such a BPS
state to the blowup equation before turning on the twisting is then
\begin{align}
  Bl_{(0,0)}(t,\eq,\et):=
  &\,\frac{Z_{(0,0)}(t+R\eq,\eq-\et,\et)
    Z_{(0,0)}(t+R\et,\eq,\et-\eq)}
    {Z_{(0,0)}(t,\eq,\et)}\nn =
  &\,\text{PE}
    \Big[\(f_{(0,0)}(q_1,q_2/q_1)q_1^R+
    f_{(0,0)}(q_1/q_2,q_2)q_2^R-f_{(0,0)}(q_1,q_2)\)Q_t \Big]\nn=
  &\,\text{PE}
    \Bigg[-\sum_{\substack{m,n\geq 0\\m+n\leq R-3/2}}q_1^{m+1/2}q_2^{n+1/2}Q_t\Bigg]=
  \prod_{\substack{m,n\geq 0\\m+n\leq R-3/2}}(1-q_1^{m+1/2}q_2^{n+1/2}Q_t),
\end{align}
where $R$ is the component of the $\und{R}$-vector defined in
\eqref{eq:R}.  If in addition the corresponding curve is a rational
($-1$) curve, after turning on the twisting $Q_t\to -Q_t$ and taking
the limit $Q_t\to \infty$ or $t\to-\ri\infty$, we find the divergent
contribution
\begin{equation}\label{eq:Bl-hat}
  \wh{Bl}_{(0,0)}(t,\eq,\et) =
  \md{e}\[\frac{(2R-1)(2R+1)}{8} t+\frac{(2R-1)R(2R+1)}{24}(\eq+\et)\]
\end{equation}
which should cancel with vanishing terms in the semiclassical
contributions.  Applying this formula to the base curve $b$ with
K\"ahler parameter $t_b$ and $R = -n-1/2$, and inserting it into
\eqref{eq:dP9-ublowup}, we obtain the following unity blowup equations
for the $E_8$ MN theory or the $dP_8$ geometry
\begin{align}
  \sum_{n\in\IZ}
  &(-1)^n\md{e}
  \[\frac{n(n+1)}{2}t_{B_8}+
    \frac{1}{2}n\und{r}_m\cdot\und{m}+P_{dP_8}(n) (\eq+\et)\]
    \wh{Z}^{\text{inst}}_{dP_8}(\und{m},t_{B_8},\eq,\et)^{-1}\nn\times
  &\wh{Z}^{\text{inst}}_{dP_8}(\und{m}+\frac{\eq}{2}\und{r}_m,
    t_{B_8}-(n+\frac{1}{2})\eq,\eq,\et-\eq)
    \wh{Z}^{\text{inst}}_{dP_8}(\und{m}+\frac{\et}{2}\und{r}_m,
    t_{B_8}-(n+\frac{1}{2})\et,\eq-\et,\et)\nn =
    \sum_{n\in\{0,-1\}}
  &(-1)^n\md{e}
  \[\frac{n(n+1)}{2}t_{B_8}
    +\frac{1}{2}n\und{r}_m\cdot\und{m}+P_{dP_8}(\und{n})(\eq+\et)\]
    \label{eq:dP8-ublowup}
\end{align}
with
\begin{equation}
  P_{dP_8}(n) = n \delta_u - \frac{n(n+1)(2n+1)}{12}
\end{equation}
where $\delta_u$ means $1$ in unity equations and $0$ in vanishing
equations.  Here because the $E_8$ symmetry is unbroken, all the 240
unity $\und{r}_m$-fields forming the Weyl orbit $\cO_{2,240}$ as well
as the lone vanishing $\und{r}_m = \und{0}$ survive the limit.  Note
that when we apply the blow-down limit to the blowup equations, we
have the liberty of focusing exclusively on the transformation of the
LHS, factoring out any terms which are independent of the summation
index $n$ along the way.  The RHS of the resulting blowup equations
can be obtained by collecting all the terms of the LHS which do not
vanish when $t_{B_8}$ (or its analogue in further blow-downs) is sent
to $\ri\infty$.  This is the strategy we will follow in the following.

To further perform blow-downs, it is convenient to change the basis of
flavor parameters from $m_i$ to the following basis associated to the
exceptional curves
\begin{equation}
  \wt{m}_i = \text{Vol}(x_i) - \frac{1}{3}\text{Vol}(h),\;\;i=1,\ldots,8
\end{equation}
It is easy to check that the corresponding homology classes
$[x]_i - 1/3[h]$ have trivial intersection numbers with the complex
surface itself.  They are related to $m_i$ by equating the RHS of
\eqref{eq:ax},\eqref{eq:ae}, and we find
\begin{equation}
  \wt{m}_1 = -\frac{1}{3}\sum_{j=1}^7 m_j + \frac{1}{3}m_8,\quad
  \wt{m}_i =  - m_i -\frac{1}{6}m_8 +\frac{1}{6}\sum_{j=1}^7 m_j,\;\;i=2,\ldots,8.
\end{equation}
The corresponding $\und{r}$-components are given by
\begin{equation}
  r_{\wt{m},1} = -\frac{1}{3}\sum_{j=1}^7 r_{m,j} + \frac{1}{3} r_{m,8},\quad
  r_{\wt{m},i} = - r_{m,i}
  -\frac{1}{6}r_{m,8}+ \frac{1}{6}\sum_{j=1}^7 r_{m,j}, \;\; i =2,\ldots,8
\end{equation}
To flow from $dP_8$ surface to $dP_d$ ($0\leq d\leq 7$), we blow down
$8-d$ exceptional curves $x_8,\ldots,x_{d+1}$ corresponding to taking
the limit
\begin{equation}\label{eq:blow-down}
  \lim_{dP_d}:\;\wt{m}_i \to -\ri\infty,\;\;i=d+1,\ldots,8,\quad
  t_{B_8}+\sum_{i=d+1}^8\wt{m}_i \;\;\text{finite}.
\end{equation}

The first thing we notice is that among the $\und{r}$-components of
mass parameters only $r_{\wt{m},i}$ ($i=1,\ldots,d$) survive, and they
organise themselves into a weight vector of $E_d$ with Dynkin
labels\footnote{Recall that the first homology group of $dP_d$
  ($d\geq 3$) contains the root lattice of $E_d$ generated by
  $x_i-x_{i+1}$ and $h-x_1-x_2-x_3$.}
\begin{equation}
  \(\frac{r_{\wt{m},1}-r_{\wt{m},2}}{2},
  \;\ldots,\;\frac{r_{\wt{m},d-1}-r_{\wt{m},d}}{2},\;
  -\frac{r_{\wt{m},1}+r_{\wt{m},2}+r_{\wt{m},3}}{2}\)
\end{equation}
if $d\geq 3$ (here $E_3,E_4,E_5$ denote respectively $SU(3)\times
SU(2), SU(5),SO(10)$) and a weight vector of $SU(2)$ with Dynkin label
\begin{equation}
  \(\frac{r_{\wt{m},1}-r_{\wt{m},2}}{2}\)
\end{equation}
if $d=2$.  We will denote this weight vector also by
$\frac{1}{2}\und{r}_m$.  The blow-up operation \eqref{eq:blow-down}
can therefore be interpreted as the decomposition of the Weyl orbit of
$E_8$ to those of $E_d$ ($d<8$).  In vanishing blowup equations
$\frac{1}{2}\und{r}_m$ is always the zero vector, while in unity
blowup equations $\frac{1}{2}\und{r}_m$ must be weight vectors in the
Weyl sub-orbits of $\cO_{2,240}$, which we summarize in
Table~\ref{tb:orbits}.

\begin{table}
  \centering
  \resizebox{\linewidth}{!}{
  \begin{tabular}{*{6}{>{$}c<{$}}}\toprule
    dP_d
    & 8
    & 7
    & 6
    & 5
    & 4 \\
    G & E_8 & E_7 & E_6 & SO(10) & SU(5)\\
    \frac{1}{2}\und{r}_m
    & \cO_{2,240}
    & \cO_{2,126},\cO_{\frac{3}{2},56}
    & \cO_{2,72},\pm\cO_{\frac{4}{3},27}
    & \cO_{2,40},\pm\cO_{\frac{5}{4},16},\pm\cO_{1,10}
    & \cO_{2,20},\pm\cO_{\frac{6}{5},10},\pm\cO_{\frac{4}{5},5}\\
    \midrule
    dP_d&\multicolumn{4}{c}{3}&2\\
    G&\multicolumn{4}{c}{$SU(3)\times SU(2)$}&SU(2)\\
    \frac{1}{2}\und{r}_m&\multicolumn{4}{c}
                          {$(\cO_{2,6},\cO_{0,1}),\pm(\cO_{\frac{2}{3},3},\cO_{\frac{1}{2},2}),(\cO_{0,1},\cO_{2,2}),\pm(\cO_{\frac{2}{3},3},\cO_{0,1}),
                          (\cO_{0,1},\cO_{\frac{1}{2},2})$}
    &\cO_{2,2},\cO_{\frac{1}{2},2},\cO_{0,1}\\
    \bottomrule
  \end{tabular}
}
  \caption{$\und{r}_m$ components of mass parameters for $dP_d$
    ($d\geq 2$).}
  \label{tb:orbits}
\end{table}

In the instanton partition function, all the terms whose power of
$Q_{\wt{m},j}=\md{e}[\wt{m}_j]$ is greater than that of
$Q_{B_8} =\md{e}[t_{B_8}]$ for any $j\in\{d+1,\ldots,8\}$ diverge,
which need to be resummed and analytically continued.  All such terms
are generated by the spin (0,0) BPS states wrapping curve classes of
$(d_B,d_{\wt{m},i}) = (1,1,\ldots,1,2)$ and all permutations of
$d_{\wt{m},i}$ for $i=d+1,\ldots,8$.  There are $8-d$ such curve
classes and they correspond to the exceptional curves
\begin{equation}
  x_{d+1},\dots,x_8.
\end{equation}
All of these curves have $-1$ intersection with $dP_8$.  We denote
their K\"ahler moduli by
\begin{equation}
  t_{x,i} = t_{B_8} + \wt{m}_i+ \sum_{j=1}^8 \wt{m}_j.
\end{equation}
The divergent contributions of these BPS states to the summand on the
LHS of \eqref{eq:dP8-ublowup} in the limit \eqref{eq:blow-down} can be
written down following \eqref{eq:Bl-hat}
\begin{equation}
  \prod_{i=d+1}^8\wh{Bl}_{(0,0)}(t_{x,i},\eq,\et) =
  \prod_{i=d+1}^8 \md{e}\[\frac{(2R_{x,i}-1)(2R_{x,i}+1)}{8}t_{x,i}+
    \frac{(2R_{x,i}-1)R_{x,i}(2R_{x,i}+1)}{24}(\eq+\et)\].
\end{equation}
where
\begin{equation}
  R_{x,i} = -n+\frac{r_{x,i}}{2} =
  -n+\frac{1}{2}(-1+r_{\wt{m},i}+\sum_{j=1}^8r_{\wt{m},j}),
\end{equation}
and it should cancel with vanishing terms in the semiclassical
contribution in \eqref{eq:dP8-ublowup}.
On the other hand, all the
other terms which do not diverge sum up to the instanton partition
function of the $dP_d$ surface \cite{Huang:2013yta}.
Therefore, we have the factorisation of the twisted partition function
\begin{equation}\label{eq:Zd}
  \lim_{dP_d}\wh{Z}^{\text{inst}}_{dP_8}(t_{B_8},\und{m},\eq,\et) =
  \bigg[\lim_{dP_d}\prod_{i=d+1}^8\wh{Bl}_{(0,0)}(t_{x,i},\eq,\et)\bigg]
  \wh{Z}_{dP_d}^{\text{inst}}(t_{B_d},\und{m},\eq,\et).
\end{equation}

The daughter theory associated to the $dP_d$ surface is parameterised
by the mass parameters $\wt{m}_i$ for $i=1,\ldots,d$ as well as the
volume of the anti-canonical class
\begin{equation}
  -K_{dP_d} = 3h - x_1 -\ldots - x_d
\end{equation}
given by
\begin{equation}
  t_{B_d} = t_{B_8}+\sum_{i=d+1}^8 t_{x,i}
  = (9-d)t_{B_8} + (8-d)\sum_{i=1}^d \wt{m}_i + (9-d)\sum_{j=d+1}^8\wt{m}_j.
\end{equation}
with the associated $\und{r}$-component
\begin{equation}
  r_{B_d} = -(9-d)+(8-d)\sum_{i=1}^d r_{\wt{m},i} + (9-d)\sum_{j=d+1}^8r_{\wt{m},j}.
\end{equation}
Since $-K_{dP_d}$ is the only true K\"ahler modulus in $dP_d$ and its
intersection with the complex surface is $-(9-d)$, we have the freedom
to bring $r_{B_d}$ within the range
\begin{equation}
  -(9-d) \leq r_{B_d} < 9-d
\end{equation}
using the equivalence relation \eqref{eq:r-equiv} without affecting
$\und{r}_m$.  We will always assume that $r_{B_d}$ falls inside this
range.  Furthermore, since the curve class $-K_{dP_d}$ is invariant
under Weyl reflections of $E_d$ as well as the Weyl transformations of
the $E_8$ lattice which are perpendicular to the root lattice of
$E_d$, its associated $\und{r}$ component $r_{B_d}$ should only depend
on the Weyl orbit of $E_d$.  In the case of $dP_2$ and $dP_1$ which
have an additional $U(1)$ factor, $r_{B_d}$ depends on the $U(1)$
charge as well.  We summarize the full $\und{r}$-fields of unity
blowup equations for the $dP_d$ theory, given by
\begin{equation}
  \und{r} = (r_{B_d}, \und{r}_m)  ,
\end{equation}
in Table~\ref{tb:r-delPezzo}.  On the other hand, the vanishing
$\und{r}$-field is always
\begin{equation}
  \und{r} = (r_B,\und{r}_m) = (-(9-d),\und{0}).
\end{equation}

Taking the limit of \eqref{eq:blow-down} on \eqref{eq:dP8-ublowup} and
using \eqref{eq:Zd}, after some algebraic gymnastics in the end we
arrive at the following blowup equations of $dP_d$
\begin{align}
  \sum_{n\in\IZ}
  &(-1)^n\md{e}\[\frac{1}{2}n(n+\frac{r_{B_d}}{9-d})t_B
    +\frac{1}{2}n\sum_{i=1}^d
    \wt{m}_i(r_{\wt{m},i}+\frac{|r_{\wt{m}}|}{9-d})
    +P_{dP_d}(n)(\eq+\et)\]\wh{Z}^{\text{inst}}_{dP_d}(t_{B_d},\und{m},\eq,\et)^{-1}
    \nn \times
  &\wh{Z}^{\text{inst}}_{dP_d}
    (t_{B_d}+\frac{\eq}{2}r_{B_d},\und{m}+\frac{\eq}{2}\und{r}_m,\eq,\et-\eq)
    \wh{Z}^{\text{inst}}_{dP_d}
    (t_{B_d}+\frac{\et}{2}r_{B_d},\und{m}+\frac{\et}{2}\und{r}_m,\eq-\et,\et)
    \nn =
  &\begin{cases}
    0, & r_{B_d} =-(9-d),\und{r}_m = \und{0}\\
    1-\md{e}\[\frac{1}{2}\sum_{i=1}^d
      \wt{m}_i(r_{\wt{m}_i}+\frac{|r_{\wt{m}}|}{9-d})
      +(\eq+\et)\], & r_{B_d} =
    -(9-d),\und{r}_m\neq \und{0}\\
    1, &\text{otherwise}
    \end{cases}
\end{align}
where $|r_{\wt{m}}| = \sum_{i=1}^d r_{\wt{m},i}$ and\footnote{This formula is derived from an identity of the orbit $9r_h^2-\sum_{i=1}^{d}(r_h+r_{\tilde{m}_i})^2=1-d.$ Here $r_h = \frac{r_{B_d}+\sum_{i=1}^d r_{\wt{m},i}}{9-d}$.}
\begin{equation}
  P_{dP_d}(n) = \(\delta_u-\frac{9-d}{12}\)n - \frac{r_{B_d}}{4}n^2 - \frac{9-d}{6}n^3.
\end{equation}
Furthermore, by taking the limit \eqref{eq:blow-down} on
\eqref{eq:E-F0}, we also find the universal form of the perpurbative
prepotential of $dP_d$, which reads\footnote{The prepotential of the
  half K3 surface \eqref{eq:E-F0} can be seen as a special case of
  this universal formula with $d=9$ if we identify
  $\wt{m}_9 = -\tau - \sum_{i=1}^8 \wt{m}_i$.}
\begin{equation}
  F_{(0,0)}^{dP_d} = \frac{9-d}{6}t_h^3 - \frac{1}{2}t_h^2
  \sum_{i=1}^d \wt{m}_i - \frac{1}{2}t_h \sum_{i=1}^d\wt{m}_i^2
\end{equation}
as well as
\begin{equation}
b_h^{\text{GV}}=-\frac{3+d}{24},\quad b_h^{\text{NS}}=-\frac{3-d}{12}.
\end{equation}
where we have defined the K\"ahler parameter
\begin{equation}
  t_h = \frac{1}{3}\text{Vol}(h) = \frac{t_{B_d}+\sum_{i=1}^d \wt{m}_i}{9-d}.
\end{equation}
These results as well as the $\und{r}$-fields given in
Table~\ref{tb:r-delPezzo} have been successfully compared with
\cite{Huang:2017mis}\footnote{In \cite{Huang:2017mis} the last
  $\und{r}$-field component other than $\und{r}_m$ is $r_h$ associated
  to $t_h$, which is related to $r_{B_d}$ by
  $r_h = \frac{r_{B_d}+\sum_{i=1}^d r_{\wt{m},i}}{9-d}$.}.

\begin{table}
  \centering%
  \resizebox{\linewidth}{!}{
    \begin{tabular}{*{5}{>{$}c<{$}}}\toprule
      & G & \#\text{vanishing} & \#\text{unity}
      & \text{unity}\;\;
        \frac{1}{2}\und{r} = (\frac{1}{2}r_{B_d},\frac{1}{2}\und{r}_m) \\
      \midrule
      dP_8 & E_8   & 1& 240&(-\frac{1}{2},\cO_{2,240}) \\
      dP_7 & E_7   & 1& 182&(-1,\cO_{2,126}),(0,\cO_{\frac{3}{2},56})\\
      dP_6 & E_6   & 1& 126&(-\frac{3}{2},\cO_{2,72}),\pm(-\frac{1}{2},\cO_{\frac{4}{3},27})\\
      dP_5 & SO(10)& 1&  82
      &(-2,\cO_{2,40}),\pm(-1,\cO_{\frac{5}{4},16}),\pm(0,\cO_{1,10})\\
      dP_4 & SU(5) & 1&  50
      &(-\frac{5}{2},\cO_{2,20}),\pm(-\frac{3}{2},\cO_{\frac{6}{5},10}),\pm(-\frac{1}{2},\cO_{\frac{4}{5},5})\\
      dP_3 & SU(3)\times SU(2)
          & 1&  28
      &
        (-3,\cO_{2,6},\cO_{0,1}),\pm(-2,\cO_{\frac{2}{3},3},\cO_{\frac{1}{2},2}),
        (-3,\cO_{0,1},\cO_{2,2}),\pm(-1,\cO_{\frac{2}{3},3},\cO_{0,1}),(0,\cO_{0,1},\cO_{\frac{1}{2},2}) \\
      dP_2 & SU(2)\times U(1)
          & 1&  14
      & (-\frac{7}{2},\cO_{2,2;0}),\pm(-\frac{5}{2},\cO_{0,1;4}),\pm(-\frac{5}{2},\cO_{\frac{1}{2},2;-3}),\pm(-\frac{3}{2},\cO_{\frac{1}{2},2;+3}),\pm(-\frac{1}{2},\cO_{0,1;-2})\\
      dP_1 & U(1)  & 1&  6
      & \pm(-3_{-3}),\pm(-2_2),\pm(-1_{-1}) \\
      \IP^1\times \IP^1 & SU(2)
          & 1&   7&
                    (-4,\cO_{2,2}),\pm(-2,\cO_{\frac{1}{2},2}),(0,\cO_{0,1})\\
      \IP^2 & -  & 1&   2& \pm(\frac{3}{2})\\
      \bottomrule
    \end{tabular}}
  \caption{$\und{r}$ fields for del Pezzo surfaces.  We denote
    $\frac{1}{2}\und{r}_m$ as an element of a Weyl orbit $\cO_{n,p}$.
    In the case of $dP_2$ and $dP_1$, we use an extra subscript to
    denote the $U(1)$ charge.}
  \label{tb:r-delPezzo}
\end{table}


There is an additional del Pezzo surface $\IF_0=\IP^1\times\IP^1$,
which does not fall into the family of $dP_d$.  On the other hand, it
is known that $dP_2$ is isomorphic to the one-point blowup of $\IF_0$.
The $\IP^1$ base $e$, the $\IP^1$ fiber $f$, and exceptional curve $x$
of the latter are identified as
\begin{equation}
  e = h-x_1, \quad f = h-x_2,\quad h-x_1-x_2.
\end{equation}
Therefore the surface $\IF_0$ can be obtained by blowing down the
exceptional curve $h-x_1-x_2$ of $dP_2$, corresponding to taking the
limit
\begin{equation}
  \wt{m}_{1,2} \to \wt{m}_{1,2} - 2\varepsilon,\quad t_{B_2} \to
  t_{B_2} - 3\varepsilon,\quad \varepsilon\to -\ri\infty.
\end{equation}
We identify the K\"ahler parameter of the anti-canonical class
$t_{B_{\IF_0}}$ and the mass parameter $\wt{m}$ of $\IF_0$ as
\begin{equation}
  t_{B_{\IF_0}} = \frac{8}{7}t_{B_2} -
  \frac{6}{7}(\wt{m}_1+\wt{m}_2),\quad
  \wt{m} = \wt{m}_1 - \wt{m}_2.
\end{equation}
Another commonly used set of K\"ahler parameters for $\IF_0$ consists
of volumes of $e$ and $f$ denoted by $t_{1,2}$, which are related to
$t_{B_{\IF_0}}$ and $\wt{m}$ through
\begin{equation}
  t_1= \frac{1}{4}t_{B_{\IF_0}} - \frac{1}{2}m,\quad
  t_2= \frac{1}{4}t_{B_{\IF_0}} + \frac{1}{2}m,
\end{equation}
with which the prepotential up to irrelevant terms reads
\begin{equation}
  F_{(0,0)} = -\frac{1}{24}(t_1^3+t_2^3) + \frac{1}{8}(t_1^2t_2 + t_1 t_2^2).
\end{equation}
The transformation of the blowup equations are analogous.  Only the
spin (0,0) BPS state wrapping the exceptional curve $h-x_1-x_2$
contribute diverge terms in the blowup equation and they can be
computed using \eqref{eq:Bl-hat}.
The $\und{r}$-fields can be computed by decomposing those of $dP_2$.
We list the unity $\und{r}$-fields in Table~\ref{tb:r-delPezzo}, and
the only vanishing $\und{r}$-field is
\begin{equation}
  \und{r} = (-8,0).
\end{equation}
They also agree with \cite{Huang:2017mis}\footnote{In
  \cite{Huang:2017mis} the $\und{r}$-field is presented by the
  components $r_z,r_m$ associated to curves $e,f-e$.  They are related
  to $r_{B_{\IF_0}},r_m$ via $r_{B_{\IF_0}} = 4r_z+2r_m$.}.  Note that
the theory of $\IF_0$ is a bit special.  Unlike $dP_d$ theories, there
is a unity equation with $\und{r}_m = 0$ in the theory of $\IF_0$.  In
addition, $r_{B_{\IF_0}}$ of the anti-canonical class is not
completely determined by $\und{r}_m$: for
$\frac{1}{2}\und{r}_m \in \cO_{\frac{1}{2},2}$, which is real,
$r_{B_{\IF_0}}$ has two possible values not equivalent to each other.

\section{From path integral to blowup equations}
\label{sc:der}


Here we try to understand the form of the elliptic blowup equations
\eqref{eq:e-blowup} for the E--,M--string theory on torus from the path
integral point of view.

6d SCFTs can be constructed by compactifying F-theory on elliptic
Calabi-Yau threeefolds with the following geometric properties of base
and fiber.  The base $B$ is a non-compact, complex two-dimensional
space.  As such, it contains 2-cycles $C^i$ which are $\mathbb{P}^1$'s
with negative intersection matrix $-\Omega^{ij} = C^i \cdot C^j$.
Furthermore, in general above each $C^i$ the elliptic fiber can
degenerate according to a Kodaira singularity.  In the resulting 6d
field theory, each 2-cycle $C^i$ in the base gives rise to a tensor
multiplet.  The bosonic components of the tensor multiplets are
denoted by $(\phi_i,B_i)$, where $\phi_i$ are real scalars and $B_i$
are 2-forms with field strengths $H_i$.  The volume of the 2-cycle
$C^i$ is proportional to $\phi^i = \Omega^{ij} \phi_j$.  Singular
elliptic fibers over $C^i$ signal the existence of gauge symmetry.
Since we only consider theories with no gauge symmetry, the Calabi-Yau
has no singular elliptic fiber.  Finally there are hypermultiplets
corresponding to isolated $(-1)$ curves in the fibral direction.  We
will also use the fact that there can be string-like objects charged
under the tensor multiplets.  A string with worldsheet $\Sigma_2$ and
tensor charges $n_i$ sources a flux of $H_i$
\begin{equation}
  \label{eq:stringsources}
  dH_i = n_i \delta_{\Sigma_2} \quad \Longleftrightarrow
  \quad n_i = \int_{\Sigma_3} H_i,
\end{equation}
where $\delta_{\Sigma_2}$ is the unit delta function localised on
$\Sigma_2$, which is linked by the three-cycle $\Sigma_3$.

When we put the 6d theory on $T^2$, the hypermultiplets become
hypermultiplets in the 4d field theory, and the tensor multiplets are
reduced to vector multiplets by
\begin{equation}
  \phi_i\to \phi_i,\quad H_i\to F_i\wedge \rd x + \star^{(4)} F_i \wedge
  \rd y + a_i \rd x\wedge \rd y,
\end{equation}
which induce a non-Abelian gauge symmetry.  Here $\star^{(4)}$ is the
4d Hodge star so that $H_i = \star H_i$ in 6d.  Each vector multiplet
includes as bosonic components a complex scalar
$\varphi_i = \phi_i+\ri a_i$ and a 1-form gauge connection $A_i$ whose
field strength is $F_i$.  Their kinetic terms in the Lagrangian are
\begin{equation}\label{eq:L-4d}
  \cL^{(4)}_{\text{Coulomb}} = -\tau
  \Omega^{ij}\(F_i\wedge\star^{(4)} F_j + \pd_\mu
  \varphi_i\pd^\mu\wb{\varphi}_j\) + \ldots
\end{equation}
Here the bare gauge coupling $\tau$ is identified with the complex
structure parameter of the torus on which the 6d SCFT is compactified.

We would like to compute the partition function of the 6d SCFT on
$M_6 = T^2\times_{\eq,\et} M_4$ with $M_4$ being $\wh{\IC}^2$, namely
the blowup of $\IC^2$ at the origin.  The four manifold $M_4$ is the
same as the total space of the $\cO(-1)$ bundle over $\IP^1$, which
can be parameterised by $z_0,z_1,z_2\in\IC$ with the equivalence
relation
\begin{equation}
  (z_0,z_1,z_2) \sim (\lambda^{-1}z_0,\lambda^1 z_1,\lambda^1
  z_2),\quad \lambda\in\IC^*.
\end{equation}
We also turn on the 6d Omega background \cite{Losev:2003py} which has
the effect of rotating the $M_4$ when one goes around 1-cycles in
$T^2$ with the $U(1)_{\eq}\times U(1)_{\et}$ action as follows:
\begin{equation}
  (z_0,z_1,z_2) \mapsto (z_0,\re^{\eq}z_1,\re^{\et}z_2).
\end{equation}
The $U(1)_{\eq}\times U(1)_{\et}$ action has two localised points at
the north pole and the south pole of the exceptional $\IP^1=S^2$ at
$(z_0,z_1,z_2)=(0,1,0),(0,0,1)$.
\begin{equation}
  \begin{aligned}
    &(z_0z_1,z_2/z_1) \mapsto (\re^{\eq}z_0z_1,\re^{\et-\eq}z_2/z_1)\\
    &(z_0z_2,z_1/z_2) \mapsto (\re^{\et}z_0z_2,\re^{\eq-\et}z_1/z_2)
  \end{aligned}
\end{equation}
Note that the $U(1)_{+}$ part of the Omega background with fugacity
$\epsilon_+ = (\eq+\et)/2$ can be identified with the $U(1)$ component
of the 6d $SU(2)_R$-symmetry, which acts non-trivially on the Wilson
lines of the 6d flavor symmetry.  This is due to the fact that upon
compactifying a 6d SCFT on $T^2$ with non-trivial Wilson lines the
conformal symmetry present in 6d is broken and the 6d conformal
stress-tensor multiplet splits into a 4d stress-tensor multiplet and a
flavor current multiplet associated with the Kaluza-Klein symmetry,
i.e.\ the momentum in the reduced direction.  This KK current then
gets identified with a linear combination of the flavor currents of
the original 6d flavor symmetry \cite{Cordova:2016xhm}.  This
identification can be thought of as an embedding of the
$SU(2)_R$-symmetry into the 6d flavor symmetry group.  As a result the
twisted flavor mass
\begin{equation}\label{eq:m-twist}
  \wh{\und{m}} := \und{m} + \wh{\und{r}}\epsilon_+
\end{equation}
with $\wh{\und{r}}$ some root vector of the flavor group, should be
invariant throughout $M_4$.
We will identify $\wh{\und{r}}$ with $\und{r}_m$ later.

There are essentially two types of configurations in the 6d field
theory with finite energy.  The first type corresponds to the string
worldsheets wrapping the torus.  They appear as point-like instantons
in $M_4$ and because of the action of $U(1)_{\eq}$ and $U(1)_{\et}$
are localised at the north pole and the south pole of the exceptional
$\IP^1$.  The path integral receives contributions from the one-loop
determinant corresponding to the elliptic genera of the strings.  In
the sector of $\und{k}'$ wrapped strings localised at the north pole
and $\und{k}''$ wrapped strings localised at the south pole of $\IP^1$
we expect the contribution of
\begin{equation}\label{eq:EE}
  \IE_{\und{k}'}(\tau,\und{m}+\frac{1}{2}\wh{\und{r}}\eq,\eq,\et-\eq)
  \IE_{\und{k}''}(\tau,\und{m}+\frac{1}{2}\wh{\und{r}}\et,\eq-\et,\et).
\end{equation}
The shifts on the flavor mass are to make sure that the twisted flavor
mass \eqref{eq:m-twist} is invariant.

The other type of finite energy configuration is the flux of the
self-dual 3-form through $S^1\times \IP^1$.  We provide two ways to
evaluate its contribution to the partition function.  The first method
is to consider the 4d field theory resulting from the torus
compactification.  Let us first consider a 6d SCFT with a single
tensor multiplet which is reduced on $T^2$ to a $U(1)$ gauge theory in
4d with gauge connection $A$.  The partition function of such a theory
on a four manifold $M_4$ is \cite{Witten:2007ct}
\begin{equation}
  Z(M_4) = \sum_{\mathcal{L}} \int D A \exp(-I(A)),
\end{equation}
where the sum runs over all the line bundles on which $A$ is a
connection.
To describe the path integral more explicitly, one
decomposes $A = A' + A^{\mathcal{L}}_h$, where $A'$ is a connection on
a trivial line bundle $\mathcal{O}$, and $A^{\mathcal{L}}_h$ is a
connection on $\mathcal{L}$ of harmonic curvature
$F^{\mathcal{L}}_h$.
The path integral then becomes
\begin{equation}
  \sum_{\mathcal{L}} \int D A \exp(-I(A)) =
  \int D A' \exp(-I(A')) \sum_{\mathcal{L}} \exp(-I(A^{\mathcal{L}}_h)).
\end{equation}
Let us now look at the sum over $\mathcal{L}$.  On the lattice
$H^2(M;\mathbb{Z})$, there is a natural, generally indefinite
quadratic form given, for $x$ an integral harmonic two-form, by
$(x,x) = \int_M x \wedge x$. There is also a positive-definite but
metric-dependent form
$\langle x, x \rangle = \int_M x \wedge \star^{(4)} x$. The indefinite
form $(x,x)$ has signature $(b_{2,+},b_{2,-})$, where $b_{2,\pm}$ are
the dimensions of the spaces of self-dual and anti-self-dual harmonic
two-forms. Setting $x= F^{\mathcal{L}}_h/2\pi$, the sum over line
bundles becomes
\begin{equation}
  \Theta = \sum_{x \in H^2(M;\mathbb{Z})} \exp
  \left(-\pi\textrm{Im}(\tau)\langle x,x \rangle
    + \ri\pi\textrm{Re}(\tau)(x,x)\right).
\end{equation}
This is a Siegel-Narain theta function with modular parameter
$\tau$. In the case where the 4d theory comes from a 6d parent theory
as in our case, the big theta function splits into a product of
holomorphic and anti-holomorphic theta functions:
\begin{equation}
  \Theta = \sum_{a} \Theta^{[a]} \wb{\Theta}^{[a]},
\end{equation}
where the sum over $a$ is a sum over the so-called ``quadratic
refinements'' and $\Theta^{[a]}$ arise from the path-integral over
self-dual parts of the 3-form $H$ while $\wb{\Theta}^{[a]}$ correspond
to the anti-self-dual parts.  In the case of $M = \wh{\IC}^2$,
$(b_{2,+},b_{2,-})=(1,0)$, the two quadratic forms $(x,x)$ and
$\vev{x,x}$ are identical and $\wb{\Theta}^{[a]}$ become trivial.

Relaxing now the condition that the tensor branch of the 6d SCFT is
only one-dimensional and using \eqref{eq:L-4d}, we deduce that in the
general case the path integral of the 4d theory produces the theta
function
\begin{equation}\label{eq:Theta-2}
  \Theta_{\Omega}^{[a]}(\tau,z) \equiv \sum_{n_i \in \mathbb{Z}}
  \md{e}[\frac{1}{2} \Omega^{ij} (n_i+a_i) (n_j+a_j) \tau
  + \Omega^{ij} (n_i+a_i) z_j].
\end{equation}
which is essentially the generalised theta function we defined in
\eqref{eq:Theta} up to the sign factor.
The $\Theta_{\Omega}^{[a]}$ can be seen as
sections of a line bundle over the torus
\begin{equation}
  \mathbb{T} \equiv \mathbb{C}^r/\left(\Omega\mathbb{Z}^r
    \oplus \tau \Omega \mathbb{Z}^r\right).
\end{equation}
We remark at this point that the number of such sections is equal to
the determinant of $\Omega$.  We also include elliptic parameters
$z_j$ which are possible source terms coupled to $F_i$.  These can be
the instanton strings localised at the north and the south poles of
$\IP^1$ or they can result from the following Green-Schwarz
counter-terms in the 6d theory
\begin{equation}
  2\pi\int_{M_6} \Omega^{ij} B_i\wedge I_j.
\end{equation}
where $I_j$ are the 't Hooft anomaly four-forms on the worldsheet of
strings.
We will write down explicit expressions of the elliptic
parameters with the second method of evaluation.

The second method of evaluation is through a holographic argument
which we outline below.  Following
\cite{Witten:1996hc,Witten:1998wy,Belov:2006jd,Shimizu:2016lbw}, there
exists a holographic action for the tensor branch of 6d SCFTs on a
seven-manifold $Y_7$ with boundary $\partial Y_7 = M_6$, whose
topological part is:
\begin{equation}\label{eq:CSaction}
  S_7^{\textrm{top}} =
  2\pi \int_{Y_7} \Omega^{ij} \( \frac{1}{2} dC_i \wedge C_j + C_i \wedge I_j\),
\end{equation}
where $C_i$ are 3-forms.  Variation with respect to $C_i$ gives
\begin{eqnarray}
  0 & = & \int_{Y_7} \left(dC_i - I_i\right) \wedge \delta C_j \Omega^{ji} \nonumber \\
  ~ & = & \int_M \left(C_i - \omega_i\right) \wedge \delta C_j \Omega^{ji} \quad \textrm{giving } C_i = \omega_i  \textrm{ on }M_6,
\end{eqnarray}
where $d \omega_i = I_i$. We want to compute the partition function
arising from the path integral
\begin{equation}\label{eq:CSpathint}
  \int \left[\prod_i \mathcal{D} C_i\right]_{C_i = \omega_i
    \textrm{ on } M_6}
  \exp\left(\ri S_7^{\textrm{top}}\right).
\end{equation}
The result will be a state in the Hilbert space $\mathcal{H}_{M_6}$
arising from the quantisation of the CS-action. Let us see what this
Hilbert space is. From the action (\ref{eq:CSaction}) we get upon
quantisation the following commutation relations:
\begin{equation}
  [ C_i(x), C_j(y)] = - 2\pi \ri~\Omega^{-1}_{ij}~ \textrm{Vol}_{M_6}~ \delta^6(x-y),
\end{equation}
Taking again $M_6$ to be of the form
$S^1_A \times S^1_B \times \widehat{\mathbb{C}^2}$, we can define
operators
\begin{equation}\label{eq:phidef}
  \Phi^A_i = \exp\(\ri \int_{S^1_A \times \mathbb{P}^1} C_i\), \quad
  \Phi^B_j = \exp\(\ri \int_{S^1_B \times \mathbb{P}^1} C_j\),
\end{equation}
These satisfy commutation relations
\begin{equation}
  \Phi_A^i \Phi_B^j = \Phi_B^j \Phi_A^i \exp\left(2\pi \ri~\Omega^{-1}_{ij}\right).
\end{equation}
This defines a Heisenberg group extension of
\begin{eqnarray}
  K \times K & \equiv & H^2(\widehat{\mathbb{C}^2},\mathbb{Z}_{\det \Omega}) \times H^2(\widehat{\mathbb{C}^2},\mathbb{Z}_{\det \Omega}) = \mathbb{Z}_{\det \Omega} \times \mathbb{Z}_{\det \Omega}: \nonumber \\
  ~ & ~ & 0 \rightarrow \mathbb{Z}_{\det \Omega} \rightarrow F \rightarrow K \times K \rightarrow 0.
\end{eqnarray}
Let us now for simplicity restrict to the case where there is only one
tensor multiplet and $\Omega^{11} = \fn$. Heisenberg groups have a
unique irreducible representation $\mathcal{R}$ such that there exists
a state $|\Omega\rangle \in \mathcal{R}$ with the property
\begin{equation}
  \Phi^A(a) |\Omega\rangle = |\Omega\rangle , \quad a \in \mathbb{Z}_\fn.
\end{equation}
From this state we obtain a basis for $\mathcal{H}$ as follows
\begin{equation}
  \Psi_a = \Phi^B(a) |\Omega\rangle, \quad \textrm{for } a \in \mathbb{Z}_\fn.
\end{equation}
We want to identity $\mathcal{R}$ with $\mc{H}_{M_6}$. There is a
natural $SL(2,\mathbb{Z})$ action on $\mc{H}_{M_6}$:
\begin{equation}
  \(\begin{array}{cc}0 & 1 \\ -1 & 0\end{array}\) \quad : \quad
  \tau \mapsto - \frac{1}{\tau}
\end{equation}
implying that $\Phi^A$ maps to ${\Phi^B }^{-1}$ and $\Phi^B$ maps to
$\Phi^A$. Under this maps $|\Omega \rangle$ maps to
\begin{equation}
  |\widetilde{\Omega}\rangle \sim \sum_{b \in \mathbb{Z}_\fn} \Phi^B(b) |\Omega\rangle,
\end{equation}
while $\Psi_a$ maps to
\begin{equation}\label{eq:thetatrf}
  \Phi^A(a) |\widetilde{\Omega}\rangle \sim \sum_{b \in \mathbb{Z}_\fn}
  \exp(2\pi \ri a b /\fn) \Phi^B(b) |\Omega\rangle.
\end{equation}
This is precisely the transformation law for a theta function.
Another way to see this is to define
\begin{equation}\label{eq:phasespace}
  \alpha^A_i \equiv \int_{S^1_A \times \mathbb{P}^1} C_i, \quad
  \alpha^B_j \equiv \int_{S^1_B \times \mathbb{P}^1} C_j,
\end{equation}
which give phase space coordinates of our 7d Chern-Simons
theory. $\alpha^A_i$ and $\alpha^B_j$ take values in $V/\Gamma$ where
\begin{equation}
  V = H^2(\mathbb{P}^1,\mathbb{R}), \quad \textrm{ and } \Gamma = H^2(\mathbb{P}^1,\mathbb{Z}).
\end{equation}
Thus we see that the phase space is a torus
\begin{equation}
  V/\Gamma \times V/\Gamma = H^3(M_6,\mathbb{R})/H^3(M_6,\mathbb{Z}) = J_{M_6},
\end{equation}
with $J_{M_6}$ being the intermediate Jacobian of $M_6$. Quantisation
amounts to finding the appropriate line bundle $\mathcal{L}$ over
$J_{M_6}$. Since the symplectic form carries a factor of
$\Omega^{ij}$, the sections are theta functions of the form
(\ref{eq:Theta-2}).  This perspective also helps resolve the apparent
discrepancy of the sign factor between \eqref{eq:Theta-2} and
\eqref{eq:Theta}.  In the case of $\fn=1$ (E--string), one has to
choose characteristic $a=1/2$ and make the shift $z\to z+\frac{1}{2}$
so that $\Theta^{[a]}_\Omega$ reduces to $\theta_1(\tau,z)$ as it is
the unique $SL(2,\IZ)$-invariant section for a line bundle with
principal polarisation, which precisely means $\fn=1$ here, over the
elliptic curve (see \cite{Witten:1996hc}).  This issue does not arise
for $\fn=2$ (M--string) and we can make the canonical choice of the
characteristic $a = i/\fn$ for $i=0,1$ without shifting $z$, whereupon
$\Theta^{[a]}_{\Omega}$ reduces to
$\theta_2(2\tau,z),\theta_3(2\tau,z)$ that furnish an irreducible
represetnation of $SL(2,\IZ)$.

A glance at (\ref{eq:CSpathint}) and (\ref{eq:phasespace}) shows that
the elliptic parameters of the theta function are given by
\begin{equation}\label{eq:thetaarg}
  z_k = \int_{S^1_A \times \mathbb{P}^1} \omega_k +
  \ri \int_{S^1_B \times \mathbb{P}^1} \omega_k.
\end{equation}
From now on, let us restrict to theories with only one tensor
multiplet and no gauge symmetry in 6d.
In this case $\Omega^{ii} = \fn$ and the anomaly four-form reads
\begin{equation}
  \Omega^{ii}I_i = \frac{\fn}{4}\Tr F_a^2 + c_2(R)-\frac{2-\fn}{4}p_1(M_6)
\end{equation}
Therefore we expect
\begin{equation}
  z = \int_{S^1\times\IP^1} \frac{\fn}{4}\omega_F + \omega_R
  - \frac{2-\fn}{4}p_1^{-1}(M_6)
\end{equation}
where
\begin{equation}
  \omega_F = \Tr\(\frac{2}{3}A_a^3+A_a\wedge\rd A_a\),\quad
  \omega_R = \Tr\(\frac{2}{3}A_R^3+A_R\wedge\rd A_R\).
\end{equation}
and $p_1^{-1}$ is the anti-derivative of the Pontryagin class.
The first term gives
\begin{equation}
  \frac{\fn}{4}\int_{S^1\times\IP^1} \omega_F \to
  \frac{\fn}{4}\Big(\int_{\IP^1} \und{F}\Big)\cdot\und{m}
\end{equation}
In order to compute the flux vector
$\frac{\fn}{4}\int_{\IP^1}\und{F}$, we make use of the fact that our
partition function should be invariant up to a sign under large gauge
transformations \cite{Witten:1996hc} (we treat the flavor symmetry as
a weakly coupled gauge symmetry)
\begin{equation}
  \und{m} \mapsto \und{m} + \alpha^\vee
\end{equation}
where $\alpha^\vee$ is an arbitrary coroot of the flavor group.
This implies that the flux vector $\frac{\fn}{4}\int_{\IP^1}\und{F}$
has to be a weight vector of the flavor group, which we also denote by
$\frac{1}{2}\und{r}_m$.
Given the embedding of the $SU(2)_R$-symmetry in the flavor symmetry,
we can assume
\begin{equation}
  \int_{S^1\times\IP^1}\omega_R \mapsto \int_{\IP^1}\und{F}_R\cdot\int_{S^1}\und{A}_R
  =\frac{1}{2}\und{r}_m\cdot\wh{\und{r}}
  \epsilon_+\quad\text{with}\;\wh{\und{r}} = \und{r}_m.
\end{equation}
Finally, if we do the replacement
\begin{equation}
  \int_{S^1\times\IP^1} p_1^{-1}(M_6) \mapsto 8c\,\epsilon_+
\end{equation}
with $c$ certain constant, we find the following elliptic parameter
\begin{equation}\label{eq:z}
  z = \frac{1}{2}\und{r}_m\cdot\und{m} +
  (\frac{1}{4}\und{r}_m\cdot\und{r}_m-(2-\fn)c) (\eq+\et)
\end{equation}
which has the same form as the elliptic parameter of the theta
function on the LHS of \eqref{eq:e-blowup} up to the last two terms
corresponding to coupling with strings.

The coupling with string sources can be added to the Chern-Simons
action (\ref{eq:CSaction}) as follows (see \cite{Shimizu:2016lbw}):
\begin{eqnarray}
  S_7^{\textrm{top}}
  & =
  & 2\pi \int_{Y_7} \Omega^{ij} \(\frac{1}{2}
    \rd C_i \wedge C_j + C_i \wedge (I_j + d_{j,1} \chi_4(N) + d_{j,2}
    \chi_4(N) )\)
    \nonumber\\  ~
  & =
  & 2\pi \int_{Y_7} \Omega^{ij} \( \frac{1}{2}
    \rd C_i \wedge C_j + C_i \wedge I_j \) + 2\pi \Omega^{ij}
    \(d_{j,1}
    \int_{T^2 \times \mathbb{R}_+}  C_i+ d_{j,2}
    \int_{T^2 \times \mathbb{R}_+} C_i\),\nonumber \\
\end{eqnarray}
where $\chi_4(N)$ is the Euler class of the normal bundle of the
string.  In the case of $\text{rk}\Omega = 1$, using
\begin{equation}
  \chi_4(N) = de_3^{(0)},
\end{equation}
where $e_3^{(0)}$ is the global angular form of the $S^3$ bundle of
the tubular neighborhood of the string, we see that the elliptic
parameter of our theta function gets shifted by
\begin{equation}
  \fn(k_1 \int_{S^1 \times \mathbb{P}^1} e_3^{(0_N)} + k_2 \int_{S^1
    \times \mathbb{P}^1} e_3^{(0_S)} )
  = \fn(\epsilon_1 k_1 + \epsilon_2 k_2),
\end{equation}
where we have inserted string sources at the north and south pole of
the exceptional $\mathbb{P}^1$.  We have thus completely reproduced
the theta function, and together with \eqref{eq:EE}, the entire LHS of
\eqref{eq:e-blowup}.

In order to obtain the RHS of \eqref{eq:e-blowup} we simply choose the
$S^3$ to surround the entire exceptional $\IP^1$ fully.
The terms in the theta function corresponding to coupling with strings
disappear because
\begin{equation}
  \fn k \int_{S^1 \times \mathbb{P}^1} e_3^{(0)} = 0
\end{equation}
and we are left only with the elliptic parameter \eqref{eq:z}.  The
contributions of string worldsheet wrapping $T^2$ merge to
\begin{equation}
  \IE_k(\tau,\und{m},\eq,\et)\quad\text{with}\;k=k_1+k_2.
\end{equation}

Finally we can invoke the modularity consistency
condition to uniquely fix $\und{r}_m$ by \eqref{eq:rm2-mod} and the
constant $c$ through \eqref{eq:y}.  Vanishing equations can arise if
there exist values of $\und{r}_m$ and $c$ so that the theta function
on the RHS vanishes identically, which happens in the case of
E--string.

\section{Conclusion}
\label{sc:con}

In this paper we propose the blowup equations for E--string and
M--string theories compactified on $T^2$, and solve their elliptic
genera and equivalently the refined BPS invariants of the associated
Calabi-Yau threefolds from these equations.  Although the elliptic
genera of these two theories have been studied via many other
approaches, for instance localization in 2d quiver gauge theories
\cite{Kim:2014dza,Haghighat:2013gba} or modular bootstrap
\cite{Gu:2017ccq} where one can write down explicit expressions for
the elliptic genera, the blowup method still sheds some new light on
this subject and gives inspiration on how to deal with more general 6d
theories.  As is well-known, the E--string and M--string are the two
simplest rank-one 6d SCFTs. Both have no gauge symmetry but only
flavor symmetry.  In the previous papers of this series, we have
established the elliptic blowup equations for rank-one pure gauge 6d
SCFTs on torus \cite{Gu:2018gmy,Gu:2019dan}.  With the inspiration
obtained in the current paper with regard to the E-- and M--string
theories, we indeed find the elliptic blowup equations for all rank
one SCFTs with both gauge and flavor symmetry \cite{IV} and even
higher rank ones \cite{V}, thus exhausting all possibilities of
untwisted torus reductions of 6d SCFTs.

We also present the elliptic blowup equations for M--string chains and
E--M string chains.  Note that the M--string chains are special cases
of the much more general ADE string chains
\cite{Gadde:2015tra}. Indeed, the $N$ M--string chain is in fact the
$A_{N-1}$ type chain without gauge symmetry.  Although the elliptic
genera for all ADE string chains can be computed from localization, it
is still interesting to consider their elliptic blowup equations and
whether their elliptic genera can be calculated from these equations.
We would like to leave the discussion of this subject to the higher
rank paper \cite{V}.

In addition, we provide a simple procedure to compute the full
perturbative prepotential with all mass parameters turned on of a
non-compact Calabi-Yau threefold from its local description as a
connected union of compact surfaces.  This paves the way for deriving
blowup equations for twisted circle reductions of 6d SCFTs from the
local description of their associated geometries given in
\cite{Bhardwaj:2019fzv}.  Relatedly, all rank one 5d SCFTs can be
obtained from circle reductions of the E--string theory by decoupling
a mass deformed hypermultiplet or from the geometric point of view by
blowing down an exceptional curve.  We demonstrate in this paper how
to perform this operation on the blowup equations of E--strings on the
torus to obtain the blowup equations of rank one 5d SCFTs.  Recently
there has been much progress towards the classification of 5d
SCFTs~\cite{Bhardwaj:2018yhy,Bhardwaj:2018vuu,Jefferson:2018irk,Bhardwaj:2019jtr,Bhardwaj:2019fzv,Apruzzi:2018nre,Apruzzi:2019vpe,Apruzzi:2019opn,Apruzzi:2019enx},
and it is conjectured that all 5d SCFTs can be obtained from untwisted
or twisted circle reductions of 6d SCFTs \cite{DelZotto:2017pti,
  Jefferson:2018irk}.  Therefore, we can write down a recipe to derive
blowup equations for all 5d SCFTs compactified on a circle from the
blowup equations of untwisted and twisted torus reductions of 6d
SCFTs.

As a byproduct of the unity blowup equations for the one E--string
elliptic genus, we obtain some novel functional equations for the
$E_8$ theta function. The unity blowup equations for more E--strings
result in more identities among $E_8$ Weyl invariant Jacobi forms. It
is desirable to know whether one can prove from the viewpoint of
Jacobi forms that our system of elliptic blowup equations only allows
for a one-dimensional solution space.

Besides, the blowup equations have connections with the bilinear
relations of isomonodromic systems \cite{Bershtein:2018zcz}. We hope
the blowup equations we find in this paper can shed some new light on
this subject. In particular, the E--string blowup equations are
expected to produce the bilinear relations of the elliptic Painlev\'e
equations \cite{Mizoguchi:2002kg}, while the M--string blowup
equations are expected to produce those of the isomonodromic system on
the one-punctured torus \cite{Bonelli:2019boe}.

Finally, a long-standing problem is a possible proof of the blowup
equations for 6d SCFTs compactified on the torus.  In this paper we
make the first attempt and derive from the path integral point of view
the elliptic blowup equations for the E--,M--string theories.  This
method can in principle be applied to derive the blowup equations for
the M--string chains and E--M string chains as well, which also have
no gauge symmetry.  We hope our derivation can inspire proofs for the
blowup equations for a generic 6d SCFT.

\section*{Acknowledgement}

We would like to thank Giulio Bonelli, Fabrizio Del Monte, Lothar
G\"ottsche, Min-xin Huang, Joonho Kim, Hiraku Nakajima, Sakura
Sch\"afer-Nameki, Alessandro Tanzini, Yi-Nan Wang and Don Zagier for
valuable discussions.
BH would like to thank the Max Planck Institute for Mathematics for
hospitality in July.
JG, AK, KS, and XW thank the organisers of ``School and Workshop on
Gauge Theories and Differential Invariants'' where part of this work
was finished, and where some results were presented.
This work has also been presented in SISSA, Oxford and USTC.
%
The work of J.G. is supported in part by the Fonds National Suisse,
subsidy 200021-175539 and by the NCCR 51NF40-182902 ``The Mathematics
of Physics'' (SwissMAP). The work of BH is supported by the National
Thousand-Young-Talents Program of China.

\appendix

\section{A Lemma}
\label{sc:lemma}

To demonstrate that the blowup equations can determine all refined BPS
invariants in Section~\ref{sc:BPS}, we need the following
lemma. Recall in section \ref{sc:41}, we defined
\begin{equation}
  f_{(j_L,j_R)}(q_1,q_2) = \frac{\chi_{j_L}(q_L)\chi_{j_R}(q_R)}
  {(q_1^{1/2}-q_1^{-1/2})(q_2^{1/2}-q_2^{-1/2})}
\end{equation}
and
\begin{equation}
  Bl_{(j_L,j_R,R)}(q_1,q_2) = f_{(j_L,j_R)}(q_1,q_2/q_1)q_1^R +
  f_{(j_L,j_R)}(q_1/q_2,q_2)q_2^R - f_{(j_L,j_R)}(q_1,q_2),
\end{equation}
where $\chi_j(q)$ is the $SU(2)$ character.
\begin{lemma}\nonumber
  $\forall R\in \IZ/2$, $Bl_{(j_L,j_R,R)}(q_1,q_2)$ are linearly
  independent with only exceptions at
  $Bl_{(0,0,1/2)}(q_1,q_2)=Bl_{(0,0,-1/2)}(q_1,q_2)=Bl_{(0,1/2,0)}(q_1,q_2)=0$.
\end{lemma}
\emph{Proof:} For a generic fixed $R$ and a finite set $J$ of spin
$(j_L,j_R)$ which satisfy $2j_L+2j_R+1\equiv 2R\mod 2$, we need to
prove if
\begin{equation}
  \sum_{(j_L,j_R)\in J}x_{(j_L,j_R)}Bl_{(j_L,j_R,R)}(q_1,q_2)=0,
\end{equation}
then all coefficients $x_{(j_L,j_R)}$ must vanish. Since $J$ is
finite, there exist maximum for $j_L$ and $j_R$, denoted as
$j_L^{\rm max}$ and $j_R^{\rm max}$. We expand $J$ to the set of all
spins on the rectangle from $(0,0)$ to
$(j_L^{\rm max},j_R^{\rm max})$. On such spin rectangle, we can define
a strict total order of $(j_L,j_R)$. Then one can use descending
method to prove the coefficients $x_{(j_L,j_R)}$ vanish one by
one. Such procedure was actually already given in section 6.2 in
\cite{Huang:2017mis}. For $R=1/2$ or $-1/2$, the lowest spin in such
order is $(0,0)$ and the value of $Bl$ function is 0, and for $R=0$,
the lowest spin in such order is $(0,1/2)$ and the value of $Bl$
function is 0 too. These are the only exceptions for linear
independence.

\section{Functional equations for theta functions of even unimodular lattices}
\label{appB}
The unity blowup equations for one E-string elliptic genus (\ref{fE1})
give a set of interesting functional equations for $E_8$ theta
function. Here we prove $E_8$ theta function is the unique solution
for such equations up to a free function of $\tau$. This statement can
be generalized to the theta function associated to any positive
definite even unimodular lattice that is generated by roots. The
generalization and proof were shown to us by Don Zagier.

\begin{prop}
Let $\Lambda$ be a positive definite even unimodular lattice that is generated by its roots, and let $f$ be a holomorphic function on $\mathfrak{H}\times\Lambda_{\IC}$ satisfying the functional equation
\be\label{B1}
\ba
\theta_1(\et)\theta_1(\alpha\cdot m &+\et)f(\tau,m+\eq\alpha)
  -\theta_1(\eq)\theta_1(\alpha\cdot m+\eq)f(\tau,m+\et\alpha)\\
    &=\theta_1(\et-\eq)\theta_1(\alpha\cdot m+\eq+\et)f(\tau,m),
\ea
\ee
for all roots $\alpha$ of $\Lambda$ and all $\epsilon_{1},\epsilon_{2}\in\IC$. Then $f$ is a multiple (depending only on $\tau$) of the theta series
\be
\theta_{\Lambda}(\tau,m)=\sum_{w\in\Lambda}e^{2\pi \ri(w\cdot w/2+ m\cdot w)}.
\ee
\end{prop}
\emph{Proof}: Fix $\tau$ and also a root $\alpha$ and a vector $m_0\in\Lambda$ with $m_0\cdot\alpha=0$, and set $F(\tau,\lambda)=f(m_0+\lambda\alpha)$, $\lambda\in\IC$. Using $\alpha\cdot\alpha=2$ and setting $h_1=\lambda+\eq$, $h_2=\lambda+\et$, $h_3=\lambda$, we can write (\ref{B1}) in a symmetric form,
\be
\sum_{i\, (\mathrm{mod}\, 3)}\theta_1(h_{i+1}-h_{i-1})\theta_1(h_{i+1}+h_{i-1})F(h_i)=0,\quad (\textrm{any}\ \{h_i\}_{i\, (\mathrm{mod}\, 3)}\in \IC^3).
\ee
Here the $\tau$ dependence is implicit. Changing $h_1$ to $h_1+1$ and $h_1+\tau$ with $h_2$ and $h_3$ fixed, we find $F(h+1)=F(h)$ and $F(h+\tau)=q^{-1}\xi^{-2} F(h)$, where $q=e^{2\pi \ri \tau}$, $\xi=e^{2\pi \ri h}$. Thus,
\be\label{B5}
f(m+\alpha)=f(m),\quad\quad f(m+\alpha\tau)=e^{-2\pi \ri (\tau+\alpha\cdot m)}f(m).
\ee
Since $\Lambda$ is even unimodular and generated by all roots $\alpha$, the first equation of (\ref{B5}) implies that we can write $f(\tau,m)$ as Fourier expansion $\sum_{w\in\Lambda}c_w(q)q^{w\cdot w/2}e^{2\pi \ri m\cdot w}$ for some coefficients $c_w(q)$. The second equation of (\ref{B5}) implies that $c_{w+\alpha}(q)=c_w(q)$ for all $w$ and $\alpha$, so $c_w(q)=c_0(q)$ and $f(\tau,m)=c_0(q)\theta_{\Lambda}(\tau,m)$.

\printindex


\bibliographystyle{JHEP}
\bibliography{blowupEM}

\end{document}